\documentclass[12pt,letterpaper]{JHEP3}

\usepackage{amsmath,epsfig,array,undertilde}
\usepackage[latin1]{inputenc}
\usepackage{amssymb,amsfonts}

\title{Quasi-conformal actions, quaternionic discrete series
and twistors: $SU(2,1)$ and $G_{2(2)}$}

\preprint{LPTENS-07-25}

\author{M.~G\"unaydin$^{a,b}$, A.~Neitzke$^b$, O.~Pavlyk$^c$
and B.~Pioline$^{\rm d, e}$\footnote{E-mail: murat@phys.psu.edu,
neitzke@ias.edu, pavlyk@wolfram.com, pioline@lpthe.jussieu.fr}\\
$^a$ Physics Department, Pennsylvania State University,\\
University Park, PA 16802, USA\\

$^b$ School of Natural Sciences, Institute for Advanced Study, Princeton,
NJ, USA\\

$^c$ Wolfram Research Inc., 100 Trade Center Dr., Champaign, IL 61820, USA\\

$^d$ Laboratoire de Physique Th\'eorique et Hautes
Energies\footnote{Unit\'e mixte de recherche du CNRS UMR 7589},~
Universit\'e Pierre et Marie Curie - Paris 6,
4 place Jussieu, F-75252 Paris cedex 05 \\

$^e$ Laboratoire de Physique Th\'eorique de l'Ecole Normale
Sup\'erieure\footnote{Unit\'e mixte
de recherche du CNRS UMR 8549}~,\\
24 rue Lhomond, F-75231 Paris cedex 05
}

\abstract{Quasi-conformal actions were introduced in the physics
literature as a generalization of the familiar fractional linear
action on the upper half plane, to Hermitian symmetric tube domains
based on arbitrary Jordan algebras, and further to arbitrary
Freudenthal triple systems.  In the mathematics literature,
quaternionic discrete series unitary representations of real reductive
groups in their quaternionic real form were constructed as degree 1
cohomology on the twistor spaces of symmetric quaternionic-K\"ahler
spaces. These two constructions are essentially identical, as we show
explicitly for the two rank 2 cases $SU(2,1)$ and $G_{2(2)}$. We 
obtain explicit results for certain principal series,
quaternionic discrete series and minimal representations of these
groups, including formulas for the lowest $K$-types in various
polarizations.  We expect our results to have applications to
topological strings, black hole micro-state counting and 
to the theory of automorphic forms.
}

\newcommand{\color}[6]{}

\newcommand{\pa}{\partial}
\newcommand{\nn}{\nonumber}
\newcommand{\eps}{\epsilon}
\newcommand{\IR}{\mathbb{R}}
\newcommand{\IN}{\mathbb{N}}
\newcommand{\IC}{\mathbb{C}}

\newcommand{\Tr}{\mbox{Tr}}

\newcommand{\tzeta}{{\tilde\zeta}}
\newcommand{\tze}{{\tilde\zeta}}
\newcommand{\ze}{{\zeta}}
\newcommand{\txi}{{\tilde\xi}}

\def\bea{\begin{eqnarray}}
\def\eea{\end{eqnarray}}
\def\be{\begin{equation}}
\def\ee{\end{equation}}
\def\ba{\begin{align}}
\def\ea{\end{align}}
\def\bse{\begin{subequations}}
\def\ese{\end{subequations}}
\def\1F1{{}_1\!F_1}
\def\2F0{{}_2\!F_0}

\newcommand{\cC}{\mathcal{C}}

\newcommand{\cZ}{\mathcal{Z}}
\newcommand{\cS}{\mathcal{S}}
\newcommand{\cN}{\mathcal{N}}

\newcommand{\fg}{\ensuremath{\mathfrak{g}}}
\newcommand{\fp}{\ensuremath{\mathfrak{p}}}

\newcommand{\eq}{\begin{equation}}
\newcommand{\en}{\end{equation}}
\newcommand{\eqn}{\begin{eqnarray}}
\newcommand{\enn}{\end{eqnarray}}
\newcommand{\beq}{\begin{equation}}
\newcommand{\eeq}{\end{equation}}

\newcommand{\PP}{{\mathbb P}}
\newcommand{\C}{{\mathbb C}}
\newcommand{\R}{{\mathbb R}}
\newcommand{\Z}{{\mathbb Z}}
\newcommand{\cO}{{\mathcal O}}

\newcommand{\N}{{\mathcal N}}
\newcommand{\half}{\ensuremath{\frac{1}{2}}}

\newcommand{\norm}[1]{\lVert#1\rVert}
\newcommand{\ti}[1]{\textit{#1}}
\newcommand{\IP}[1]{\langle#1\rangle}
\newcommand{\cL}{{\mathcal L}}
\newcommand{\bas}{\backslash}

\newcommand{\moment}[1]{\utilde{#1}}

\begin{document}
%\maketitle \setcounter{tocdepth}{2}
%\tableofcontents

% delete this before submitting
% {\footnotesize{\tt{\begin{verbatim} $Id: quasimin.tex 1927 2009-02-16 22:36:41Z boris $ \end{verbatim}}}}

\numberwithin{equation}{section}

\section{Introduction and Summary}
Despite much recent progress, classifying the unitary representations of
a real reductive group $G$ remains a challenging task,
which has only been addressed in a few, typically low-rank cases,
including for example
$SL(2,\IR)$ \cite{MR0021942,MR0017282,MR0047055},
$SL(3,\IR)$ \cite{MR0222215,MR0383989},
$SU(2,1)$ \cite{Bars:1989bb}, $SU(2,2)$ \cite{MR645645}
and $G_{2(2)}$ \cite{MR1253210}.
Even in well-charted cases, the available mathematical description is
often not directly useful to physicists, who are in general more
interested in explicit differential operator realizations than
abstract classifications. The goal of this paper is to give an
explicit description of aspects of principal and discrete series
representations (and continuations of the discrete series) which arise
when $G$ is a quaternionic real form of a semi-simple group.  For the
sake of explicitness, we restrict to the rank 2 case -- so $G=SU(2,1)$
or $G=G_{2(2)}$.  Quaternionic real forms arise as symmetries of
supergravity theories with 8 supercharges in three
dimensions~\cite{Gunaydin:1983rk,Gunaydin:1983bi,Cecotti:1989qn,deWit:1990na,
deWit:1992wf}, and therefore as spectrum-generating symmetries for
black holes in four dimensional supergravity; a detailed discussion of
this relation, which indeed motivated our interest in the first place,
can be found in \cite{Gunaydin:2005mx,Gunaydin:2006bz,Pioline:2006ni}
and references therein.

The simplest discrete series representations\footnote{Discrete series representations arise as discrete summands in the
spectral decomposition of $L^2(G)$ under the left action of $G$.
A basic result of Harish-Chandra \cite{MR0219666,MR0219665}
states that $G$ admits discrete series representations if and only if
its maximal compact subgroup $K$ has the same rank as $G$ itself.}
arise when $K=U(1)\times M$, so that $G/K$ is a Hermitian
symmetric domain.
The simplest example is $G/K = SL(2,\IR)/U(1)$: $G$ acts holomorphically
on the upper half-plane by fractional linear transformations
$\tau\to (a\tau+b)/(c\tau+d)$, which
preserve the K\"ahler potential $K=-\log[(\tau-\bar\tau)]$ up to
K\"ahler transformations.
This construction can be extended to all
Hermitian symmetric tube domains using the language of Jordan algebras.
Indeed, consider the ``upper half plane'' $\tau\in J+iJ^+$,
where $J$ is a Euclidean
Jordan algebra $J$ of degree $n$ and $J^+$  is the domain of positivity
of $J$, equipped with the K\"ahler
potential $K=-\log \mathcal{N}(\tau-\bar\tau)$, where
$\mathcal{N}$ is the norm form of $J$ \footnote{ The generalized upper half planes associated to Jordan algebras are sometimes called K\"ocher half-spaces since K\"ocher pioneered their
study \cite{MR1718170}. In particular, he introduced the linear fractional groups of Jordan algebras 
\cite{MR0214630}, which were interpreted as conformal groups of generalized
spacetimes defined by Jordan algebras and extended to Jordan superalgebras in the physics literature \cite{Gunaydin:1975mp,Gunaydin:1979df,Gunaydin:1989dq}. The terminology stems from the well-known action 
of the conformal group $SO(4,2)$ on Minkowski spacetime, which arises
when $J$ is the Jordan algebra of $2\times 2$
Hermitian matrices \cite{Gunaydin:1975mp}. Choosing $J=\IR$
leads instead to our original example $G/K=SL(2,\IR)/U(1)$.
}. The corresponding metric is invariant
under a non-compact group $G={\rm Conf}(J)$, the ``conformal
group''
associated to $J$,
acting holomorphically by generalized fractional linear
transformations on $\tau$
\cite{Gunaydin:1992zh, Gunaydin:2000xr}.
The resulting space is the Hermitian symmetric
tube domain $G/K={\rm Conf}(J)/\widetilde{{\rm Str}}(J)$, where
$\widetilde{{\rm Str}}(J)=U(1)\times \widetilde{{\rm Str}}_0(J)$
is the compact real form of the
structure group\footnote{The structure group ${\rm Str}(J)$
leaves the norm form $\mathcal{N}$ of the Jordan algebra $J$ invariant up to an overall scaling. It decomposes
as an Abelian factor $\IR$ times the reduced structure group ${\rm Str}_0(J)$ which leaves the norm invariant.}
of $J$, which coincides with
the maximal compact subgroup of $G$.
The action of $G$ on sections of holomorphic
vector bundles over $G/K$ leads to the
holomorphic discrete series representations.

The next simplest case, of interest in this paper, arises when $G$ is in
its quaternionic real form, such that $K=SU(2)\times M$, and its Lie algebra
decomposes as
$\fg=\mathfrak{su}(2)\oplus \mathfrak{M}\oplus (2,V)$, where $V$
is a pseudo-real representation of $M$. The symmetric
space $G/K$, of real dimension $4d$, is now a quaternionic-K\"ahler
space, and does not generally admit a $G$-invariant complex
structure. The twistor space $\cZ = G/U(1)\times M$, a bundle over
$G/K$ with fiber $\C\PP^1 = SU(2)/U(1)$, does however carry a
$G$-invariant complex structure. In \cite{MR1421947} this complex
structure was exploited to construct a family of representations
$\pi_k$ of $G$, labeled by $k \in \Z$: namely, $\pi_k$ is the sheaf
cohomology $H^1(\cZ,\cO(-k))$ of a certain line bundle $\cO(-k)$ over
$\cZ$.\footnote{For a definition of sheaf cohomology see \ti{e.g.} \cite{MR95d:14001}.}
 $\pi_k$ is a representation of $G$, with Gelfand-Kirillov
(functional) dimension $2d+1$, equal to the complex dimension of
$\cZ$.  For $k \ge 2d+1$, $\pi_k$ are discrete series unitary
representations of $G$, called ``quaternionic discrete series.''

The $\pi_k$ for $k<2d+1$ are also of interest.  In \cite{MR1421947}
special attention was paid to quaternionic groups of type
$G_2,D_4,F_4,E_6,E_7,E_8$, such that $d=3f+4$ with $f=-\frac{2}{3}, 0,
1, 2, 4, 8$.  For these groups, it was shown that
$\pi_k$ is irreducible and unitarizable even for $k \geq d+1$ (although
it no longer belongs to the quaternionic discrete series for $k < (2d+1)$).
Moreover, for selected smaller values of $k$, namely $k=3f+2$,
$k=2f+2$ and $k=f+2$, $\pi_k$ was shown to be reducible but to admit a
unitarizable submodule $\pi'_k$, of smaller Gelfand-Kirillov
dimension, $2d=6f+8,\, 5f+6,\, 3f+5=d+1$, respectively.
The smallest of these representations is the ``minimal''
or ``ladder'' representation of $G$; the latter name refers to
the structure of its $K$-type decomposition\footnote{We 
remind the reader of a few definitions.  Suppose $\rho$ is
a representation of a real Lie group $G$ on a vector space $V$.
Let $V_K$ denote the space of $K$-finite vectors in
$V$ (i.e. those which generate a finite-dimensional
subspace of $V$ under the action of the maximal compact
subgroup $K\subset G$).  Then any representation of $K$ which occurs
in $V_K$ is called a $K$-type of $\rho$.  Parameterizing the $K$-types by
their highest weights $\mu$, a ``lowest $K$-type'' is one
with the minimal value of $\norm{\mu + 2 \rho_K}^2$, where $\rho_K$ is half 
the sum of positive roots of $K$.  A discrete
series representation always has a unique lowest $K$-type.
If the lowest $K$-type is the trivial representation,
then it is also called the spherical vector and
the representation $\rho$ is deemed spherical.  The Gelfand-Kirillov dimension
measures the growth of the multiplicities of the $K$-types; morally it counts the number of variables
$x_i$ needed to realize $V$ as a space of functions. Finally, if all $K$-types
in $\rho$ occur with multiplicity 1 and lie along a ray in the weight space of $K$,
then $\rho$ is called a ladder representation.};
a review of minimal representations can be found in \cite{minrep-review}.

Independently of these mathematical developments, it was shown
in \cite{Gunaydin:2000xr}
that the conformal realization of the group ${\rm Conf}(J)$ attached to
a Jordan algebra $J$ of degree 3, preserving a generalized
light-cone $\mathcal{N}_3(\tau-\bar\tau) = 0$,
could be extended to an action of a larger non-compact
group,  preserving a ``quartic light-cone''
\be
N_4\equiv I_4(\Xi-\bar\Xi)+(\alpha-\bar\alpha+ \langle \Xi,\bar\Xi
\rangle)^2 = 0\ ,
\ee
where $\Xi=(\xi^I,\txi_I)=(\xi^0,\xi^\Lambda,\txi_\Lambda,\txi_0)$ 
is an element of the Freudenthal triple system
$\mathcal{F}=\IC \oplus J_{\IC} \oplus J_{\IC} \oplus \IC$ associated to $J$,
$\langle \Xi,\bar\Xi \rangle$ a symplectic pairing invariant under
the linear action of  ${\rm Conf}(J)$ on $\Xi$, $I_4$ a quartic polynomial
invariant under this same action\footnote{$I_4$ is 
expressible in terms of $\mathcal{N}_3$ via
$8I_4(\Xi)=\left( \xi^0\txi_0 - \xi^\Lambda \txi_\Lambda\right)^2 
-4 \xi^\sharp_\Lambda \txi_\sharp^\Lambda + 4 \xi^0
\,\cN_3(\xi^\Lambda) + 4 \txi_0\,\cN_3(\txi_\Lambda)$, where
$\xi_\sharp$ is related to $\xi$ by the (quadratic) adjoint map, defined by 
$(\xi^\sharp)^\sharp = \cN(\xi)\,\xi$.}
and  $\alpha$
an additional complex variable of homogeneity degree two.
This larger group was called the ``quasi-conformal group'' ${\rm QConf}(J)$
attached to the Jordan algebra $J$, or more appropriately
to the Freudenthal triple system $\mathcal{F}$; its
geometric action on $(\Xi,\alpha)$ was called the ``quasi-conformal realization''.
When $J$ is Euclidean, ${\rm QConf}(J)$
is a non-compact group in its quaternionic real form; other
real forms can be similarly obtained from Jordan algebras of indefinite
signature~\cite{Gunaydin:2000xr}.
Moreover, it was observed in \cite{Gunaydin:2001bt} for $G=E_{8(8)}$,
and generalized to other simple groups in \cite{Gunaydin:2004md,
Gunaydin:2005zz,Gunaydin:2006vz}, that this quasi-conformal
action on $2d+1$ variables could be reduced to a representation
on functions of $d+1$ variables, obtained by first adding one more variable
(symplectizing) and then quantizing the resulting $2d+2$-dimensional symplectic space.
This smaller representation was identified as the minimal representation
of $G={\rm QConf}(J)$.

Let us now comment briefly on the physics.
Euclidean Jordan algebras $J$ of degree three made
their appearance in supergravity a long time
ago \cite{Gunaydin:1983rk,Gunaydin:1983bi,Gunaydin:1984ak}.
Maxwell-Einstein supergravity theories with $\cN=2$ supersymmetry in $D=5$
and a symmetric moduli space $G/H$ such that $G$ is a symmetry group of the action are in one-to-one correspondence
with Euclidean Jordan algebras $J$ of degree three. Their
symmetry group in $D=5$ is simply the reduced structure group
${\rm Str}_0(J)$ of $J$. Upon reduction to $D=4$ and $D=3$,
the symmetry groups are extended to the conformal
${\rm Conf}(J)$ and quasi-conformal groups ${\rm QConf}(J)$, respectively.
The corresponding moduli spaces are given by the quotient of the
respective symmetry group by its maximal compact subgroup, and are
special real, special K\"ahler and quaternionic-K\"ahler manifolds,
respectively~\cite{Gunaydin:1983rk,Gunaydin:1983bi,deWit:1992wf}.
An explicit description of the quasi-conformal action
${\rm QConf}(J)$ of $D=3$ Maxwell-Einstein supergravity theories with
symmetric target spaces was obtained in \cite{Gunaydin:2005zz}.
Finally, it was observed in \cite{Gunaydin:2006bz} that the minimal
unitary representation of ${\rm QConf}(J)$ is closely related to the
vector space to which the topological string partition function naturally
belongs.

The minimal representation has also appeared in another physical context:
in \cite{Gunaydin:2007vc}, a connection between the
harmonic superspace (HSS) formulation of $\cN=2$, $d=4$ supersymmetric
quaternionic K\"ahler sigma models that couple to $\cN=2$ supergravity
and the minimal unitary representations of their isometry groups was
established. In particular, for $\cN=2$ sigma models with quaternionic
symmetric target spaces of the form ${\rm QConf}(J)/\widetilde{\rm
Conf}(J)\times SU(2)$ there exists a one-to-one mapping between the
quartic Killing potentials that generate the isometry group ${\rm
QConf}(J)$ under Poisson brackets in the HSS formulation, and the
generators of the minimal unitary representation of ${\rm QConf}(J)$.
It would be important to understand
physically how the minimal representation may arise by quantizing the
sigma-model in harmonic superspace.

\medskip

The main goal of the present work is to explain the relation between
the twistorial construction of the quaternionic discrete series in
\cite{MR1421947} and the quasi-conformal actions discovered in
\cite{Gunaydin:2000xr}, and moreover to elucidate the sense in which
the minimal representation is obtained by ``quantizing the
quasi-conformal action''. While our results are unlikely to cause
any surprise to the informed mathematician, we hope that our exegesis
of \cite{MR1421947} will be useful to physicists, \ti{e.g.} in subsequent
applications to supergravity and black holes, and possibly to
mathematicians too, \ti{e.g.} in obtaining explicit formulae for
automorphic forms along the lines of \cite{MR2094111}.

As the main body of
this paper is fairly technical, we summarize its content below,
including some open problems and possible applications:

\begin{enumerate}
\item  The key observation is that the variables $(\Xi,\alpha)$
of the quasi-conformal realization have a natural interpretation as
\textit{complex} coordinates on the twistor space $\cZ=G/U(1)\times M$
over the quaternionic-K\"ahler space $G/SU(2)\times M$, adapted
to the action of the Heisenberg algebra in the nilpotent radical
of the Heisenberg parabolic subgroup of $G$. The
logarithm of the ``quartic norm'' $N_4$ provides
a K\"ahler potential \eqref{kzxi} for the
$G$-invariant Einstein-K\"ahler metric on $\cZ$.\footnote{This follows
by specializing analysis
performed in \cite{Neitzke:2007ke} of general ``dual'' quaternionic-K\"ahler
spaces to symmetric spaces. It will be rederived below using purely group
theoretic methods (in fact, this approach was a
useful guide for the general analysis in \cite{Neitzke:2007ke}).}.
Using the Harish-Chandra decomposition, we also construct the complex
coordinates adapted to another Heisenberg algebra related by
a Cayley-type transform, whose center is a compact generator rather
than a nilpotent one. These coordinates are the analogue of the
Poincar\'e disk coordinates for $SL(2,\IR)/U(1)$, and it would
be interesting to give a Jordan-type description of
the corresponding K\"ahler potential, given in \eqref{kzpq} for $G = SU(2,1)$.

\item It is also possible to view $(\Xi,\alpha)$  as
\textit{real} coordinates on
$G/P$, where $P$ is a parabolic subgroup of $G$ with Heisenberg radical.
This $G/P$ arises as a piece of the boundary of $\cZ$ or of $G/K$.
The ``quartic light-cone'' then provides the causal structure
on the boundary, as shown in Section \ref{causala2} for $G = SU(2,1)$.

\item The quasi-conformal action admits
a continuous deformation by a parameter $k \in \C$, corresponding to the action of
$G$ on sections of a line bundle over $G/P$ induced from a character
of $P$. This action provides a degenerate principal series representation,
which is manifestly unitary for $k \in 2 + i\IR$ ($SU(2,1)$) or $k \in 3+ i \IR$ ($G_{2(2)}$).
We tabulate the formulas for the infinitesimal action of $\fg$ in this representation,
and determine the spherical vector.

\item When $k$ is an integer, one can also consider sections on $G/P$
which can be (in an appropriate sense) extended holomorphically into $\cZ$.
Thus, the quasi-conformal action with $\Xi, \alpha$ complex as in i)
leads to a differential operator realization of the action of $G$ on these sections.
When $G = SU(2,1)$ and $k \ge 3$, as explained in \cite{MR1421947}, this action
gives quaternionic discrete
series representations; we explicitly compute the $K$-finite vectors of this submodule
in the principal series.  When $G = G_{2(2)}$ we similarly study the $K$-finite vectors
of the principal series and identify a natural subset which has the same $K$-type decomposition 
as the quaternionic discrete series.  Moreover, in this case we
study also the more intricate picture described in \cite{MR1421947} for smaller values of $k$;
while the results there do not strictly apply to $G_{2(2)}$, we nevertheless find constraints 
defining ``small'' invariant submodules of the principal series at the expected values of $k$.

\item Given a class in $H^1(\cZ,\cO(-k))$, the Penrose transform
produces a section of a vector bundle over the quaternionic-K\"ahler
base $G/K$ annihilated by some ``quaternionic'' differential operators
\cite{MR664330,MR1038279, Neitzke:2007ke}. For $G = SU(2,1)$ and $k$ even,
we argue formally but explicitly in Section \ref{matpenrose} that computing the
Penrose transform of $\Psi$ is equivalent to evaluating a matrix
element between $\Psi$ and the lowest $K$-type of the quaternionic
discrete series. A similarly explicit understanding of the Penrose
transform for $k$ odd remains an open problem, which would require a
proper prescription for dealing with the branch cuts in the formula of
\cite{Neitzke:2007ke}.

\item The quasi-conformal action on $\cZ$ can be lifted to a
tri-holomorphic action on the hyperk\"ahler cone (or Swann space) $\cS
= \IR \times G/M$, locally isomorphic to the smallest nilpotent coadjoint
orbit of $G_\C$. In
particular, the action of $G$ on $\cS$ preserves the holomorphic
symplectic form.  The minimal representation of $G$ can be viewed as
the ``holomorphic quantization'' of $\cS$. We show explicitly in
Section \ref{minclasa2} for $SU(2,1)$ and \ref{hkcliftg2} for $G_{2,2}$ that
the leading differential symbols of the generators of the minimal
representation are equal to the holomorphic moment maps of the action
of $G$ on $\cS$, and identify the corresponding semi-classical limit.

\item We determine explicitly the lowest $K$-type of the minimal
representation, generalizing the analysis of \cite{Kazhdan:2001nx} to
these two quaternionic groups. For $G_{2(2)}$, the lowest $K$-type
in the real polarization \eqref{g2minreplkr} bears strong similarities
to the result found in \cite{Kazhdan:2001nx} for simply laced split
groups, while the wave function in the complex (upside-down)
polarization \eqref{g2minreplkr} is analogous to the topological
string wave function.  (Such a relation is not unexpected, given the
results of \cite{Gunaydin:2006bz}, which showed that the holomorphic
anomaly equations of the topological string can be naturally explained
in terms of the minimal representation.)  We show further that the
semi-classical limit of the lowest $K$-type wave function yields the
generating function for a holomorphic Lagrangian cone inside the
holomorphic symplectic space $\cS$, invariant under the holomorphic
action of $G$\footnote{This Lagrangian cone was already instrumental
in \cite{MR2094111}, in extending the real spherical vector found in
\cite{Kazhdan:2001nx} to the adelic setting.}.  It would be very
interesting to formulate the hyperk\"ahler geometry of $\cS$ in terms
of this Lagrangian cone, by analogy to special K\"ahler geometry.

\item Deviating slightly from the main subject of this paper, we also give an
explicit construction of
a degenerate principal series of $G_{2(2)}$ induced from a
parabolic subgroup $P_3$ (different from the Heisenberg parabolic),
with nilpotent radical of order 3. We also describe a polarization
of the minimal representation adapted to this parabolic subgroup.
We expect that this construction will have applications to the physics
of supersymmetric black holes in 5 dimensions,
along the lines discussed in \cite{Gao:2007mi,Gao:2008hw}.
\end{enumerate}

The organization of this paper is as follows: In Sections 2 and 3,
for the two rank 2 quaternionic groups $G = SU(2,1)$ and $G = G_{2(2)}$
successively, we give explicit parametrizations of the quaternionic-K\"ahler
homogeneous spaces $G/K$ and their twistor spaces, provide explicit differential operator
realizations of the principal series, quaternionic discrete series and minimal
representations, and compute their
spherical or lowest $K$-type vectors.

Throughout this paper we work mostly at the level of the Lie algebras.
We are not careful about the discrete factors in any of the various
groups that appear.  We are also not careful about the precise
globalizations of the representations we consider (so we do not worry
about whether we study $L^2$ functions, smooth functions,
hyperfunctions etc); for the most part we are really considering only
the $(\fg, K)$-modules consisting of the $K$-finite vectors.  We
emphasize concrete formulas even if they are formal, with the idea
that these formal manipulations may be related to ones frequently
carried out by physicists studying the topological string, in light of
the relations between group representations and topological strings
identified in \cite{Gunaydin:2006bz}.

\medskip

It is our pleasure to thank A.~Waldron for collaboration at an initial
stage of this project, and W.~Schmid, P.~Trapa, D.~Vogan and
M.~Weissman for correspondence and discussions. M.G. and B.P. express
their gratitude to the organizers of the program ``Mathematical
Structures in String Theory'' that took place at KITP in the Fall of
2005, where this study was initiated. The research of B.P. is
supported in part by the EU under contracts MTRN--CT--2004--005104,
MTRN--CT--2004--512194, by ANR (CNRS--USAR) contract No
05--BLAN--0079--01.  The research of A.~N. is supported by the Martin
A. and Helen Chooljian Membership at the Institute for Advanced Study
and by NSF grant PHY-0503584. The research of M.G.  was supported in
part by the National Science Foundation under grant number PHY-0555605
and the support of the Monell Foundation during his sabbatical stay at
IAS, Princeton, is gratefully acknowledged. Any opinions, findings and
conclusions or recommendations expressed in this material are those of
the authors and do not necessarily reflect the views of the National
Science Foundation.

\section{$SU(2,1)$}

\subsection{Some group theory}
The non-compact Lie group $G=SU(2,1)$ is defined as
the group of unimodular transformations of $\C^3$ which
preserve a given hermitian metric $\eta$
with signature $(+,+,-)$. It is convenient to choose
\be
\label{eta21}
\eta=\begin{pmatrix}&&1\\&1&\\1&&\end{pmatrix}
\ee
The Lie algebra $\fg= su(2,1)$ consists of traceless matrices such that
$\eta X + X^{\dagger} \eta=0$. This condition is solved by
\be
\label{su21X}
X = \begin{pmatrix}
\underline{H} + i \underline{J}/3 &
\underline{E}_p - i \underline{E_q}, & i \underline{E} \\
\underline{F}_p + i \underline{F}_q & -2i \underline{J}/3 & -(
\underline{E}_p + i \underline{E}_q)\\
-i \underline{F} & -(\underline{F}_p - i \underline{F}_q) & -\underline{H}
+ i \underline{J}/3
\end{pmatrix} \equiv \underline{X}_i\, X_i
\ee
where $\{\underline{X}_i\}=\{\underline{H},\underline{J},\underline{E}_p,
\underline{E}_q,\underline{E},\underline{F}_p,\underline{F}_q,\underline{F}\}$
are the real coefficients of the generators $\{X_i\}=\{H,J,{E_p},{E_q},{E}$,${F_p},{F_q},{F}\}$
in $\fg$. The latter are represented
by anti-hermitean operators in any unitary representation.
They obey the commutation relations (consistent with the
matrix representation above)
\be
\begin{array}{l!{\hspace{10mm}}l!{\hspace{10mm}}l!{\hspace{10mm}}l}
\left[{E_p},{E}\right]=0, &
\left[{E_q},{E}\right]=0, &
\left[{F_p},{F}\right]=0, &
\left[{F_q},{F}\right]=0, \\
\left[{E_p},{E_q}\right]= -2  {E}, &
\left[{F_p},{F_q}\right] =  2{F}, &
\left[J,{E_p}\right]  =- {E_q}, &
\left[J,{F_p}\right] =- {F_q}, \\
\left[J,{E_q}\right]  = {E_p}, &
\left[J,{F_q}\right] = {F_p}, &
 \left[H,{E_p}\right]  = {E_p}, &
\left[H,{F_p}\right] =- {F_p}, \\
 \left[H,{E_q}\right] = {E_q}, &
\left[H,{F_q}\right] =- {F_q}, &
 \left[H,{E}\right]  = 2{E},   &
\left[H,{F}\right] =- 2{F}, \\
 \left[{E_p},{F_p}\right] = H,    &
\left[{E_q},{F_q}\right] = H, &
 \left[{E_p},{F_q}\right] = 3J,   &
\left[{E_q},{F_p}\right]=- 3J, \\
 \left[{E},{F_p}\right] = {E_q},   &
\left[{F},{E_p}\right] =- {F_q}, &
 \left[{E},{F_q}\right] = -{E_p},  &
\left[{F},{E_q}\right] = {F_p}, \\
 \left[J,{F}\right] =0,      &
\left[J,{E}\right]  =0, &
 \left[{E},{F}\right] =  H, & \left[H,J\right] =0
\end{array}
\ee
The center of the universal enveloping algebra $U(\fg)$ is generated by the
quadratic Casimir
\be
C_2 = \frac14 H^2 - \frac34 J^2 + \frac14
( E_p F_p + F_p E_p + E_q F_q + F_q E_q ) + \frac12 (EF+FE)
\ee
and cubic Casimir
\be
\begin{split}
C_3 =& H^2\,J + J^3 -E\, (F_p^2 +F_q^2) + F(E_p^2 + E_q^2) \\
&+ (F_q\,E_p-F_p\,E_q)H +(4 E\,F-F_p\,E_p-F_q\,E_q-2)J
\end{split}
\ee
The Casimir operators take values \cite{Bars:1989bb}
\be
\label{cpq}
C_2=p+q+\frac13(p^2+pq+q^2)\ ,\quad
C_3=\frac{4i}{27}(p-q)(p+2q+3)(q+2p+3)
\ee
in a finite-dimensional representation of $SU(2,1)$ corresponding to a tensor
$T^{i_1\dots i_p}_{j_1\dots j_q}$ with $p$ upper and $q$ lower
indices (in particular, $(C_2,C_3)=(3,0)$ and $(4/3,80 i/27)$ for the adjoint
and fundamental representations, respectively). It is convenient to
continue to use the same variables $(p,q)$ defined in \eqref{cpq},
no longer restricted to integers, to label the infinite-dimensional
representations of $SU(2,1)$. Alternatively, one may define
the ``infinitesimal character''
\be
x_1=-\frac13(p+2q+3)\ ,\quad x_2=\frac13(p-q)\ ,\quad
x_3=\frac13(2p+q+3)
\ee
with $x_1+x_2+x_3=0$, such that the Weyl group of $SU(2,1)$ acts by
permutations of $(x_1,x_2,x_3)$. In a unitary representation, either
all $x_i$ are real, or one is real and the other two are complex
conjugates \cite{Bars:1989bb}. In the former case
one may choose to order $x_1\leq x_2\leq x_3$, corresponding to
$p\geq -1,q\geq -1$. We shall be particularly interested in
representations where $p=q$, such that the infinitesimal character
is proportional to the Weyl vector $(-1,0,1)$.

The generators $H,J$ generate a Cartan subalgebra of $\fg$. The
spectrum of the adjoint action of
\be
\mbox{Spec}(J)=\{0,\pm i\}\ ,\quad \mbox{Spec}(H)=\{0,\pm 1,\pm2\}
\ee
shows that $H$ and $J$ are non-compact and compact, respectively.
In fact, the generator $H$ gives rise to a ``real non-compact'' 5-grading
\be
\label{su21-5gnc}
\fg =
{F}\vert_{-2} \oplus
\{ {F_p}, {F_q} \}\vert_{-1} \oplus
\{ H, J \}\vert_0  \oplus
\{ {E_p}, {E_q} \}\vert_{1} \oplus
{E}\vert_{2}
\ee
where the subscript denotes the eigenvalue under $H$. In this
decomposition, each subspace is invariant under hermitian conjugation.
Moreover $J\oplus \{E,H,F\}$ generate a $U(1) \times SL(2,\IR)$
(non-compact) maximal subgroup of $G$. The remaining roots
arrange themselves into a pair of doublets of
$SL(2,\IR)$ with opposite charge under $J$,
\be
\begin{pmatrix}
{F_p} - i {F_q} & {E_p} - i {E_q} \\
{F_p} + i {F_q} & {E_p} + i {E_q} \end{pmatrix}
\ee
as shown on the root diagram \ref{su21rootdiag}.
The parabolic subgroup $P=L N$ with Levi $L=\IR \times U(1)$ generated by
$\{H,J\}$ and unipotent radical $N$ generated by $\{F_{p},F_{q},
F\}$, corresponding to the spaces with zero and negative grade in the
decomposition \eqref{5gradg2}, is known as the Heisenberg
parabolic subgroup, and will play a central r\^ole in all
constructions in this paper.

\FIGURE{\begin{picture}(0,0)%
\includegraphics{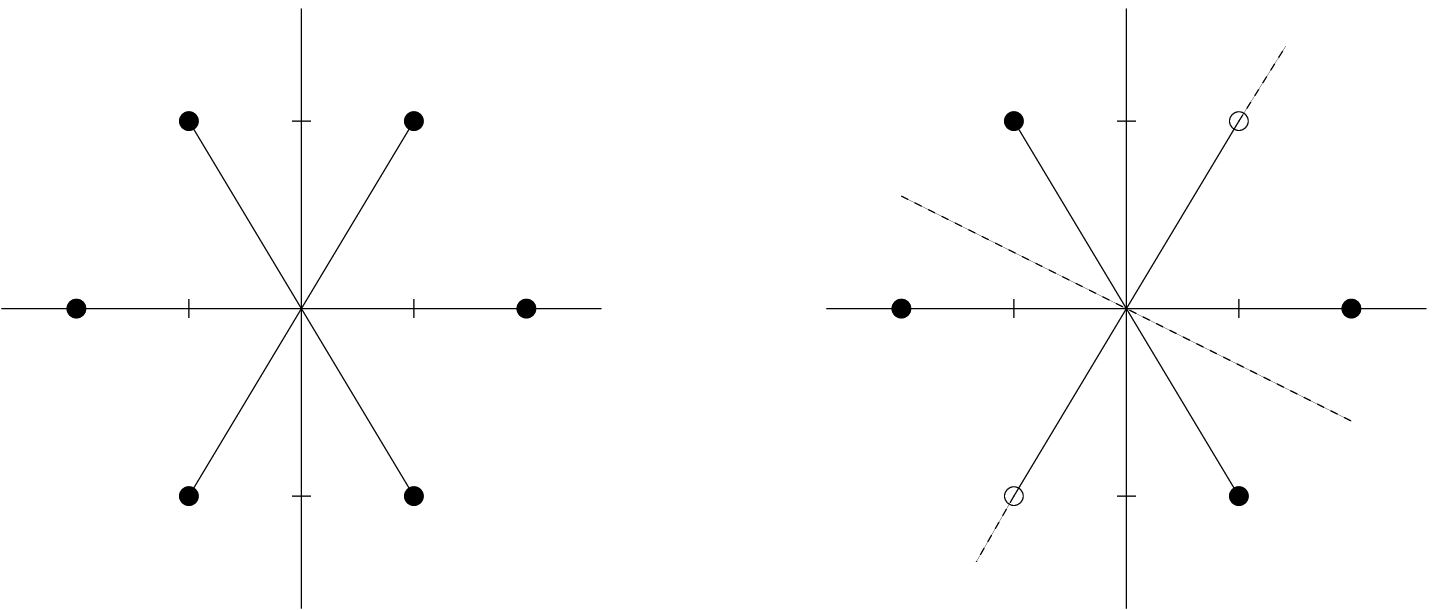}%
\end{picture}%
\setlength{\unitlength}{2368sp}%
\begingroup\makeatletter\ifx\SetFigFont\undefined%
\gdef\SetFigFont#1#2#3#4#5{%
  \reset@font\fontsize{#1}{#2pt}%
  \fontfamily{#3}\fontseries{#4}\fontshape{#5}%
  \selectfont}%
\fi\endgroup%
\begin{picture}(11424,4824)(1189,-6373)
\put(12376,-5161){\makebox(0,0)[lb]{\smash{{\SetFigFont{11}{13.2}{\rmdefault}{\mddefault}{\updefault}{\color[rgb]{0,0,0}$S$}%
}}}}
\put(1726,-4486){\makebox(0,0)[lb]{\smash{{\SetFigFont{11}{13.2}{\rmdefault}{\mddefault}{\updefault}{\color[rgb]{0,0,0}$F$}%
}}}}
\put(5326,-4486){\makebox(0,0)[lb]{\smash{{\SetFigFont{11}{13.2}{\rmdefault}{\mddefault}{\updefault}{\color[rgb]{0,0,0}$E$}%
}}}}
\put(5851,-3661){\makebox(0,0)[lb]{\smash{{\SetFigFont{11}{13.2}{\rmdefault}{\mddefault}{\updefault}{\color[rgb]{0,0,0}$H$}%
}}}}
\put(3826,-1861){\makebox(0,0)[lb]{\smash{{\SetFigFont{11}{13.2}{\rmdefault}{\mddefault}{\updefault}{\color[rgb]{0,0,0}$J$}%
}}}}
\put(10501,-1786){\makebox(0,0)[lb]{\smash{{\SetFigFont{11}{13.2}{\rmdefault}{\mddefault}{\updefault}{\color[rgb]{0,0,0}$J$}%
}}}}
\put(8701,-2461){\makebox(0,0)[lb]{\smash{{\SetFigFont{11}{13.2}{\rmdefault}{\mddefault}{\updefault}{\color[rgb]{0,0,0}$K_-$}%
}}}}
\put(11476,-2461){\makebox(0,0)[lb]{\smash{{\SetFigFont{11}{13.2}{\rmdefault}{\mddefault}{\updefault}{\color[rgb]{0,0,0}$J_+$}%
}}}}
\put(11401,-5761){\makebox(0,0)[lb]{\smash{{\SetFigFont{11}{13.2}{\rmdefault}{\mddefault}{\updefault}{\color[rgb]{0,0,0}$K_+$}%
}}}}
\put(8776,-5761){\makebox(0,0)[lb]{\smash{{\SetFigFont{11}{13.2}{\rmdefault}{\mddefault}{\updefault}{\color[rgb]{0,0,0}$J_-$}%
}}}}
\put(8251,-4486){\makebox(0,0)[lb]{\smash{{\SetFigFont{11}{13.2}{\rmdefault}{\mddefault}{\updefault}{\color[rgb]{0,0,0}$L_-$}%
}}}}
\put(11851,-4486){\makebox(0,0)[lb]{\smash{{\SetFigFont{11}{13.2}{\rmdefault}{\mddefault}{\updefault}{\color[rgb]{0,0,0}$L_+$}%
}}}}
\put(12301,-3586){\makebox(0,0)[lb]{\smash{{\SetFigFont{11}{13.2}{\rmdefault}{\mddefault}{\updefault}{\color[rgb]{0,0,0}$L_0$}%
}}}}
\put(11551,-1861){\makebox(0,0)[lb]{\smash{{\SetFigFont{11}{13.2}{\rmdefault}{\mddefault}{\updefault}{\color[rgb]{0,0,0}$J_3$}%
}}}}
\put(11926,-3811){\makebox(0,0)[lb]{\smash{{\SetFigFont{7}{8.4}{\rmdefault}{\mddefault}{\updefault}{\color[rgb]{0,0,0}$2i$}%
}}}}
\put(10351,-5461){\makebox(0,0)[lb]{\smash{{\SetFigFont{7}{8.4}{\rmdefault}{\mddefault}{\updefault}{\color[rgb]{0,0,0}$-i$}%
}}}}
\put(1726,-3811){\makebox(0,0)[lb]{\smash{{\SetFigFont{7}{8.4}{\rmdefault}{\mddefault}{\updefault}{\color[rgb]{0,0,0}$-2$}%
}}}}
\put(5326,-3811){\makebox(0,0)[lb]{\smash{{\SetFigFont{7}{8.4}{\rmdefault}{\mddefault}{\updefault}{\color[rgb]{0,0,0}$+2$}%
}}}}
\put(3751,-2536){\makebox(0,0)[lb]{\smash{{\SetFigFont{7}{8.4}{\rmdefault}{\mddefault}{\updefault}{\color[rgb]{0,0,0}$+i$}%
}}}}
\put(3751,-5536){\makebox(0,0)[lb]{\smash{{\SetFigFont{7}{8.4}{\rmdefault}{\mddefault}{\updefault}{\color[rgb]{0,0,0}$-i$}%
}}}}
\put(2626,-3811){\makebox(0,0)[lb]{\smash{{\SetFigFont{7}{8.4}{\rmdefault}{\mddefault}{\updefault}{\color[rgb]{0,0,0}$-1$}%
}}}}
\put(4426,-3811){\makebox(0,0)[lb]{\smash{{\SetFigFont{7}{8.4}{\rmdefault}{\mddefault}{\updefault}{\color[rgb]{0,0,0}$+1$}%
}}}}
\put(9226,-3811){\makebox(0,0)[lb]{\smash{{\SetFigFont{7}{8.4}{\rmdefault}{\mddefault}{\updefault}{\color[rgb]{0,0,0}$-i$}%
}}}}
\put(11101,-3811){\makebox(0,0)[lb]{\smash{{\SetFigFont{7}{8.4}{\rmdefault}{\mddefault}{\updefault}{\color[rgb]{0,0,0}$i$}%
}}}}
\put(8326,-3811){\makebox(0,0)[lb]{\smash{{\SetFigFont{7}{8.4}{\rmdefault}{\mddefault}{\updefault}{\color[rgb]{0,0,0}$-2i$}%
}}}}
\put(1501,-2461){\makebox(0,0)[lb]{\smash{{\SetFigFont{11}{13.2}{\rmdefault}{\mddefault}{\updefault}{\color[rgb]{0,0,0}$F_p+i F_q$}%
}}}}
\put(4801,-2461){\makebox(0,0)[lb]{\smash{{\SetFigFont{11}{13.2}{\rmdefault}{\mddefault}{\updefault}{\color[rgb]{0,0,0}$E_p+i E_q$}%
}}}}
\put(1501,-5761){\makebox(0,0)[lb]{\smash{{\SetFigFont{11}{13.2}{\rmdefault}{\mddefault}{\updefault}{\color[rgb]{0,0,0}$F_p-i F_q$}%
}}}}
\put(4801,-5761){\makebox(0,0)[lb]{\smash{{\SetFigFont{11}{13.2}{\rmdefault}{\mddefault}{\updefault}{\color[rgb]{0,0,0}$E_p-i E_q$}%
}}}}
\end{picture}%

\caption{Root diagram of $SU(2,1)$ with respect to the mixed
Cartan torus $H,J$ (left) and the compact Cartan torus $L_0,J$ (right).
The compact (resp. non-compact) roots are indicated by a white
(resp. black) dot.\label{su21rootdiag}}}

%\EPSFIGURE{su21root,height=6cm}{
%Root diagram of $SU(2,1)$ with respect to the mixed
%Cartan torus $H,J$ (left) and the compact Cartan torus $L_0,J$ (right).
%The compact (resp. non-compact) roots are indicated by a white
%(resp. black) dot.\label{su21rootdiag}}

For later purposes, it will be useful to introduce another basis of $\fg$ adapted
to a maximal {\it compact} subgroup $K = SU(2)\times U(1)$. We first
go to a compact basis for the $SL(2,\IR)$ factor generated by $E,F,H$:
\bea
L_0 &=& \frac12 (F-E) \ ,\quad K_0 = \frac14(2L_0+3J)\ ,\quad
L_{\pm} = -\frac{1}{2\sqrt{2}} \left(E + F \pm i H\right) \\
K_{\pm} &=& -\frac14 \left[ {E_p} \pm i {E_q} + ({F_p} \pm i {F_q})\right]\ ,
\quad
J_{\pm} = -\frac{1}{2\sqrt{2}} \left[{E_p}\mp i {E_q} - ({F_p} \mp i {F_q})\nn
\right]
\eea
Then $\{L_+,L_0,L_-\}$ and $\{K_+,K_0,K_-\}$ make two $SL(2,\IR)$ subalgebras,
with compact Cartan generators $L_0$ and $K_0$, respectively,
\bea
\left[ L_0, L_\pm \right] = \pm i L_\pm \ ,&\quad& [L_+,L_-]= - i L_0 \\
\left[ K_0, K_\pm \right] = \pm i K_\pm \ ,&\quad& [K_+,K_-]= - i K_0
\eea
Since $\mbox{Spec}(L_0)=\{0,\pm \frac{i}{2},\pm i\}$, $(L_0,J)$ now form
a Cartan torus, and $L_0$ gives rise to a new 5-grading
\be
\label{su21-5gh}
\fg =
L_- \vert_{-i} \oplus
\{ K_-, J_-  \}\vert_{-\frac{i}{2}} \oplus
\{ L_0, J \}\vert_0  \oplus
\{ J_+, K_+ \}\vert_{\frac{i}{2}} \oplus
L_+ \vert_{i}
\ee
where the subscript denotes the eigenvalue under $L_0$.
Unlike the 5-grading \eqref{su21-5gnc}, in \eqref{su21-5gh}
hermitean conjugation exchanges the positive and negative grade spaces.

Next, we perform a $\pi/3$ rotation of the root diagram, and define
\bea
J_3 &=& \frac14(F-E-3J)\ ,\quad
S = \frac34 (F-E+J) \\
J_{\pm\frac12,\pm\frac32} &=& \pm 2\sqrt{2} i L_{\pm}\ ,\quad
J_{\pm\frac12,\mp\frac32} = \sqrt{2} K_\mp
\eea
Then $\{J_+,J_3,J_-\}$ and $S$ generate
the compact subgroup $K = SU(2) \times U(1)$,\footnote{To be precise, 
$K = (SU(2) \times U(1)) / \Z_2$.  We abuse notation by suppressing
this $\Z_2$ in most of what follows.}
\be
\left[ J_3, J_\pm \right] = \pm i J_\pm \ ,\quad [J_+,J_-]= 2 i J_3
\ee
In particular, $J_3$ induces  the ``compact 5-grading''
\be
\label{su21-5gc}
J_- \vert_{-i} \oplus
\{ L_-, K_+  \}\vert_{-\frac{i}{2}} \oplus
\{ J_3, S \}\vert_0  \oplus
\{ K_-, L_+ \}\vert_{\frac{i}{2}} \oplus
J_+ \vert_{i}
\ee
where the subscript now denotes the $J_3$ eigenvalue. The remaining
roots can then be arranged as a pair of doublets under $SU(2)$
with opposite $U(1)_S$ charges,
\be
\begin{pmatrix}
J_{-\frac12,-\frac32} & J_{\frac12,-\frac32} \\
J_{-\frac12,\frac32} & J_{\frac12,\frac32}
\end{pmatrix}
\ee
as shown on the second root diagram on Figure \ref{su21rootdiag}.

In order to represent the generators in the compact basis by
pseudo-hermitian matrices, it is
convenient to change basis and diagonalize the hermitian metric \eqref{eta21},
\be
\label{cayleysu21}
\eta' \equiv C \eta C^t = \begin{pmatrix} 1 &&\\ &-1& \\ &&1 \end{pmatrix}
\ ,\quad
C= \begin{pmatrix}
\frac{1}{\sqrt{2}} & 0 & \frac{1}{\sqrt{2}} \\
\frac{1}{\sqrt{2}} & 0 & -\frac{1}{\sqrt{2}} \\
0&1&0
\end{pmatrix}
\ee
The generators preserving the metric $\eta'$ can now be parametrized as
\be
\label{cayleysu21X}
\begin{pmatrix}
-\frac{i}{2}(\underline{J}_3+\underline{S}) &
2\underline{J}_{-\frac12,-\frac32} & -\underline{J}_- \\
2\underline{J}_{\frac12,\frac32} & i \underline{S} &
-2\underline{J}_{-\frac12,\frac32} \\
\underline{J}_+ & - 2\underline{J}_{\frac12,-\frac32} & \frac{i}{2}
(\underline{J}_3-\underline{S})
\end{pmatrix}
\ee
where $\underline{J}_3,\underline{S}$ are real, while $\underline{J}_\pm = \underline{J}_\mp^*$,
$\underline{J}_{\pm \frac12,\pm \frac32} = \underline{J}_{\mp \frac12,\mp \frac32}^*$,
$\underline{J}_{\pm \frac12,\mp \frac32} = \underline{J}_{\mp \frac12,\pm \frac32}^*$.
By analogy with the $SL(2,\IR)=SU(1,1)$ case, we shall refer to
the matrix $C$ as a Cayley rotation.

\subsection{Quaternionic symmetric space}

We now describe the geometry of the quaternionic-K\"ahler symmetric space
$K \bas G=(SU(2)\times U(1)) \bas SU(2,1)$. This space is
well known in the string theory literature as the tree-level
moduli space of the universal hypermultiplet (see e.g. \cite{Cecotti:1989qn,
Antoniadis:2003sw,Anguelova:2004sj,Rocek:2006xb} for some useful
background). It is in the class of  ``dual quaternionic manifolds'',
in the sense that it can be constructed by the $c$-map
procedure \cite{Ferrara:1989ik,deWit:1990na} from the trivial
zero-dimensional special K\"ahler manifold
with quadratic prepotential $F=-i (X^0)^2 / 2$.

In order to parameterize this space, we use the Iwasawa decomposition of $G$
\be
\label{su21iwa}
g = k \cdot e^{-U H} \cdot
e^{\tzeta {E_p}-\zeta {E_q}} \cdot
e^{\sigma {E}}
\ee
where $k$ is an element of the maximal compact subgroup $K=SU(2)\times U(1)$.
In terms of the fundamental representation, this is $g=k\cdot e_{QK}$ where
$e_{QK}$ is the coset representative
\begin{equation}
\label{vielb}
e_{QK}= \begin{pmatrix}e^{-U}&&\\&1&\\&&e^{U}\end{pmatrix}\cdot
\begin{pmatrix}1& ~\tzeta+i\zeta~ &
i\, \sigma -\frac12 (\tzeta^2+\zeta^2) \\ &1
&-(\tzeta-i\zeta)\\&&1\end{pmatrix}\ ,
\end{equation}
The right-invariant $\fg$-valued 1-form is then
\be
\theta = de_{QK}\, e_{QK}^{-1} = \begin{pmatrix}
-dU & e^{-U} (d\tzeta + i d\zeta)
& i e^{-2U} (d\sigma+\tzeta d\zeta - \zeta d\tzeta) \\
0 & 0 & - e^{-U} (d\tzeta - i d\zeta)  \\
0 & 0 & dU
\end{pmatrix}
\ee
Expanding $\theta$ on the compact basis of $\fg$, its 1-form components are
\be
\label{su21-qv}
\begin{pmatrix}
\underline{J}_{-\frac12,\frac32}  & \underline{J}_{\frac12,\frac32} \\
\underline{J}_{-\frac12,-\frac32} & \underline{J}_{\frac12,-\frac32}
\end{pmatrix}
=
-\begin{pmatrix}
\bar u &v \\  \bar  v & u
\end{pmatrix}
\ ,
\ee
\be
\label{su21-cv}
\begin{pmatrix}
\underline{J}_- \\ \underline{J}_3 \\ \underline{J}_+
\end{pmatrix}
=-\frac12
\begin{pmatrix}
\bar u \\
\frac{1}{4i}(v-\bar v) \\
u
\end{pmatrix}\ ,\quad
\underline{S} = \frac{3i}{8} (v-\bar v)
\ee
where we defined the 1-forms
\be
\label{uvsu21}
u = -\frac{1}{\sqrt{2}} e^{-U} (d\tilde\zeta+i d\zeta) \ ,\quad
v = dU +\frac{i}{2} e^{-2U}
(d\sigma + \tzeta d\zeta - \zeta d\tzeta)
\ee
The non-compact components \eqref{su21-qv} give the quaternionic viel-bein
of the invariant metric, while the compact components \eqref{su21-cv} give
the spin connection, with restricted holonomy $SU(2)\times U(1)$.
The invariant metric on $K \bas G$ is thus given by
\be
\label{ds2su21}
\begin{split}
ds^2 &= 2 (u\bar u + v \bar v) \\
&= 2(dU)^2 + e^{-2U} \left(d\tzeta^2 + d\zeta^2\right)
+\frac12 e^{-4U} \left( d\sigma+\tzeta d\zeta - \zeta d\tzeta \right)^2
\end{split}
\ee
An exception among quaternionic-K\"ahler, this metric is K\"ahler in the
complex structure induced from the  $U(1)$ generator $S$. A  K\"ahler
potential is given by $-\log( s+\bar s-2 c \bar c)$, where $s,c$
are the complex coordinates
\be
s = e^{2U} + \zeta^2 + \tzeta^2 + i \sigma\ ,\quad c=\tzeta+i \zeta\ .
\ee

The group $G$ acts on the coset space $K\backslash G$
by right multiplication,
followed by a left action of the maximal compact $K=SU(2)\times U(1)$
so as to maintain the Iwasawa gauge $k=1$ in \eqref{su21iwa}. The metric
is invariant under this action. This gives an action of $SU(2,1)$
by Killing vectors on $K \backslash G$:
\begin{subequations}
\begin{eqnarray}
\label{su21krep}
E^{QK} &=& \pa_\sigma\ ,\quad
E_p^{QK} = \pa_{\tzeta} - \zeta \pa_\sigma \ ,\quad
E_q^{QK} = -\pa_{\zeta} - \tzeta \pa_\sigma \ ,\quad\\
J^{QK} &=& \zeta\pa_{\tzeta} - \tzeta\pa_{\zeta}\ ,\quad
H^{QK} = -\pa_U - \zeta\pa_{\zeta} - \tzeta\pa_{\tzeta}
- 2 \sigma\pa_\sigma \\
{F_p}^{QK} &=& -\tzeta \pa_U -(\sigma+2\zeta\tzeta)\pa_{\zeta}
+ \left[ e^{2U} + \frac12 ( 3 \zeta^2 -\tzeta^2) \right]\pa_{\tzeta} \nn \\
&& + \left[ \zeta \left( e^{2U}+ \frac12(\zeta^2 + \tzeta^2)\right)
- \sigma\tzeta\right]
\pa_\sigma \\
{F_q}^{QK} &=& \zeta \pa_U
- \left[ e^{2U} + \frac12 ( 3 \tzeta^2 -\zeta^2) \right]\pa_{\zeta}
-(\sigma-2\zeta\tzeta)\pa_{\tzeta} \nn \\
&& + \left[ \tzeta \left( e^{2U}+ \frac12(\zeta^2 + \tzeta^2)\right)
+ \sigma\zeta\right]
\pa_\sigma \\
{F}^{QK} &=&  - \sigma \pa_U
+ \left[ \left( e^{2U} + \frac12 (\zeta^2+\tzeta^2) \right)^2
- \sigma^2 \right]
\pa_\sigma - \left[ \tzeta \left(e^{2U}
+ \frac12 (\zeta^2+\tzeta^2)\right)+ \sigma\zeta \right] \pa_{\zeta} \nn \\
&& + \left[   \zeta  \left( e^{2U}+ \frac12(\zeta^2 + \tzeta^2)
\right)- \sigma \tzeta \right]
\pa_{\tzeta}
\end{eqnarray}
\end{subequations}

The action of $G$ on $K \backslash G$ also induces a representation of
$G$ on $L^2(K \bas G)$.  The quadratic Casimir in this representation is
proportional to the Laplace-Beltrami operator of the metric \eqref{ds2su21},
while the cubic Casimir vanishes identically:
\be
C_2 = \frac14 \Delta = \frac14 \frac{1}{\sqrt{g}}
\pa_i {\sqrt{g}} g^{ij} \pa_j\ ,\quad C_3=0
\ee

\subsection{Twistor space and Swann space}

The twistor space $\cZ$ is a $\C\PP^1 = U(1) \bas SU(2)$ bundle over
the quaternionic-K\"ahler space $(SU(2)\times U(1)) \bas SU(2,1)$,
which carries a K\"ahler-Einstein metric. The fibration is such
that the $SU(2)$ ``cancels''~\cite{MR0185554}, so that $\cZ$ is an homogeneous (but
not symmetric) space,
\be
\label{zsu21}
\cZ = (U(1)_S \times U(1)_{J_3}) \bas SU(2,1)
\ee
Let $H$ denote the subgroup $U(1)_S \times U(1)_{J_3}$.
In the following we construct two canonical sets of
complex coordinates on $\cZ$, adapted to two different Heisenberg algebras,
and relate them to the coordinates
$U,\zeta,\tzeta,\sigma$ on the base and the stereographic
coordinate $z$ on the sphere.\footnote{Of course, these ``coordinates on $\cZ$'' 
really cover only an open dense subset; they become
singular at the (canonically defined) north and south poles of the $\C\PP^1$ fibers.}

\subsubsection{Harish-Chandra coordinates \label{hchandsu21}}

The complex structure on $\cZ$ can be constructed by using
the Borel embedding $H \bas G \hookrightarrow P'_{\IC} \bas G_\C$
where $P'_{\IC}$ is the parabolic (Borel) subgroup of the 
complexified group $G_{\IC}$,
generated by the positive roots $J_{\frac12,\pm\frac32}, J_+$ and
Cartan generators $J_3,S$.  To obtain complex coordinates from this
embedding we go to the Cayley-rotated matrix 
representation \eqref{cayleysu21X},
and perform a $\bar N' A' N'$ decomposition,
\be
\label{nban}
C e_\cZ C^{-1} =
\begin{pmatrix} 1&&\\ \tilde p&1& \\ \tilde k&\tilde q &1 \end{pmatrix} \cdot
\begin{pmatrix} a b&&\\ &1/a^2& \\ &&a/b \end{pmatrix} \cdot
\begin{pmatrix} 1&p&k\\ &1&q \\ &&1 \end{pmatrix}
\ee
The entries $p,q,k$ in the upper-triangular matrix
by construction provide holomorphic coordinates on $\cZ$. The lower-triangular
and diagonal matrices are then expressed in terms of $p,q,k$ and
their complex conjugates $\bar p,\bar q, \bar k$ by requiring
that \eqref{nban} is an element of $SU(2,1)$ rather than of
its complexification:
\bse
\be
\tilde p = \frac{{\bar p}-{\bar k} q}{\sqrt{\Sigma}}\ ,\quad
\tilde q = \frac{{\bar k} p-{\bar p} {\bar q} p
+{\bar q}}{\sqrt{k {\bar k}-p   {\bar p}+1}} \ ,\quad
\tilde k = \frac{({\bar p} {\bar q}-{\bar k})\sqrt{k {\bar k}-p {\bar p}+1}
}{\sqrt{\Sigma}}
\ee
\be
a= \frac{\Sigma^{1/4}}{(k {\bar k}-p   {\bar p}+1)^{1/4}}\ ,\quad
b=\frac{1}{(k {\bar k}-p {\bar p}+1)^{1/4} \Sigma^{1/4}}
\ee
\ese
where
\be
\Sigma= 1 + k \bar k -q \bar q -{\bar k} p q - k {\bar p} {\bar q}
+p {\bar p} q {\bar q}
\ee
These coordinates are adapted to the holomorphic action of the
Heisenberg algebra generated by $J_{-\frac12,\pm\frac32},J_-$, in the
sense that
\be
J_{-\frac12,\frac32} = -2 ( \pa_q + p \pa_k ) \ ,\quad
J_{-\frac12,-\frac32} = 2 \pa_p \ ,\quad
J_-=-\pa_k
\ee
It is useful to record the action of the other generators in the
compact basis,
\bse
\bea
J_3&=&\frac{i}{2}(p\pa_p+q\pa_q+2k\pa_k)\ ,\quad
S= \frac{3i}{2}(p\pa_p-q\pa_q)\ ,\\
J_+&=&- k p \pa_p - (k-p q) q \pa_q - k^2 \pa_k\\
J_{\frac12,\frac32} &=&-2 p^2 \pa_p -2 (k-p q) \pa_q - 2 k p \pa_k \ ,\quad
J_{\frac12,-\frac32} = - 2 k \pa_p + 2 q^2 \pa_q
\eea
\ese
A $G$-invariant metric on $\cZ$ can be constructed in the usual way,
by applying an $L$-invariant quadratic form on $\fg$ to the $\fg$-valued $1$-form
$\theta=de_\cZ\cdot e_\cZ^{-1}$.
In contrast to the case of $K \backslash G$, this quadratic form
is not unique up to scalar multiple, but has parameters
$(\alpha,\beta,\gamma)\in \IR^3$:
\be
\label{dsabc}
\alpha \underline{J}_+ \underline{J}_-
+ \beta \underline{J}_{\frac12,\frac32} \underline{J}_{-\frac12,-\frac32}
+ \gamma \underline{J}_{\frac12,-\frac32} \underline{J}_{-\frac12,\frac32}
\ee
These parameters can be fixed by requiring
that the resulting metric on $\cZ$ is Einstein-K\"ahler.
In particular, the K\"ahler potential must be proportional to the
volume element. This uniquely fixes  $\alpha=-2,\beta=\gamma=4$ (up
to rescalings) and gives the K\"ahler potential
\be
\label{kzpq}
K_\cZ=\frac12\log\left[(1 + k \bar k - p\bar p)(1 +  |k-pq|^2 - q \bar q)\right]
\ee
This reproduces Eq.~7.28 in \cite{deWit:2001dj}
upon identifying $k=\zeta/2, p=v, q=-u$. Under the action of
$J_{\frac12,\frac32}+J_{-\frac12,-\frac32}$,
$J_{\frac12,-\frac32}+J_{-\frac12,\frac32}$,
$J_++J_-$, $K_\cZ$ transforms by a K\"ahler transformation
$K_\cZ \to K_\cZ + f + \bar{f}$, with $f$
proportional to $p, q, k-\frac12 pq$, respectively.

\subsubsection{Iwasawa coordinates}

We now exhibit the twistor space $\cZ$ as an $S^2$ fibration over
$K \bas G$, such that the $S^2$ fiber over any point on the base
is holomorphically embedded in $\cZ$. For this purpose, we
recall that
the complex structure on $S^2 = U(1) \bas SU(2)$
can be constructed using the Borel embedding
$U(1) \bas SU(2) \hookrightarrow B_\C \bas SL(2,\IC)$.
Here $SL(2,\C) \subset G_\C$ is generated by $J_+, J_3, J_-$
and $B_\IC$ is the Borel subgroup generated by $J_+, J_3$.
The embedding is simply obtained by starting with a coset representative $e \in U(1) \bas SU(2)$
and viewing it instead as a representative in $B_\C \bas SL(2,\IC)$ (this is consistent
since $B_\C \cap SU(2) = U(1)$.)  The same class in $B_\C \bas SL(2,\IC)$ is also
represented by $\exp z J_-$ for some $z \in \C$
(with one exception corresponding to $z = \infty$), giving the desired complex coordinate.

Now we use the same idea for $\cZ$.
So we return to the original matrix representation \eqref{su21X},
and parameterize $\cZ$ by the coset representative
\bea
\label{ezsu21}
e_\cZ&=& e^{-\bar z J_+}\ (1+z\bar z)^{-i J_3} \ e^{-z J_-} \ e_{QK} \\
&=&\frac{1}{\sqrt{1+z\bar z}} \begin{pmatrix}
\frac12(1+\sqrt{1+z\bar z}) & \frac{z}{\sqrt{2}}
& \frac12(1-\sqrt{1+z\bar z}) \\
-\frac{1}{\sqrt{2}} \bar z  & 1 &-\frac{1}{\sqrt{2}} \bar z \\
\frac12(1-\sqrt{1+z\bar z}) & \frac{z}{\sqrt{2}} &
\frac12(1+\sqrt{1+z\bar z}) \\
\end{pmatrix}
\cdot e_{QK}
\eea
where $e_{QK}$ is a representative for $K \bas G$ in the Iwasawa decomposition \eqref{vielb}.
The coordinate $z$ is then a stereographic coordinate
on the $S^2$ fiber over each point of $K \bas G$.

By Cayley rotating \eqref{ezsu21} and performing the Harish-\-Chandra
decomposi\-tion \eqref{nban}, we can now relate the complex coordinates
$p,q,k$ and their complex conjugates to the coordinates $U,\zeta,\tzeta,\sigma$
on the base and the coordinate $z$ on the fiber:
\bse
\bea
\label{hcsu21}
p&=& \frac{4}{2+2 i \sigma+2 e^{2  U}-\zeta^2-\tzeta^2
+2 i \sqrt{2} e^U z (\zeta+i \tzeta)}-1\\
q&=& \frac{2 \sqrt{2} e^U (i   \zeta+\tzeta)
-\left(\zeta^2+\tzeta^2+2 i \sigma+2-2 e^{2 U} \right) z}
{2 i \sqrt{2} z (\zeta+i \tzeta)+4 e^U}\\
k&=& \frac{2 \left(2 e^U z+\sqrt{2} (i\zeta+\tzeta)\right)}
{2+2 i \sigma+2 e^{2  U}-\zeta^2-\tzeta^2
+2 i \sqrt{2} e^U z (\zeta+i \tzeta)}
\eea
\ese
Rather than obtaining the metric on $\cZ$ in the
coordinates $U,\zeta,\tzeta,\sigma,z,\bar z$ from the K\"ahler potential
\eqref{kzpq} by following the change of variables, we can simply
decompose the invariant form $\theta_\cZ=de_\cZ e_\cZ^{-1}$
and plug into \eqref{dsabc} using the values of $\alpha,\beta,
\gamma$ that were determined above. The components of
$\theta_\cZ$ along $(U(1) \times U(1)) \bas SU(2,1)$ read
\bse
\be
\begin{pmatrix}
\underline{J}_{-\frac12,\frac32}  & \underline{J}_{\frac12,\frac32} \\
\underline{J}_{-\frac12,-\frac32} & \underline{J}_{\frac12,-\frac32}
\end{pmatrix}
= -\frac{1}{\sqrt{1+z\bar z}}
\begin{pmatrix}
\bar u + \bar z v &
v-z \bar u \\
\bar v-\bar z u &
u + z \bar  v 
\end{pmatrix}
\ee
\bea
\underline{J}_+ &=&
-\frac{1}{1+z\bar z}\left( d\bar z + \bar u
+ \frac12 \bar z (v-\bar v)+ \bar z^2 u\right)
\equiv -\frac{D\bar z}{1+z\bar z}\\
\underline{J}_- &=&
-\frac{1}{1+z\bar z}\left( dz + u - \frac12 z (v-\bar v)+ z^2 \bar u\right)
\equiv -\frac{D z}{1+z\bar z}
\eea
\ese
leading to the K\"ahler-Einstein metric
\be
ds^2_\cZ = u  \bar u + v  \bar v -2\frac{Dz D\bar z}{(1+z\bar z)^2}
\ee
with signature $(4,2)$.
The connection term in $Dz$ is recognized as the projective
$SU(2)$ connection,
\bea
Dz&=&  dz - \frac12(A_1+i A_2) + i A_3 z - \frac12(A_1-i A_2) z^2\\
&=&  dz + u - \frac12 z (v-\bar v) + z^2 \bar u
\eea
where
\be
A_1 = - (u+\bar u)\ ,\quad A_2 = i(u-\bar u)\ ,
\quad A_3=\frac{i}{2}(v-\bar v)
\ee
are the components of the $SU(2)$ spin connection
computed in \eqref{su21-cv}. A basis of holomorphic (1,0) forms
providing a holomorphic viel-bein of $\cZ$ is given by the
components of $\theta_{\cZ}$ with negative weight under $J_3$,
\be
\mathcal{V} = \frac{1}{\sqrt{1+z\bar z}} \,
\left( \sqrt{2} Dz , u+z  \bar v, v- z \bar u \right)
\ee
The K\"ahler form can be written as
\be
\omega_\cZ =-
\frac{Dz\ D\bar z}{(1+z\bar z)^2}
+ i\ x_a \omega^a
\ee
where $x^a$ is the unit length vector with stereographic coordinate $z$,
\be
x_1 = \frac{z+\bar z}{1+z \bar z} \ ,\quad
x_2 = \frac{i(z-\bar z)}{1+z \bar z}\ ,\quad
x_3 = \frac{1-z\bar z}{1+z \bar z}
\ee
and $\omega_a$ are the quaternionic 2-forms on the base,
\be
\omega^1 = \frac{1}{2i}(u  v-\bar u \bar v)\ ,\quad
\omega^2 = \frac{1}{2}(u  v+\bar u \bar v)\ ,\quad
\omega^3 = \frac{1}{2i}(u \bar u+v \bar v)
\ee
It may be checked that this triplet of 2-forms satisfies the constraints
from quaternionic-K\"ahler geometry,
\be
d\omega^i + \eps_{ijk} A_j \wedge \omega^k = 0\ ,\quad
dA_i + \eps_{ijk} A_j \wedge A_k = 2 \omega_i
\ee

\subsubsection{Complex $c$-map coordinates}

While the complex coordinates $p,q,k$ are adapted to the action
of the generators $J_{-\frac12,\frac32},J_{-\frac12,-\frac32},J_-$,
in the sequel it will be useful to have complex coordinates $\xi,\txi,\alpha$
adapted to the ``non-compact'' Heisenberg algebra $E_p,E_q,E$, i.e. such that
the action of these generators takes the canonical form
\be\label{epqka}
E_p = \pa_{\txi} + \xi \pa_\alpha \ ,\quad
E_q = -\pa_{\xi} + \txi \pa_\alpha \ ,\quad
E = -\pa_\alpha
\ee
(Since $G$ acts holomorphically, we have abused notation by writing only
the holomorphic part; the real vector fields would be obtained by adding 
the complex conjugates, in \eqref{epqka} and below.)

The change of variables can be found by diagonalizing the action
of these generators in the $p,q,k$ variables, leading to
\bse
\label{xitop}
\be
\xi=\frac{i \left(k^2-(p+1) q k+p+1\right)}{\sqrt{2} (p+1) (-k+p q+q)}\ ,\quad
\txi=\frac{-k^2+(p+1) q k+p+1}{\sqrt{2} (p+1)(-k+p q+q)}
\ee
\be
\alpha=\frac{i(q+kp-p^2 q)}{(p+1) (k-p q-q)}
\ee
\ese
or, conversely,
\be
\label{pqktoxi}
p= -\frac{\xi^2+\txi^2-2 i\alpha+2}
{\xi^2+\txi^2-2 i\alpha-2}\ ,\quad
q=\frac{i\left(\xi^2+\txi^2+2 i\alpha+2 \right)}
{2 \sqrt{2} (\xi+i \txi)}\ ,\quad
k= -\frac{2 i \sqrt{2} (\xi-i \txi)}
{\xi^2+\txi^2-2 i\alpha-2}
\ee
The full action of $G$ is then given by
\begin{subequations}
\begin{eqnarray} \label{su21holaction}
{E}^{QC} &=& -\pa_\alpha\ ,\quad
{E_p}^{QC} = \pa_\txi +\xi \pa_\alpha\ ,\quad
{E_q}^{QC} = -\pa_\xi + \txi \pa_\alpha\ , \\
H^{QC}&=& -\txi \pa_\txi -\xi \pa_\xi -2 \alpha \pa_\alpha \ ,\quad
J^{QC}= - \txi\pa_\xi + \xi \pa_\txi\\
{F_p}^{QC}&=&\frac12 (3 \xi^2-\txi^2)\pa_\txi +(\alpha - 2 \txi \xi)\pa_\xi
- \frac12\left[\xi( \txi^2+ \xi^2)+2\alpha \txi\right] \pa_\alpha \\
{F_q}^{QC}&=&-\frac12(3\txi^2-\xi^2)\pa_\xi +(\alpha + 2 \txi \xi)\pa_\txi
- \frac12\left[\txi( \txi^2+ \xi^2)-2\alpha \txi\right] \pa_\alpha \\
{F}^{QC}&=&\frac12 \left[ \xi(\txi^2+\xi^2)+2\alpha \txi\right]\pa_\txi
-\frac12 \left[ \txi(\txi^2+\xi^2)-2\alpha \xi\right]\pa_\xi \nn \\
&&-\frac14 \left[ (\txi^2+\xi^2)^2-4 \alpha^2 \right] \pa_\alpha
\end{eqnarray}
\end{subequations}

In the new coordinate system $\xi,\txi,\alpha$, the K\"ahler
potential \eqref{kzpq} (after a K\"ahler transformation by
$f = (p+1)(k-q-pq)$) is
\be
\label{kzxi}
K_\cZ=\frac12\log N_4 = \frac12\log  \left(
\left[ (\xi-\bar\xi)^2+(\txi-\bar\txi)^2 \right]^2
+4 (\alpha-\bar \alpha+ \xi \bar\txi - \bar\xi \txi)^2 \right)
\ee
Here we note that $N_4(\xi,\txi,\alpha;\bar\xi,
\bar\txi,\bar\alpha)$ is the quartic distance function of quasi-conformal geometry.
 Since $G$ acts by isometries on $\cZ$, it leaves the K\"ahler potential
\eqref{kzxi} invariant up to K\"ahler transformations. Equivalently,
the quartic norm $N_4(\xi,\txi,\alpha;\bar\xi,
\bar\txi,\bar\alpha)$ defined by \eqref{kzxi} transforms multiplicatively
by a factor $f(\xi,\txi,\alpha) \bar f(\bar \xi, \bar\txi,\bar\alpha)$,
where the holomorphic function $f$ depends on the generator under
consideration. In particular, the ``quartic light-cone'' $N_4=0$ is invariant under the full
action of $SU(2,1)$, which
motivated the appellation ``quasi-conformal action''
in \cite{Gunaydin:2000xr}.

Such a coordinate system adapted to the holomorphic action
of a Heisenberg group exists for any $c$-map space and  was used heavily in \cite{Neitzke:2007ke}.
Moreover, the result \eqref{kzxi} agrees with a general formula for the K\"ahler potential
given there.

\subsubsection{Twistor map \label{sectwimapsu21}}

Combining the changes of variables \eqref{hcsu21} and \eqref{xitop},
we find that the complex coordinates $(\xi,\txi,\alpha)$
are expressed in terms of the Iwasawa coordinates $U,\zeta,\tzeta,\sigma,
z,\bar z$
as
\bse
\label{zetatoxi}
\bea
\xi&=&\zeta - \frac{i}{\sqrt{2}} e^{U}
\left( z + z^{-1} \right) \label{twixi}\\
\txi&=&\tzeta + \frac{1}{\sqrt{2}} e^{U} \left( z - z^{-1} \right)\\
\alpha &=& \sigma + \frac{1}{\sqrt{2}} e^{U}
\left[  z ( \zeta+i\tzeta) + z^{-1} (-\zeta+i \tzeta) \right]
\eea
\ese
Again $(\xi,\txi,\alpha)$ is a holomorphic (rational) function
of $z$, at fixed values of  $U,\zeta,\tzeta,\sigma$, so that the twistor fiber
over any point is a rational curve in $\cZ$. Conversely,
\bse
\bea
e^{2U}&=&-\frac{N_4}{8[(\xi-{\bar\xi})^2+(\txi-{\bar\txi})^2]}\\
\zeta&=&\frac{2 (\alpha-{\bar\alpha})
   (\txi-{\bar\txi})+(\xi-{\bar\xi})
\left(\xi^2-{\bar\xi}^2+\txi^2-{\bar\txi}^2\right)}{2[(\xi-{\bar\xi})^2+(\txi-{\bar\txi})^2]}\\
\tzeta&=&\frac{2({\bar\alpha}-\alpha) (\xi-{\bar\xi})
+(\txi-{\bar\txi}) \left(\xi^2-{\bar\xi}^2+\txi^2-{\bar\txi}^2\right)}
{2[(\xi-{\bar\xi})^2+(\txi-{\bar\txi})^2]}\\
\sigma &=&
\frac{\alpha+{\bar\alpha}}{2}-\frac{(\alpha-\bar \alpha
+ \xi \bar\txi - \bar\xi \txi)
\left(\xi^2-{\bar\xi}^2+\txi^2-{\bar\txi}^2\right)}{2[(\xi-{\bar\xi})^2+(\txi-{\bar\txi})^2]}\\
z &=&
\sqrt{\frac{\xi-\bar\xi-i(\txi-\bar\txi)}{\xi-\bar\xi+i(\txi-\bar\txi)}}
\times \sqrt{\frac{(\xi-\bar\xi)^2+(\txi-\bar\txi)^2
- 2i(\alpha-\bar \alpha+ \xi \bar\txi - \bar\xi \txi)}
{(\xi-\bar\xi)^2+(\txi-\bar\txi)^2
+ 2i(\alpha-\bar \alpha+ \xi \bar\txi - \bar\xi \txi)}}
\eea
\ese
These formulae are in agreement with the general results
in \cite{Neitzke:2007ke}.

Using the twistor map \eqref{zetatoxi},
one finds that the action of $G$ on $\cZ$
reproduces the action on the base \eqref{su21krep},
plus an action along the fiber:
\bea
E^{QC} + \bar E^{QC} &=& E^{QK}\ ,\quad
E_{p}^{QC}+\bar E_{p}^{QC} =E_{p}^{QK}\ ,\quad
E_{q}^{QC}+\bar E_{q}^{QC} =E_{q}^{QK}\ ,\nn\\
H^{QC} + \bar H^{QC} &=& H^{QK}\ ,\quad
J^{QC}+\bar J^{QC}   = J^{QK}  - i (z \pa_z - \bar z \pa_{\bar z}) \nn\\
F_{p}^{QC}+\bar F_{p}^{QC} &=& F_{p}^{QK} -\sqrt{2} e^U
\left[(1+z^2)\pa_z+(1+\bar z^2)\pa_{\bar z}\right]
-3i \zeta   ( z \pa_z - \bar z \pa_{\bar z}) \nn \\
F_{q}^{QC}+\bar F_{q}^{QC} &=& F_{q}^{QK} + i \sqrt{2} e^U
\left[(1-z^2)\pa_z-(1-\bar z^2)\pa_{\bar z}\right]
-3i \tzeta   ( z \pa_z - \bar z \pa_{\bar z}) \nn \\
F^{QC}+\bar F^{QC} &=& F^{QK}
+\left( i e^{2U} - 3 (\zeta^2+\tzeta^2) \right)
( z \pa_z - \bar z \pa_{\bar z}) \nn \\
&& -\sqrt{2} e^U
\left[ \left( (1+z^2)\zeta-i(1-z^2)\tzeta\right)\pa_z
+\left( (1+{\bar z}^2)\zeta+i(1-{\bar z}^2)\tzeta\right)\pa_{\bar z}\right]
\nn\\
\eea
These formulae will be useful in Section \ref{matpenrose} when we discuss the
Penrose transform.

At this stage, we note that the twistor map relations \eqref{zetatoxi}
can be obtained more directly by the following trick.  Recall that $P$ is the
parabolic subgroup of lower-triangular matrices in the matrix
representation \eqref{su21X}, and let $\bar{P}$ be its opposite subgroup
consisting of upper-triangular matrices.  Then we construct a new
embedding $H \bas G \hookrightarrow P_\C \bas G_\C$ by choosing a
different coset representative: we use the formula \eqref{ezsu21} for
$e_\cZ$ but make an analytic continuation, regarding $z$ and $\bar{z}$
as independent and then taking a limit
\begin{equation} \label{singlim}
\bar{z} \to -1/z
\end{equation}
If we ignore for a moment the fact that the limit \eqref{singlim} of
\eqref{ezsu21} is singular, formally it defines an element of $G_\C$.
Moreover, the locus of elements in $G_\C$ so obtained is formally
invariant under $G$, essentially because \eqref{singlim} is
constructed from the antipodal map (real structure) on $\C\PP^1$,
which is $SU(2)$-invariant.  This locus will give the desired copy of
$H \bas G$ inside $P_\C \bas G_\C$.
We now define the coordinates $(\xi, \txi, \alpha)$ in $P_\C \bas G_\C$
using the $N_\C A \bar{N}_\C$ decomposition,
\be
\label{fakehcdec}
e_\cZ = \begin{pmatrix} 1&&\\ *&1& \\ *&*&1 \end{pmatrix} \cdot
\begin{pmatrix} *&&\\ &*& \\ &&* \end{pmatrix} \cdot
\begin{pmatrix} 1&\txi+i \xi&i\alpha-\frac12(\xi^2+\txi^2)\\
&1&-\txi+i\xi \\ &&1 \end{pmatrix}
\ee
Then a direct matrix computation gives the coordinates of $e_\cZ$ as
\bse
\bea
\label{zetatoxi2}
\xi&=&\zeta - \frac{i}{\sqrt{2}}
\left[\frac{-1+\sqrt{1+z\bar z}}{z\bar z}\right] e^{U}
\left( z - \bar z \right) \\
\txi&=&\tzeta + \frac{1}{\sqrt{2}}
\left[\frac{-1+\sqrt{1+z\bar z}}{z\bar z} \right]
e^{U}
\left( z +\bar  z \right)\\
\alpha &=& \sigma + \frac{1}{\sqrt{2}}
\left[\frac{-1+\sqrt{1+z\bar z}}{z\bar z}\right] e^{U}
\left[  z ( \zeta+i\tzeta) - \bar z (-\zeta+i \tzeta) \right]
\eea
\ese
recovering the desired result \eqref{zetatoxi} after setting $\bar z=-1/z$.
The result shows that the limit \eqref{singlim} is in fact regular
in $N_\C A \backslash G_\C$ (although not in $G_\C$).

\subsubsection{Swann space} \label{secswann}

There is an important complex line bundle over $\cZ$, which we call $\cO(2)$.
It may be defined in two different ways.  One way is to use $\cZ = H \bas G$.  Then $\cO(2)$ is
determined by the character $\exp i \underline{J_3}$ of $H$ --- in other words, a section of $\cO(2)$ is
a function on $G$ which transforms under $H$ by this character.\footnote{By definition, the character ``$\exp i\underline{J_3}$'' takes the value $\exp i\underline{J_3}$ on $\exp(\underline{J_3} J_3 + \underline{S} S) \in H$; we will use this notation for characters frequently, always implicitly with respect to some basis of the 
corresponding Lie algebra.}
This definition makes it easy to prove that $\cO(2)$ admits a Hermitian structure.

On the other hand, one can also define $\cO(2)$ using the Borel
embedding $\cZ \hookrightarrow P_\C \bas G_\C$.  In that case we would define
$\cO(2)$ as the pullback to $\cZ$ of the line bundle determined by the character $\exp \underline{H}$
of $P_\C$.  This definition makes it easy to see that $\cO(2)$ is a holomorphic line bundle; it is 
what we will use in the discussion of quaternionic discrete series below.

More generally, define $\cO(m) = \cO(2)^{\otimes \frac{m}{2}}$ for $m$ even; this is similarly a 
Hermitian holomorphic line bundle over $\cZ$.

The Swann space $\cS$, also known as the hyperk\"ahler cone over $K \bas G$, is the total
space of $\cO(-2)$ over $\cZ$; locally it could be paramaterized by complex coordinates on $\cZ$ plus one additional
coordinate in the fiber of $\cO(-2)$.  It is naturally a hyperk\"ahler 
manifold, with an $SU(2)$ isometry rotating the
complex structures into one another \cite{MR1096180,deWit:2001dj}.

The circle in the unit circle bundle of $\cO(-2)$ ``cancels against'' the $U(1)$ in the denominator of \eqref{zsu21}.
So this circle bundle is the homogeneous 3-Sasakian space $U(1) \bas SU(2,1)$,
which can be parameterized by the coset representative
\be
e_{3S} = e^{i\phi J_3} e_\cZ
\ee
$\cS$ is then a real cone over this homogeneous space,
\be
\cS = \IR^+ \times U(1) \bas SU(2,1)
\ee

The right-invariant form $\theta_{3S}=de_{3S} e_{3S}^{-1}$
has $J_3$ component
\be
\underline{J}_3 = d\phi + \frac{i}{1+z\bar z} \left(
\bar z dz - z d\bar z + 2 (\bar z u - z \bar u) + \frac12 (v-\bar v)
(1-z \bar z) \right):= D\phi
\ee
while the other components are identical to those of $\theta_\cZ$
except for a rotation under $U(1)_{J_3}$.
Note that we can rewrite
\be
\underline{J}_3 = d\phi + \frac{i}{1+z\bar z} \left(
\bar z dz - z d\bar z\right)
+x_a A^a
\ee
The metric of the 3-Sasakian space is obtained by adding $\underline{J_3}^2$
to \eqref{dsabc}, with the appropriate coefficient to enforce
$SU(2)$ (left) invariance. The metric on $\cS$ is therefore
\be
ds_{\cS}^2 = - \left[ dr^2 + r^2 \left( D\phi^2 - ds_\cZ^2 \right) \right]\ ,
\ee
with indefinite signature (4,4).

\subsection{Quasi-conformal representations}

\subsubsection{Principal series} \label{secprincipal}

An interesting family of ``principal series'' representations of $G$
are obtained by induction from the parabolic subgroup $P$
generated by $\{ {F}, {F_p}, {F_q}, H, J\}$, using the
character $\chi_k = e^{-k \underline{H}/2}$ of $P$
for some $k \in \C$.
We now briefly recall the definition of induction, which we will use many times in this paper;
see \ti{e.g.} \cite{MR1880691} for more.

The representation space of the induced representation consists of functions $f$ on $G$ which obey
\begin{equation} \label{rep-twist}
f(g p) = \chi_k(p) f(g).
\end{equation}
These functions can also be thought of as representing sections of a homogeneous line bundle over $P \bas G$
defined by the character $\chi_k$.  We represent them concretely by choosing specific representatives of $P \bas G$;
a simple choice is to use elements of the opposite nilpotent radical $\bar{N}$ generated by
$E_p, E_q, E$, \ti{i.e.} upper-triangular matrices.

So to compute concretely the action of some $E \in \fg$, we act on the upper-triangular matrix
\begin{equation}
\begin{pmatrix} 1&\tzeta+i \zeta&i\sigma-\frac12(\zeta^2+\tzeta^2)\\
&1&-\tzeta+i\zeta \\ &&1 \end{pmatrix}
\end{equation}
by $E$ and then act from the left by a suitable $X \in \fp$ to put the result back in upper
triangular form.  This gives a differential operator acting on $(\zeta, \tzeta, \sigma)$,
to which we must add $\chi_k(X)$ reflecting the twist by \eqref{rep-twist}.
The result is
\begin{subequations}
\label{su21quasi}
\begin{eqnarray}
{E}^{QC} &=& -\pa_\sigma\ ,\quad
{E_p}^{QC} = \pa_\tzeta +\zeta \pa_\sigma\ ,\quad
{E_q}^{QC} = -\pa_\zeta + \tzeta \pa_\sigma\ , \\
H^{QC}&=& -\tzeta \pa_\tzeta -\zeta \pa_\zeta -2 \sigma \pa_\sigma -k\ ,\quad
J^{QC}= - \tzeta\pa_\zeta + \zeta \pa_\tzeta\\
{F_p}^{QC}&=&\frac12 (3 \zeta^2-\tzeta^2)\pa_\tzeta +(\sigma - 2 \tzeta \zeta)\pa_\zeta
- \frac12\left[\zeta( \tzeta^2+ \zeta^2)+2\sigma \tzeta\right] \pa_\sigma -k \tzeta \\
{F_q}^{QC}&=&-\frac12(3\tzeta^2-\zeta^2)\pa_\zeta +(\sigma + 2 \tzeta \zeta)\pa_\tzeta
- \frac12\left[\tzeta( \tzeta^2+ \zeta^2)-2\sigma \tzeta\right] \pa_\sigma +k \zeta \\
{F}^{QC}&=&\frac12 \left[ \zeta(\tzeta^2+\zeta^2)+2\sigma \tzeta\right]\pa_\tzeta
-\frac12 \left[ \tzeta(\tzeta^2+\zeta^2)-2\sigma \zeta\right]\pa_\zeta\\
&&-\frac14 \left[ (\tzeta^2+\zeta^2)^2-4 \sigma^2 \right] \pa_\sigma + k  \sigma \nn
\end{eqnarray}
\end{subequations}
The quadratic and cubic Casimirs are constants,
\be
\label{c23qcsu21}
C_2= -k(4-k)/4\ ,\quad C_3 = 0
\ee
corresponding to $p=q=(k-4)/2$.

If we choose $k = 2 + is$ for $s \in \R$, then this representation is
infinitesimally unitary with respect to the $L^2$ inner product
with measure $d\zeta~d\tzeta~d\sigma$.

\subsubsection{Quaternionic discrete series \label{secqdisc}}

Above we constructed the principal series representations using the action of
$G$ on appropriate sections of line bundles over $P \bas G$.  There is a complex-analytic
analogue of this construction, described in \cite{MR1421947}, which uses instead the
action of $G$ on $\cZ$.

Formally the construction is easy to understand.  We gave the action of $G$
on holomorphic functions on $\cZ$ above in \eqref{su21holaction}.  It may be slightly generalized
to give formally a representation on holomorphic sections of the homogeneous line bundle $\cO(-k)$
over $\cZ$,
\begin{subequations}
\label{su21quasiQuat}
\begin{eqnarray}
{E}^{QC} &=& -\pa_\alpha\ ,\quad
{E_p}^{QC} = \pa_\txi +\xi \pa_\alpha\ ,\quad
{E_q}^{QC} = -\pa_\xi + \txi \pa_\alpha\ , \\
H^{QC}&=& -\txi \pa_\txi -\xi \pa_\xi -2 \alpha \pa_\alpha -k\ ,\quad
J^{QC}= - \txi\pa_\xi + \xi \pa_\txi\\
{F_p}^{QC}&=&\frac12 (3 \xi^2-\txi^2)\pa_\txi +(\alpha - 2 \txi \xi)\pa_\xi
- \frac12\left[\xi( \txi^2+ \xi^2)+2\alpha \txi\right] \pa_\alpha -k \txi \\
{F_q}^{QC}&=&-\frac12(3\txi^2-\xi^2)\pa_\xi +(\alpha + 2 \txi \xi)\pa_\txi
- \frac12\left[\txi( \txi^2+ \xi^2)-2\alpha \txi\right] \pa_\alpha +k \xi \\
{F}^{QC}&=&\frac12 \left[ \xi(\txi^2+\xi^2)+2\alpha \txi\right]\pa_\txi
-\frac12 \left[ \txi(\txi^2+\xi^2)-2\alpha \xi\right]\pa_\xi\\
&&-\frac14 \left[ (\txi^2+\xi^2)^2-4 \alpha^2 \right] \pa_\alpha + k  \alpha \nn
\end{eqnarray}
\end{subequations}
The differential operators \eqref{su21quasiQuat} are of course simply related to the
ones we wrote above in \eqref{su21quasi}, by replacing the real coordinates $(\zeta, \tzeta, \sigma)$
on $P \bas G$ by the complex $(\xi, \txi, \alpha)$ on $P_\C \bas G_\C$.

This representation is formally unitary with respect to the inner product 
\be \label{quatdiscksu21}
\langle f_1 | f_2 \rangle = \int_\cZ d\xi\, d\txi\, d\alpha\, d\bar\xi\, d\bar\txi\, d\bar\alpha\,
e^{(k-4) K_{\cZ}}\, f_1^*(\bar\xi,\bar\txi,\bar\alpha)\, f_2(\xi,\txi,\alpha) 
\ee 
Indeed, the invariant volume form on $\cZ$ is $e^{-4 K_{\cZ}} |d\xi d\txi d\alpha|^2$ (more generally $-4$
in the exponent would be replaced by $-2d-2$, where $d$ is the dimension of $K \bas G$)
while the factor $e^{k K_{\cZ}}$ comes from the Hermitian norm in $\cO(-k)$\cite{Neitzke:2007ke}.

However, this representation would appear to be trivial, since (for $k>0$) there are no
global sections of this line bundle, i.e. the zero-th cohomology $H^0(\cZ, \cO(-k))$ is empty.  In
\cite{MR1421947} the desired representation space is identified instead as $H^1(\cZ, \cO(-k))$.
Here we work directly with holomorphic sections possessing some singularities, which should be
understood as Cech representatives for classes in $H^1(\cZ, \cO(-k))$ with respect to coverings by
two open sets, in the spirit of the early literature on twistor theory.  It is not obvious that one
obtains all classes in $H^1(\cZ,\cO(-k))$ in this way, but in most of our considerations we
restrict ourselves to these, and indeed we simply write $f \in H^1(\cZ, \cO(-k))$ where $f$ is a
section with singularities.

The formal inner product \eqref{quatdiscksu21} then has to be
carefully interpreted, since $f_1$ and $f_2$ are Cech representatives
and hence only well defined up to certain shifts by holomorphic
functions; in order to get a well defined inner product, one must
interpret \eqref{quatdiscksu21} in a way that involves only contour
integrals.  Here we give only a formal heuristic account, which will
be adequate for our purposes in Section \ref{matpenrose}.  We begin by
analytically continuing $K_\cZ$ to a function on $\cZ \times
\bar{\cZ}$, obtained by considering the holomorphic and
antiholomorphic dependence independently.  Then for any $f_1 \in
H^1(\cZ, \cO(-k))$ we define $\hat{f_1} \in H^1(\bar{\cZ},
\bar\cO(k-4))$ by
\begin{equation} \label{def-tt}
\hat{f_1} = \int d\xi\, d\txi\, d\alpha\, e^{(k-4) K_\cZ} f_1,
\end{equation}
where the integral runs over some contour in $\cZ \times \bar{\cZ}$.
This construction is an analogue of the ``twistor transform''
discussed in \ti{e.g.} \cite{MR610183} for the case $\cZ = \C\PP^3$.
In \cite{MR610183} it is argued that this transform is involutive, $\hat{\hat{f}} = f$;
in Section \ref{matpenrose} we will assume that the same is true in
the present case.  (This is an analogue of the fact that on a Hermitian symmetric space
the K\"ahler potential behaves as a reproducing kernel for holomorphic functions.)

Complex conjugation gives $\bar{\hat{f_1}} \in H^1(\cZ, \cO(k-4))$,
which can be paired with $f_2 \in H^1(\cZ, \cO(-k))$ and
contour-integrated, so that
\begin{equation}
\IP{f_1 | f_2} = \int d\xi\, d\txi\, d\alpha\, \bar{\hat{f_1}} f_2.
\end{equation}
This is our interpretation of \eqref{quatdiscksu21}.
It is still formal, since we did not specify the contours of
integration.  In the example we consider below there will be a natural
choice.  In general, however, it is difficult to make sense of this
formal prescription, much less to check that it is positive definite;
one instead checks the existence of a positive definite norm by a
purely algebraic computation on a special basis of ``elementary
states'' (in our context, these are the $K$-finite vectors).

We do not perform such an analysis here, but rely on the results of
\cite{MR1421947}.  There one finds that for $k \ge 2$ the
spaces $H^1(\cZ, \cO(-k))$ are irreducible and unitarizable
representations of $G$.  For $k \ge 3$ they belong to the
discrete series and are called quaternionic discrete series
representations.  The representation at $k = 2$ is a limit
of the quaternionic discrete series.

\subsubsection{Quaternionic discrete series as subquotients of principal series}

These quaternionic discrete series representations are expected to be obtained as
subquotients of the principal series which we discussed above.  To understand why this
happens, we recall the simpler case of the unitary representations of $SL(2,\R)$.  There
one has a continuous principal series realized in a space of sections of a line bundle $\cL_k$
($k \in \C$) over the equator of $\C\PP^1 = B_+ \bas SL(2,\C)$.
If $k \in \Z$, then $\cL_k$ extends holomorphically over the whole of $\C\PP^1$.  
One then gets the holomorphic and antiholomorphic discrete series by looking at sections
which admit analytic continuation from the equator over respectively the northern or southern
hemisphere, then dividing out by the space of sections which can be continued over the whole
of $\C\PP^1$.

In the quaternionic case the situation is similar.  Firstly, the space $P \bas G$ occurs as part of
the boundary of $\cZ$ in an appropriate sense; in terms of our coordinates $(\xi, \txi, \alpha)$
on $\cZ$, this boundary is the locus where all coordinates become real (then they are identified
with our coordinates on $P \bas G$ by $\xi \to \zeta$, $\txi \to \tzeta$, $\alpha \to \sigma$.)
So for discrete values of $k$, one might expect to
obtain a submodule of the principal series representation by considering those
sections which are boundary values of holomorphic objects on $\cZ$, and then obtain a
unitary representation as some quotient thereof.\footnote{The notion of ``boundary value''
in this case is somewhat subtler than it was in the case of $SL(2,\R)$, because on $\cZ$ we deal not with
functions but with cohomology classes.  The reason is that the structure of $\cZ$ near
the boundary is more complicated than that of the upper half-plane; it is not contained in any
convex tube domain, essentially because of the circle parameterized by the phase of $z$, which winds
around the boundary.  The correct notion of boundary value in this case should involve integration over
this circle, as described in \ti{e.g.} \cite{MR1343164}.}
We will see this expectation realized in the algebraic discussion
of the next subsection.

\subsubsection{$K$-type decomposition \label{kdecsu21}}

We now discuss the decomposition of the principal series representation
under the maximal compact subgroup $K = SU(2) \times U(1)$, which we 
recall is generated by $J_\pm,J_3$ and $S$.

We begin by constructing the spherical vector, invariant under $K$.
For this purpose consider the action of $K$ from the right on our coset representative for $P \bas G$,
\begin{equation} \label{coset-gp}
n = \begin{pmatrix} 1&\tzeta+i \zeta&i\sigma-\frac12(\zeta^2+\tzeta^2)\\
&1&-\tzeta+i\zeta \\ &&1 \end{pmatrix}.
\end{equation}
$K$ acts on the three rows $v_i$, preserving their Hermitian norms
$\norm{v_i}^2$.  On the other hand, the action of $P$ from the left mixes the rows.  Since $P$ is lower-triangular, though,
its action on the top row is simple:  it just acts by the character $e^{\underline{H}+i\underline{J}/3}$.
Now consider the function
\be \label{sv-principal}
f_K = \norm{v_1}^{-k/2} = \left(1+ \tzeta^2 +\zeta^2 + \sigma^2 + \frac14 (\tzeta^2+\zeta^2)^{2}\right)^{-k/2}
\ee
as an element in the principal series representation.  By definition, to compute the action of $k \in K$ on $f_K$, we first
transform $n$ by $k$ acting from the right, then act by a compensating element $p \in P$ from the left to restore the
form \eqref{coset-gp}.  This modifies \eqref{sv-principal} by a factor $e^{- k \underline{H}(p) / 2}$.  However, we also have to
include the explicit factor $e^{k \underline{H}(p) / 2}$ from the definition of the principal series.  So altogether we find that
\eqref{sv-principal} is $K$-invariant.

More generally, the highest weight vectors of $SU(2)_J$
(vectors annihilated by $J_+$) are given by
\be
\label{su21ktype}
\begin{split}
f_{j,s}=&
\left[\frac{ \tzeta + i \zeta }
{1 + i \sigma + \frac12 (\zeta^2+\tzeta^2)} \right]
^{j-\frac13 s}
\left[\frac
{ 1 + i \sigma - \frac12 (\zeta^2+\tzeta^2)}
{ 1 - i \sigma + \frac12 (\zeta^2+\tzeta^2)} \right]
^{j+\frac13 s}\\
&\times\,\left[1+ \tzeta^2 +\zeta^2 + \sigma^2 + \frac14
(\tzeta^2+\zeta^2)^2 \right]^{-k/2}
\end{split}
\ee
with eigenvalues $j i$ and $ s i$ for $J_3$ and $S$, respectively.

In section \ref{lifthkc} below,
we shall see that these states have a simple expression
in terms of the moment maps of the action of $G$
on the symplectization of $P \bas G$.
For now, we observe that
the highest weight states $f_{j,s}$ are mapped to each other
by the action of the raising operators $J_{\frac12,\pm\frac32}$:
\bea
J_{\frac12,\frac32}\cdot f_{j,s} &=&
\frac{1}{3} (3k+6 j+2 s)\,f_{j+\frac12,s+\frac32}\nn\\
J_{\frac12,-\frac32}\cdot f_{j,s} &=&
\frac{2\sqrt{2}}{3} (3k+6 j-2 s)\,f_{j+\frac12,s-\frac32}
\eea
Applying the lowering operators  $J_{-\frac12,\pm\frac32}$ gives a linear
combination of the highest weight state $f_{j-\frac12,s\pm\frac32}$
and a descendant of $f_{j+\frac12,s\pm\frac32}$:
\bea
J_{-\frac12,\frac32}\cdot f_{j,s} &=&
\frac{(3j-s)(6+6j-3k-2s)}{18\sqrt{2}(2j+1)}\,
f_{j-\frac12,s+\frac32}
+ \frac{3k+6 j+2 s}{3(2 j+1)}\, J_{-1}\cdot
f_{j+\frac12,s+\frac32}\nn\\
J_{-\frac12,-\frac32}\cdot f_{j,s} &=&
\frac{(3j+s)(6+6j-3k+2s)}{9(2j+1)}\,
f_{j-\frac12,s-\frac32}
+ 2\sqrt{2} \, \frac{3k+ 6 j-2 s}{3(2 j+1)}\, J_{-1}\cdot
f_{j+\frac12,s-\frac32}\nn\\
\eea

Using these equations, we may now study the structure of the
module generated by $f_{j,s}$ and its descendants; it is pictured in
Figure \ref{su21module}.

For generic $k$ this module is irreducible and not manifestly unitarizable.
When $k$ is an integer, the situation is more interesting.  For even integers $k \ge 2$, there is an
irreducible submodule generated by $f_{j=(k-2)/2,s=0}$.  Its $K$-type
decomposition coincides with that of the representation labeled by
$p=q=(k-4)/2$ in the parameterization of \cite{Bars:1989bb}, namely,
\be
\bigoplus_{m=k-2}^{\infty} \, \bigoplus_{s=-3m/2}^{3m/2}\,
[j=m/2]_{s}
\ee
We identify it as the quaternionic discrete series with index $k$ (or limit discrete
series for $k=2$) \cite{MR1421947}; in particular, it is unitarizable.
It has no spherical vector unless $k=2$.

For (possibly negative) even integers $k \le 2$, we can similarly obtain the representation with
$p=q=-k/2$, this time as a quotient instead of a submodule; namely, we divide out the submodule
consisting of all states with $3j-|s|<3(k-2)/2$.  The resulting representations
are equivalent to the ones just discussed.

\EPSFIGURE{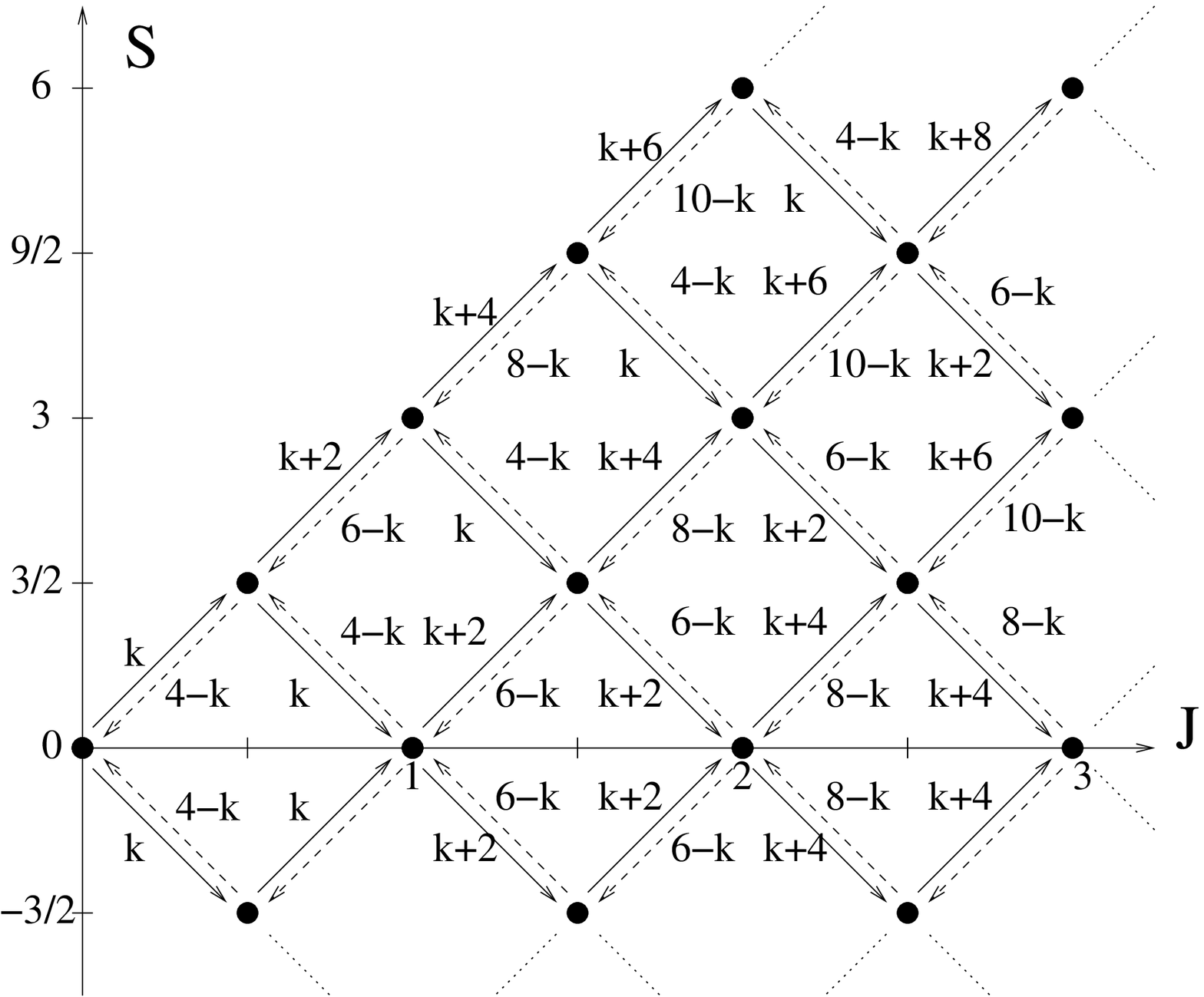,height=8cm}{Structure of the module generated by the
highest weights $f_{j,s}$. The solid (resp. dotted) arrows denote the
action of the raising (resp. lowering) operators, with coefficient
proportional to the indicated function of $k$. \label{su21module}}

\FIGURE{
\centerline{
\includegraphics[height=5cm]{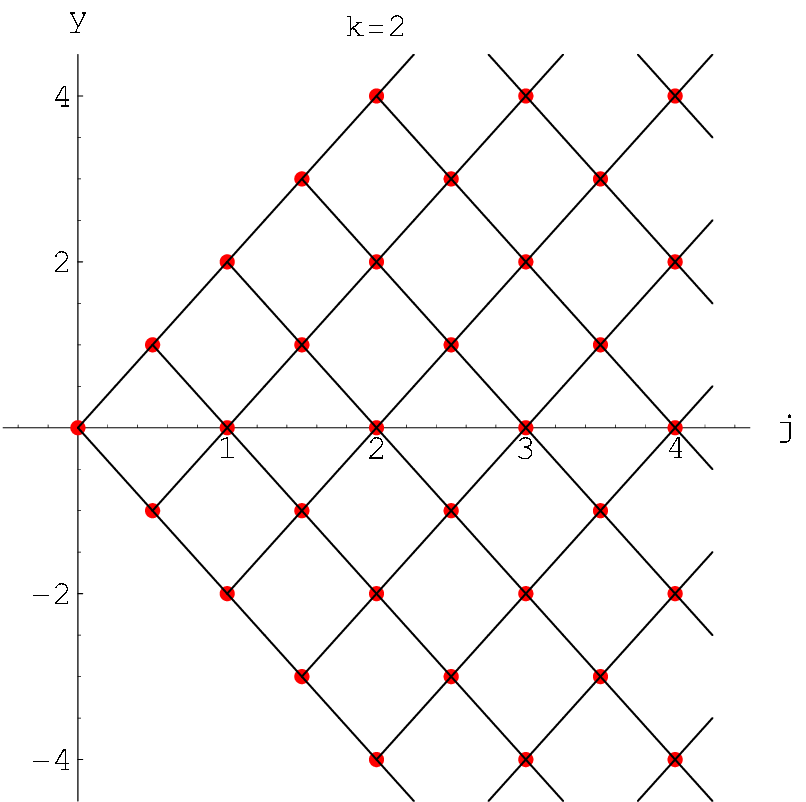}\hfill
\includegraphics[height=5cm]{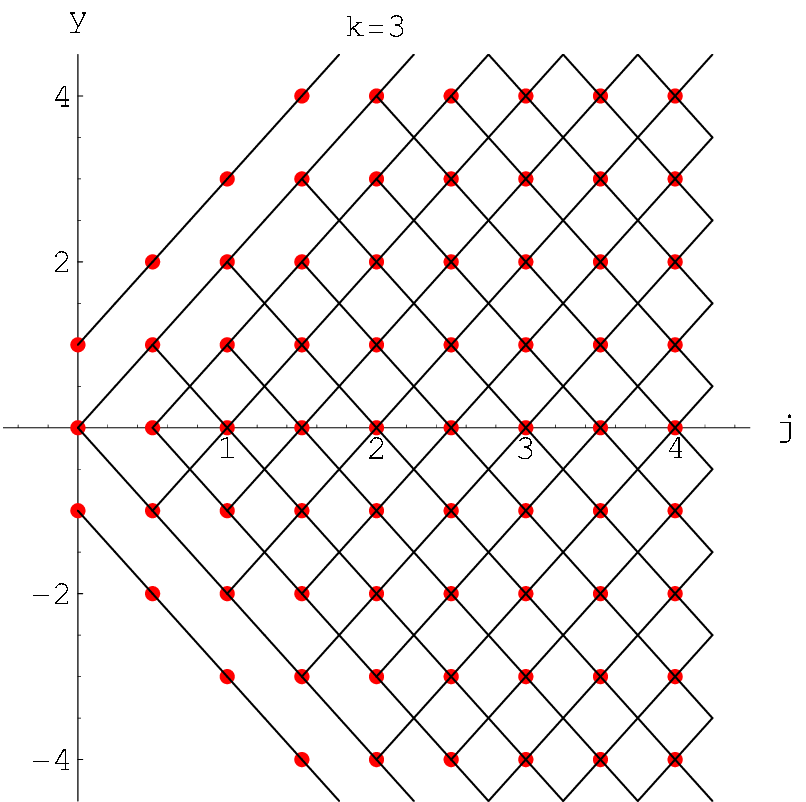}\hfill
\includegraphics[height=5cm]{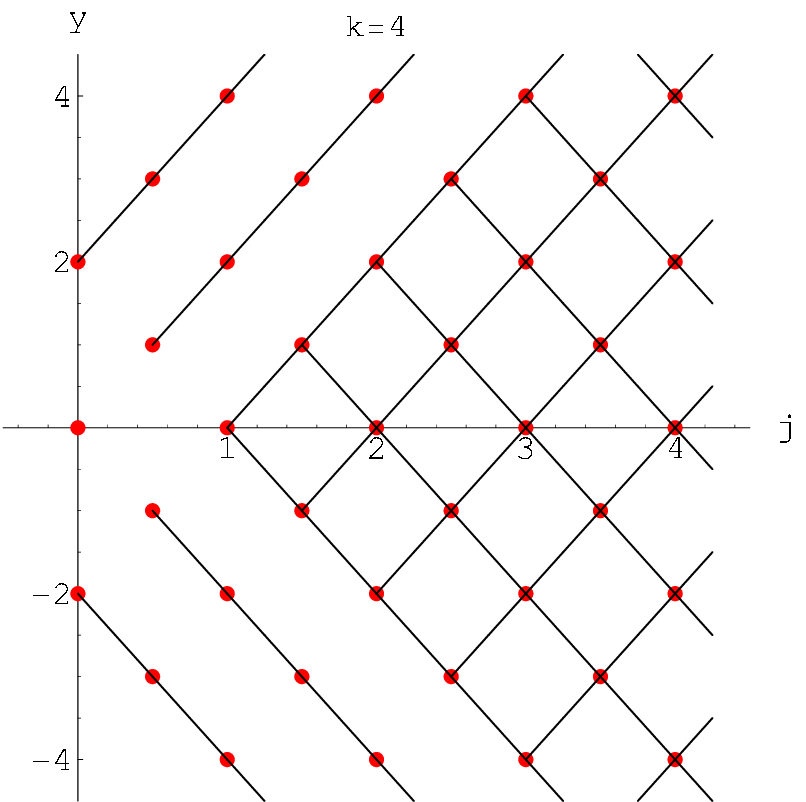}
}\\[1cm]
\centerline{
\includegraphics[height=5cm]{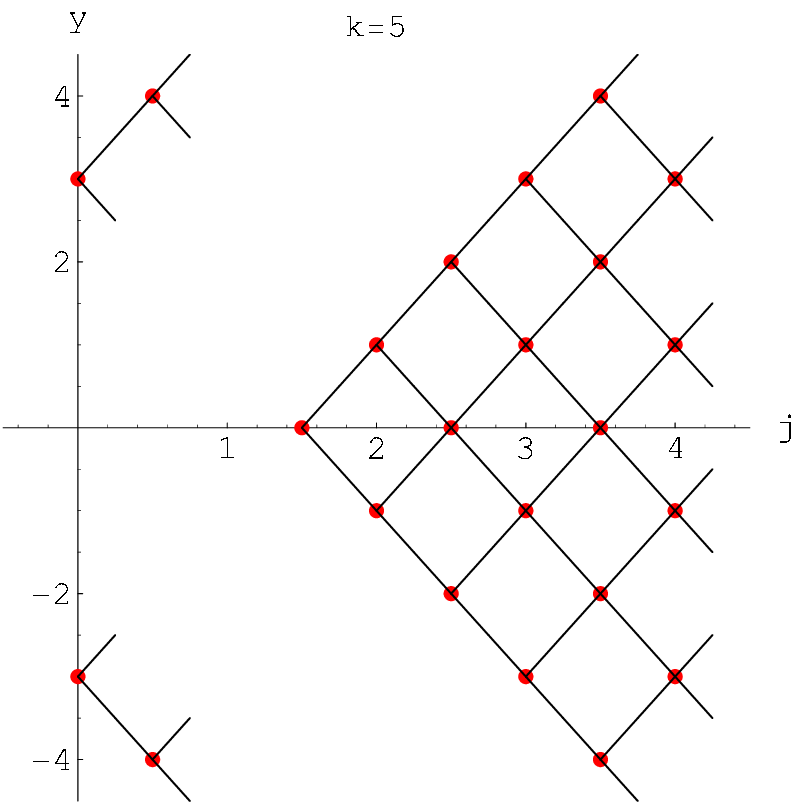}\hfill
\includegraphics[height=5cm]{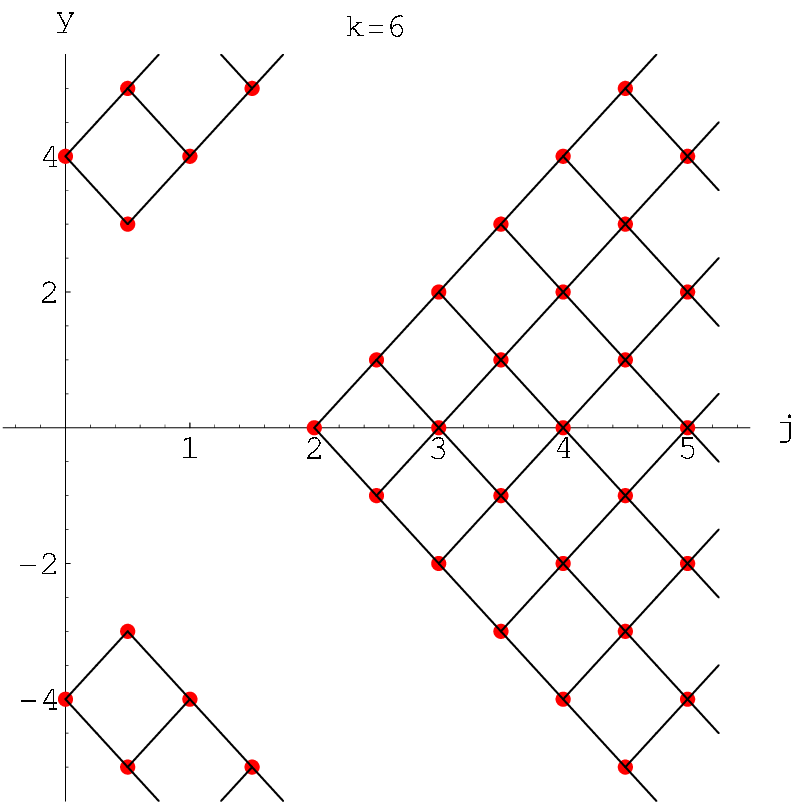}\hfill
\includegraphics[height=5cm]{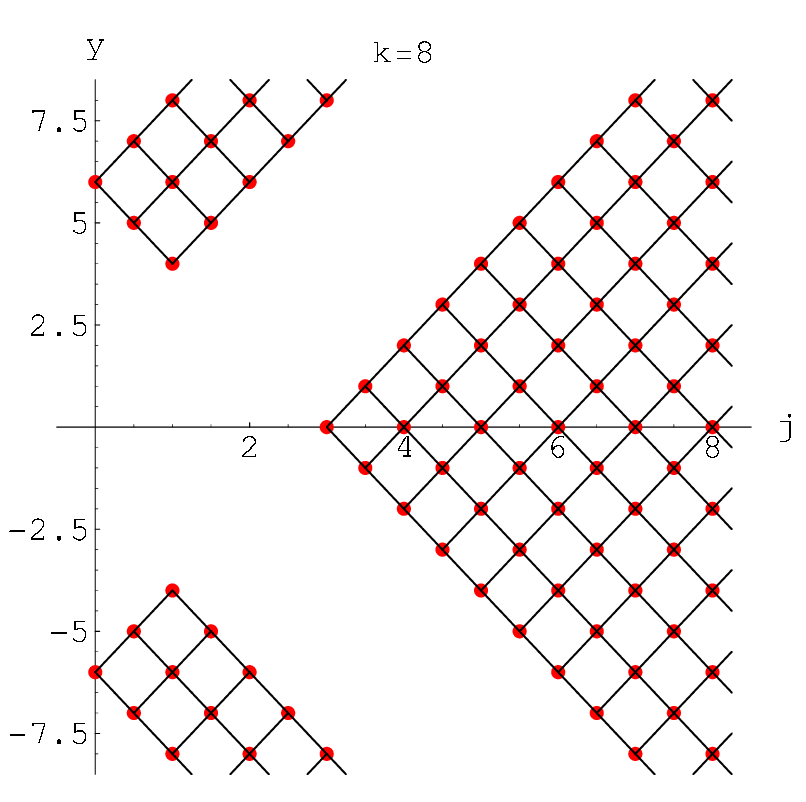}}
\caption{$K$-type decomposition of the discrete series representations
of $SU(2,1)$ in the $(j,y=2/3s)$ plane, for low
values of $k=(p-4)/2=(q-4)/2$. The quaternionic discrete series corresponds
to the ``$p+q$ discrete branch'' in the terminology of \cite{Bars:1989bb}.
The ``$p$-discrete'' branch for $k=3$
corresponds to the minimal representation, see Section \ref{su21kdec}.
The second ``$p$-discrete branch'' for $k=4$,
starting at $(j,s)=(1/2,\pm 3/2)$,
corresponds to the deformed minimal representation at $\nu=\pm 1$.
For $k=5$ and higher odd values of
$k$, there are no ``$p$-discrete'' or ``$q$-discrete'' branches, as
the lowering operators $J_{\frac12,\pm\frac32}$ map into
forbidden regions, as illustrated by the open links.\label{barsfig}}
}

In \cite{MR1421947} one also finds quaternionic discrete series representations 
for odd $k$; one might wonder why we did not encounter those above.  
The answer is that strictly speaking they are not representations of $G = SU(2,1)$ but
rather of its double cover.  Correspondingly, they do not appear as subrepresentations 
of the principal series we considered here, but of a closely related ``non-spherical 
principal series'' representation of the double cover.

\subsubsection{Matrix elements and Penrose transform \label{matpenrose}}

Suppose $\rho$ is a spherical unitary representation of $G$, with spherical
vector $f_K$.  Then $\rho$ can be realized in the space of functions on $K\backslash G$,
by mapping any state $f$ to the matrix element
\be
\label{emqcqk}
\varphi(e_{QK}) = \langle f | \rho(e_{QK}^{-1}) f_K \rangle
\ee
$\varphi$ is a function on $K\backslash G$ because the left action
of $k \in K$ on $e_{QK}$ becomes
a right action on $e_{QK}^{-1}$, hence a left action on $f_K$, which is
trivial because $f_K$ is spherical.  Moreover, the $\varphi$ so obtained
obeys differential equations determined by $\rho$; in particular, it is
an eigenfunction of the Laplacian on $K\backslash G$ with eigenvalue
$2 C_2(\rho)$.

Even if $\rho$ is not spherical one may still apply this construction
replacing $f_K$ by any $K$-finite vector, and thus embed $\rho$ into a
space of sections of a homogeneous vector bundle over $K \backslash G$,
induced from the representation of $K$ in which $f_K$ transforms.
The sections so obtained have particularly good properties if $f_K$ is
in the lowest $K$-type (\ti{e.g.} for holomorphic discrete series representations
they turn out to be holomorphic sections).

We now apply this construction to the quaternionic discrete series
representations, beginning with the special case $k=2$.
Using the Iwasawa decomposition \eqref{su21iwa}
and the Baker-Campbell-Hausdorff formula $e^A e^B=e^B e^A e^{[A,B]}$
when $[A,B]$ is central,
\be
e_{QK}^{-1} =
e^{-(\sigma-\zeta\tzeta) {E}} \cdot
e^{\zeta {E_q}} \cdot
e^{-\tzeta {E_p}}\cdot
e^{-U H}
\ee
and
\be
\rho(e_{QK}^{-1}) =
e^{-(\sigma-\txi\zeta+\xi\tzeta)\pa_\alpha} \cdot
e^{-\zeta \pa_\xi} \cdot
e^{-\tzeta \pa_\txi}\cdot
e^{U (\pa_U + \xi \pa_\xi + \txi \pa_\txi + 2 \alpha \pa_\alpha + k)}.
\ee
Applying this $\rho(e_{QK}^{-1})$ to the spherical vector $f_K=f_{0,0}$ yields
\be
\label{sphsu21nu}
\begin{split}
\Psi(U,\zeta,\tzeta,\sigma; \xi, \txi, \alpha) =&
\left\{
e^{2U} + (\txi-\tzeta)^2
+ (\xi-\zeta)^2 \right. \\ & \left.
+  e^{-2U} \left[
(\sigma+\alpha+\txi\zeta-\xi\tzeta)^2
+ \frac14 \left(  (\txi-\tzeta)^2  + (\xi-\zeta)^2 \right)^2 \right]
\right\}^{-1}
\end{split}
\ee
Since $C_2(\rho) = -1$, taking the inner product \eqref{emqcqk}
between $f$ and \eqref{sphsu21nu} yields
an eigenfunction of the conformal Laplacian on $K \bas G$,
\be
\label{conflap}
\left[ \Delta_{QK} + 2 \right]\, \varphi = 0
\ee
Thus the transformed spherical vector \eqref{sphsu21nu} is the kernel for an
integral operator which intertwines between the quaternionic discrete series
and a subspace of the space of functions on $K\backslash G$. In physics parlance, it is related to
the boundary to bulk propagator for \eqref{conflap}, upon taking the coordinates $\xi$, $\txi$, $\alpha$
to be real --- in that case they lie on a part of the boundary of $\cZ$, which may be identified with a boundary
of $K \backslash G$. (See \cite{Britto-Pacumio:1999sn} for an attempt to set up a bulk-boundary
correspondence on this space.)

Since $K \backslash G$ is a quaternionic-K\"ahler manifold
there is an {\it a priori} different way to construct a
function on $K\backslash G$ from $f(\xi,\txi,\alpha) \in H^1(\cZ,\cO(-2))$, which is
to apply the quaternionic Penrose transform \cite{MR1165872,quatman}.
Some details of this correspondence for quaternionic-K\"ahler spaces
obtained by the $c$-map construction were worked out in \cite{Neitzke:2007ke},
where a simple contour integral formula for the Penrose transform was
established:
\be
\label{twisu21}
\mbox{Penrose}[f] = \varphi(U,\zeta,\tzeta,\sigma) =
2\,e^{2U}\, \oint
\frac{dz}{z}\, f\left[ \xi(z), \txi(z), \alpha(z) \right]
\ee

The resulting $\varphi$ is annihilated by the conformal Laplacian $\Delta_{QK} + 2$.
This agrees with \eqref{conflap} in the case of $K \bas G$, and
in this context it can be checked directly by acting with
$\Delta_{QK}$ on the integrand of \eqref{twisu21} and using \eqref{su21quasi}:
\be
\left[ \Delta_{QK}+ 2 -  \pa_z \left( \frac{z^2}{2} \pa_z +\frac12
z \pa_z+ 2 i e^{2U} z \pa_\alpha\right) \right] \,
z^{-1} f\left[ \xi(z), \txi(z), \alpha(z) \right] =0
\ee
Integration over $z$ eliminates the total derivative, leading to \eqref{conflap}.

We now argue that these two constructions are in fact identical, i.e.
\begin{equation}
\langle f | \rho(e_{QK}^{-1}) f_K \rangle = \mbox{Penrose}[f].
\end{equation}
According to our discussion of the inner product above, this is equivalent to
\begin{equation} \label{penrose-inner}
\int d\xi_0\,d\txi_0\,d\alpha_0\,f\,\Psi = \mbox{Penrose}[\hat{f}],
\end{equation}
where $\Psi = \rho(e^{-1}_{QK}) f_K$; then using \eqref{def-tt} and the
explicit form of $K_\cZ$, \eqref{penrose-inner} is
equivalent to requiring that $\Psi$ arises as the Penrose transform of
\be
\Phi(\xi,\tilde\xi,\alpha) = \left[ (\alpha-\alpha_0
+\txi_0 \xi - \txi \xi_0)^2
+ \frac14 \left[(\xi-\xi_0)^2+(\txi-\txi_0)^2\right]^2 \right]^{-1}
\ee
But this we can evaluate directly: the contour integral \eqref{twisu21} defining
$\mbox{Penrose}[\Phi]$ has poles at $z=z_\pm$,
where
\be
z_+ = \frac{2i\sqrt{2} e^U\ \left[(\zeta-\xi_0)-i(\tzeta-\txi_0)\right]}
{2i(\sigma-\alpha_0+\txi_0 \zeta - \tzeta \xi_0)
+ (\zeta-\xi_0)^2+(\tzeta-\txi_0)^2-2 e^{2U}}
\ee
and $z_-=-1/{\bar z_+}$. The residue at $z=z_\pm$ yields
\be
\begin{split}
\Psi =& \left\{ e^{2U} + (\zeta-\xi_0)^2+(\tzeta-\txi_0)^2 \right.\\
&\left.
+ e^{-2U} \left[ (\sigma-\sigma_0+\txi_0 \zeta - \tzeta \xi_0)^2
+ \frac14 \left[(\zeta-\xi_0)^2+(\zeta-\xi_0)^2\right]^2\right]
\right\}^{-1}
\end{split}
\ee
which indeed agrees with the formula \eqref{sphsu21nu} for $\rho(e^{-1}_{QK}) f_K$ as desired.

Similar considerations apply for other even values of $k$.  In that
case the Penrose transform gives a section of $Sym^{k-2}(H)$;
for $c$-map spaces, in a natural trivialization of $H$, the formula
is given in \cite{Neitzke:2007ke} as
\begin{equation}
\varphi_m(U,\zeta,\tzeta,\sigma) =
2\,e^{kU}\, \oint
\frac{dz}{z}\, z^{\half m}\,f\left[ \xi(z), \txi(z), \alpha(z) \right]
\end{equation}
where $m = -k+2, \dots, k-2$ labels the $2k-3$ components of $\varphi$.
This turns out to agree with the matrix element construction, where we
now use the $(2k-3)$-dimensional lowest $K$-type of the quaternionic
discrete series.
Establishing a similar correspondence for $k$ odd would require
a better understanding of the branch cuts appearing in the contour
integral \eqref{twisu21}.

\subsubsection{Causal structure and quartic light-cone\label{causala2}}

A general fact about twistor spaces of four-dimensional conformally
self-dual manifolds is that two points $x$, $x'$ are lightlike
separated if and only if their corresponding twistor lines $L_x$, $L_{x'}$
intersect in $\cZ$.  (Since $K \bas G$ has Euclidean signature, we must
of course allow $x$ and $x'$ to belong to its complexification if this
condition is to be satisfied.)

Using the twistor map \eqref{zetatoxi},
the condition for $(U,\zeta,\tzeta,\sigma)$
and $(U',\zeta',\tzeta',\sigma')$ to be light-like separated is therefore
that there exist $z$ and $z'$ such that
\bea
\label{elimzzp}
\zeta - \frac{i}{\sqrt{2}} e^{U} \left( z + z^{-1} \right)
&=&\zeta' - \frac{i}{\sqrt{2}} e^{U'} \left( z' + z^{'-1} \right)
\nn \\
\tzeta + \frac{1}{\sqrt{2}} e^{U} \left( z - z^{-1} \right)
&=& \tzeta' + \frac{1}{\sqrt{2}} e^{U'} \left( z' - z^{'-1} \right)
\\
\sigma + \frac{1}{\sqrt{2}} e^{U}
\left[  z ( \zeta+i\tzeta) + z^{-1} (-\zeta+i \tzeta) \right]
&=&\sigma' + \frac{1}{\sqrt{2}} e^{U'}
\left[  z' ( \zeta'+i\tzeta') + z^{'-1} (-\zeta'+i \tzeta') \right] \nn
\eea
Eliminating $z$ and $z'$ from the first two equations, we find
\bse
\bea
z&=& \frac{i\ e^{-U}}{2 \sqrt{2} (\Delta\zeta+i \Delta\tzeta)}
 \left(2 (e^{2 U'} - e^{2 U}) + \Delta\zeta^2 -\Delta\tzeta^2 +
  2 \sqrt{\Delta_4} \right) \\
z'&=& \frac{i\ e^{-U'}}{2 \sqrt{2} (\Delta\zeta+i \Delta\tzeta)}
 \left(2 (e^{2 U} - e^{2 U'}) +\Delta\zeta^2 + \Delta\tzeta^2 +
  2 \sqrt{\Delta_4} \right)
\eea
\ese
where we denoted
\be
\label{defdel}
\Delta\zeta=\zeta'-\zeta\ ,\quad\Delta\tzeta=\tzeta'-\tzeta\ ,\quad
\Delta \sigma = \sigma'-\sigma +\zeta' \tzeta - \zeta \tzeta'
\ee
and
\be
\Delta_4 = \frac14(\Delta\zeta^2+\Delta\tzeta^2)^2
+ (e^{2 U} + e^{2 U'})(\Delta\zeta^2 +\Delta\tzeta^2)
+ (e^{2 U} - e^{2 U'})^2
\ee
Reinserting in the third equation in \eqref{elimzzp}, we find
\be
\Delta\equiv (\Delta \sigma)^2+\frac{1}{4}
   \left(\Delta\zeta^2+\Delta\tzeta^2\right)^2+
(e^{2 U}+e^{2U'})
\left(\Delta\zeta^2+\Delta\tzeta^2\right) + (e^{2U}-e^{2U'})^2 = 0
\ee
For $U,U' \to-\infty$, we recognize the familiar
quartic norm $N_4$ from \eqref{kzxi}, now evaluated at real values
of its arguments. If only $U'$ is sent to $-\infty$, we recover
the ``transformed spherical vector'' \eqref{sphsu21nu}.
Expanding to quadratic order in the variations \eqref{defdel},
we find
\be
\Delta\sim 2 e^{4U} \left[
2\,dU^2 + e^{-2U} \left(d\tzeta^2 + d\zeta^2\right)
+\frac12 e^{-4U} \left(d\sigma+\tzeta d\zeta - \zeta d \tzeta\right)^2 \right]
\ee
which indeed vanishes for infinitesimal light-like displacements
under the metric \eqref{ds2su21}.

Since we are discussing issues of causal structure and group theory, we make a side comment here.
The groups $SU(2,n)$ are the only quaternionic groups that admit positive energy
unitary representations. The $U(1)$ generator in the maximal compact subgroup $U(1)\times SU(n)$ is the generator whose spectrum is bounded from below for such representations and hence can be taken as the Hamiltonian for a causal
time  evolution. The other quaternionic groups do not have any generators whose spectrum is bounded
from below for any unitary representation. Hence for these groups the Hamiltonian with a bounded spectrum that describes causal evolution can not be one of the generators.

\subsection{Minimal representation}

For any real Lie group $G$ there is a notion of
``minimal unitary representation''
introduced in \cite{MR0342049} and much studied thereafter
(see \ti{e.g.} \cite{minrep-review} for a recent review.)  For many $G$
the minimal representation can be characterized as the unitary representation
of smallest Gelfand-Kirillov dimension.  For $G = SU(2,1)$ the situation is
somewhat degenerate and there are many representations sharing this minimal
dimension, including the holomorphic and antiholomorphic discrete series \footnote{ In fact, this is true for the entire quaternionic family $SU(n,2)$.}.
We focus here on the representation constructed
explicitly in \cite{Gunaydin:2000xr} by truncation of the minimal unitary realization
of $E_{8(8)}$  and whose structure
parallels that of the minimal representation of higher rank groups.
With a slight change of
notation and normalization relative to this  reference,
the generators can be written as
\bse
\label{gknsu21}
\bea
E_p = i x u \ ,&\quad&
F_p = i \pa_u \pa_x +\frac{i}{2x} \left( u^3 - \pa_u u \pa_u \right) \\
E_q = x \pa_u \ ,&\quad&
F_q = u \pa_x +\frac{i}{2x} \left(i \pa_u^3 - i u \pa_u u  \right) \\
E   = \frac{i}{2} x^2 \ ,&\quad&
F   = \frac{i}{2} \pa_x^2 +  \frac{i}{8x^2}
\left[ \left(u^2 - \pa_u^2 \right)^2 -1 \right]\\
H   = x \pa_x + \frac12 \ ,&\quad&
J   = \frac{i}{2} \left( \pa_u^2 - u^2 \right)
\eea
\ese
acting on functions of two variables $(u,x)$.
Equivalently, by defining $y= x^2$ and
$x_0=x u$, we reach a presentation analogous to the one
used in \cite{Kazhdan:2001nx}
for split groups,\footnote{The split real form of $SU(2,1)$ is $SL(3,\mathbb{R})$.}
\bse
\label{kpwsu21}
\bea
E_p = i x_0 \ ,&\quad&
F_p = \frac{i}{2} x_0 \pa^2_0+2 i y \pa_0 \pa_y
+ \frac{i x_0^3}{2 y^2}+ \frac{i}{2} \pa_0 \\
E_q = y \pa_0 \ ,&\quad&
F_q = -\frac12 y \pa_0^3 + 2 x_0 \pa_y  + \frac{3 x_0^2}{2y} \pa_0
+ \frac{x_0}{2y}\\
E   = \frac{i}{2} y \ ,&\quad&
H   = x_0 \pa_0 + 2 y \pa_y + \frac12 \ ,\quad
J   = \frac{i}{2} ( y \pa_0^2 - \frac{x_0^2}{y} )
\eea
\be
F   =2 i y \pa_y^2 - i \frac{x_0^4}{8 y^3}+\frac{i x_0}{2y} \pa_0
+\frac{3i x_0^2}{4y} \pa_0^2-\frac{i}{8} y \pa_0^4
+i \pa_y+2 i x_0 \pa_y \pa_0+\frac{3i}{8y}
\ee
\ese
Any minimal representation is annihilated by the Joseph ideal in the universal enveloping
algebra of $\fg$.  This means that the generators of the minimal
representations satisfy certain quadratic identities, \ti{e.g.}
\begin{subequations}
\label{opadsu21}
\bea
H^2 + 2 (E F + F E) + J^2 +1 &=& 0 \\
E_p^2+E_q^2+4 J E &=& 0\\
C_2(J) + \frac19 S^2 +\frac14 &=& 0.
\eea
\end{subequations}
These hold in addition to the Casimir identities
\be
\label{c2minrepsu21}
C_2 = -\frac34\ ,\quad C_3 = 0
\ee
corresponding to the parameters $(p,q)=(-\frac12,-\frac12)$ in the classification of
\cite{Bars:1989bb}.

The generators \eqref{gknsu21} or \eqref{kpwsu21}
are antihermitean with respect to the inner product
\be
\label{su21minrepinner}
\langle f_1 | f_2 \rangle = \int y^{-1} dy\, dx_0\, f_1^*(y,x_0) f_2(y,x_0) =
\int dx\,du f_1^*(u,x) f_2(u,x) \ ,
\ee
so this representation is unitary.
For later reference, we note that the (non-normalizable) states
\be
y^{\frac12}\,\exp\left( \pm \frac{x_0^2}{2y} \right)
= x\, \exp\left( \pm \frac12 u^2 \right)
\ee
are invariant under the nilpotent radical $N$
generated by $F,F_p,F_q$, and carry charges $(3/2,\pm i/2)$ under
the Cartan generators $(H,J)$.

\subsubsection{Induction from the maximal parabolic and deformation}

The minimal representation, in fact a one-parameter deformation thereof,
can be obtained by induction from the maximal parabolic subgroup
$Q \subset G_\C$ generated by $\{F,F_p,F_q,H,J,E_p+i E_q\}$.
For this purpose,
decompose any element of $G_{\IC}$ into a product
\be
g = \begin{pmatrix} * & 0 & 0 \\
* & * & *\\
* & * & *
\end{pmatrix}
\cdot
\begin{pmatrix}
1& z & i a \\
& 1 & 0\\
& & 1
\end{pmatrix}
\ee
Then induction from the character $\exp[\tau(\underline{H}
+ i\underline{J}/3)]$ of $Q$ gives the action of $G$ on sections $f(z,a)$
over $Q \backslash G_\C$ by first order differential operators,
\bse
\label{su21za}
\bea
E_p = \pa_z + i z \pa_a &,&F_p = -(ia+z^2)\pa_z- a z \pa_a + \tau z\\
E_q = -i \pa_z - z \pa_a &,& F_q = -(a+i z^2)\pa_z -i a z \pa_a + i \tau z\\
E = \pa_a &,& F = -a z \pa_z - a^2 \pa_a + \tau a\\
H = -z\pa_z -2 a \pa_a + \tau  &,& J = -i z \pa_z+ i\frac{\tau}{3}
\eea
\ese
Set $\nu=-(2\tau+3)/3$.
Passing from $f(z,a)$ to $f(u,x)$ by the intertwining operator
\be
f(z,a) =
\int du dx\, x^{1+4 \nu}\,
e^{-\frac12 u^2 + i u x z + \frac14 x^2 z^2 + \frac{i}{2} a x^2}\, f(u,x)
\ee
the action of $G$ on $f(u,x)$ is given by a one-parameter
deformation\footnote{The fact that the minimal representation is not isolated
is a peculiarity of the $A$ series.}
of the minimal representation \eqref{gknsu21},
\bse
\label{gknsu21def}
\bea
E_p^{(\nu)}&=&E_p\ ,\quad E_q^{(\nu)}=E_q\ ,\quad E_k^{(\nu)} = E_k \\
H^{(\nu)}&=&H+\frac52\nu\ ,\quad J^{(\nu)}=J- \frac{i}{2}\nu\\
F_p^{(\nu)}&=&F_p+\nu\frac{i}{2x}(3u+5\pa_u)\ ,\quad
F_q^{(\nu)}=F_q + \nu\frac{1}{2x}(5u+3\pa_u)\ ,\\
F^{(\nu)}&=&F + \nu\frac{i}{4x^2} \left[ 3 \pa_u^2+10 x\pa_x+3 (1-u^2) \right]
+\frac{2i}{x^2}\nu(\nu-1)
\eea
\ese
with Casimirs
\be
\label{c2minrepsu21gen}
C_2 = \frac34 (\nu^2-1)\ ,\quad C_3 = i\,\nu (1-\nu^2)\ ,
\ee
corresponding to
\be
(p,q)=\left( -\frac12 (1-3\nu), -\frac12(1+3\nu) \right)
\ee
in the notation of \cite{Bars:1989bb}.
The resulting representation is not obviously unitary for $\nu\neq 0$,
as the inducing character $\exp[\tau (\underline{H}+ i\underline{J}/3)]$ is in general not
unitary. We shall however find evidence in the next section
that it is unitarizable at $\nu=\pm 1$.

The annihilator of the $\nu$-deformed minimal
representation is deformed to
\bse
\label{defjoseph}
\bea
H^2 + 2 (E F + F E) + J^2 -2i \nu \, J +(1-\nu^2) &=& 0 \label{defjoseph1}\\
E_p^2+E_q^2 + 4 J\, E + 2 i \nu\, E&=& 0 \label{defjoseph2}\\
C_2(J) + \frac19 S^2 +3 i \nu \, S + \frac14(1-\nu^2) &=& 0\label{defjoseph3}
\eea
\ese
while the vectors invariant under the nilpotent radical $N$
are deformed to
\be
\label{pinvsu21}
y^{\frac{1-\nu}{2}}\,\exp\left( -\frac{x_0^2}{2y} \right)\ ,\quad
y^{\frac{1-4\nu}{2}}\,\exp\left( \frac{x_0^2}{2y} \right)
\ee
carrying charges $(\frac32(1+\nu),-\frac{i}{2}(1+\nu))$ and
$(\frac32(1-\nu),\frac{i}{2}(1-\nu))$ under $(H,J)$, respectively.

\subsubsection{$K$-type decomposition \label{su21kdec}}
For completeness, we now review the $K$-type decomposition of the minimal
representation,
i.e. the decomposition under the maximal compact subgroup $SU(2)\times U(1)$,
as discussed in \cite{Gunaydin:2000xr} \footnote{ We should note that the positive and negative grade generators we define in this paper are opposite to those of \cite{Gunaydin:2000xr}.}. For this purpose, we change
polarization to oscillator representation for both the $u$ and $x$ variable,
i.e. define $a_u$, $a_u^\dagger$, $N$ by
\be
\frac{1}{\sqrt{2}} (u - \pa_u) = a_u^\dagger\ ,\quad
\frac{1}{\sqrt{2}} (u + \pa_u) = a_u\ ,\quad N_u = a^\dagger_u a_u
\ee
and similarly for $x$. The compact generator $S$ is
manifestly positive,
\be
S = \frac38 \left( 2 N_x + 1 + x^{-2} N_u ( N_u+1) + 2 N_u + 1 \right)
\ee
so the representation is of lowest-weight type. The positive grade
generators in the 3-grading by $S$ read
\bse
\bea
K_+ &=& a_u a_x + \frac{\sqrt{2}}{x} (N_u+1) a_u \\
L_+ &=& \frac12 a_x^2 - \frac{1}{4x^2} N_u(N_u+1)
\eea
\ese
The only normalizable state annihilated by these two generators is
$a^\dagger_x |0_{u,x}\rangle$, or, in the real polarization,
\be
\label{su21mink}
f_{K} = x\,\exp\left[-\frac12(u^2+x^2) \right]
= y^{1/2}\, \exp\left[ -\frac12 \left(y + \frac{x_0^2}{y} \right) \right]
\ee
It is easy to check that this generator is a singlet of $SU(2)_{J_\pm,J_3}$,
but carries a non-zero charge $-3i/2$ under $S$.
By acting with the raising operators $K_-$ and $L_-$, we generate
the complete $K$-type decomposition of the minimal representation,
\be
\bigoplus_{m=0}^{\infty} \left[ \frac{m}{2} \right]_{-\frac{3i}{2}(m+1)}
\ee
where the term in bracket is the spin of the $SU(2)_J$ representation,
and the subscript indicates the $S$ charge. This agrees with the
$K$-type decomposition of the ``$p$-discrete'' module at $p=q=-1/2$, as seen on
Figure \ref{barsfig}, for $k=3$.

Let us now briefly discuss the $\nu$-deformed minimal representation.
It is easy to check that
\be
\label{su21minkVector}
f_{0} = x^{1-\nu}\,\exp\left[-\frac12(u^2+x^2) \right]
\ee
is a singlet of $SU(2)_J$, with charge $-\frac{3i}{2}(1+\nu)$ under
$S$, and annihilated by the deformed generators $K_+,L_+$. Acting with deformed generators $K_-,L_-$ produces
a $SU(2)_J$ doublet with $S_3=-\frac{3i}{2}(2+\nu)$,
\be
f_{\frac12}= -i\,u\, x^{2-\nu} \,\exp\left[-\frac12(u^2+x^2) \right]\ ,\quad
f_{-\frac12}=\frac{i}{2\sqrt2}( 3 + 3\nu-2 y^2 )\,
\exp\left[-\frac12(u^2+x^2) \right]
\ee
For $\nu=-1$, both of these states are annihilated by $K_+, L_+$,
so generate a module corresponding to the semi-infinite
line $s=-3/2 j$ in the diagram
on Figure \ref{barsfig} for $k=4$. Thus, the $K$-type decomposition of the
$\nu$-deformed minimal representation at $\nu=-1$ has a ladder structure
\be
\bigoplus_{m=1}^{\infty} \left[ \frac{m}{2} \right]_{-\frac{3i}{2}m}
\ee

As usual, we can use the lowest $K$-type to embed the minimal representation
into the space of sections of a vector bundle on $K \bas G$, in this case
a line bundle with $-3/2$ units of charge under $S$.
For this purpose, as in \eqref{emqcqk}, we let
the coset representative $e_{QK}$ act on the  lowest $K$-type,
and construct the intertwiner
\bea
\Psi(U,\zeta,\tzeta,\sigma;u,x)&=&
e^{U(x\pa_x+\frac12)}\cdot
e^{-i \tzeta u x}\cdot
e^{\zeta\ x\pa_u}\cdot
e^{-\frac{i}{2} (\sigma-\zeta\tzeta) x^2}\cdot f_{SU(2)}\\
&=& x \exp\left[
\frac{3U}{2} - \frac12 u^2 -\frac12 (e^{2U}+i \sigma) x^2
-\frac12 x(2u+x\zeta)(\zeta+i\tzeta) \right]\nn
\eea
The overlap
\be
\Phi_f = \int dx\, du\, f^*(u,x)\, \Psi(U,\zeta,\tzeta,\sigma;u,x)
\ee
is then an eigenmode of the Laplacian twisted by $S$,
\be
\left[ \Delta_{QK} -\frac12 e^{2U} \pa_\sigma - \frac{15}{16}  \right]
\Phi_f = 0\ .
\ee

\subsubsection{As a submodule of the principal series representation \label{emsu21}}

In this section, we investigate to what extent the minimal representation
(or its $\nu$-deformation) may be viewed as a submodule of the principal series
representation.

To that purpose, we first observe that the
Casimirs \eqref{c2minrepsu21gen} and \eqref{c23qcsu21} agree for $k=1,3$
$(\nu=0)$ or $k=0,4$ $(\nu=\pm 1)$. Second, the equation
\eqref{defjoseph2} in the annihilator of the deformed minimal
representation, when expressed in terms of the generators of the
quasi-conformal action, becomes
\be
\label{c0su21}
C_0 \equiv (\pa_\zeta + \tzeta \pa_\sigma)^2 + (\pa_{\tzeta}-\zeta \pa_{\sigma})^2
-2 i \nu \pa_\sigma = 0
\ee
The physicist will recognize $C_0$ as the Hamiltonian of a charged particle
on the plane $(\zeta,\tzeta)$, with a constant magnetic field proportional
to $i\pa_\sigma$.  The spectrum of
$C_0$ (for the usual $L^2$ norm on the plane) consists of the usual
infinitely degenerate Landau levels. Defining
\bse
\label{defnabla}
\bea
\nabla&=&\pa_\zeta + \tzeta \pa_\sigma + i (\pa_{\tzeta}-\zeta \pa_{\sigma}) \\
\bar\nabla&=&\pa_\zeta + \tzeta \pa_\sigma 
- i (\pa_{\tzeta}-\zeta \pa_{\sigma})
\eea
\ese
one may rewrite \eqref{c0su21} as
\be
C_0 = \nabla \bar\nabla -2 i (\nu+1) \pa_\sigma = 0
\ee
The lowest Landau level corresponds to functions annihilated by
$\bar\nabla$. This constraint commutes with the action of $G$ for
$k=0,\nu=-1$:  this is evidently so for the
positive root generators $E_{p},E_{q},E$ (the former two being
the generators of magnetic translations on the plane), and it suffices
to check invariance under the action of the lowest root generator $F$,
\be
\left[F,\bar\nabla\right]= -\frac{i}{2}[ 3 (\zeta^2+\tzeta^2) - 2i\sigma ]
\bar\nabla - k (\tzeta + i\zeta)\ ,\quad
\ee
which indeed vanishes on the subspace annihilated by $\bar\nabla$
when $k=0$. Solutions to $\bar\nabla=0$ are of the form
\footnote{After Fourier transforming over $\sigma$,
one recovers the usual form $f=g_K(\tzeta+ i \zeta)\,e^{-\frac12 K(\zeta^2+\tzeta^2)-i
K \sigma}$ of the lowest Landau level wave functions.}
\be
f(\zeta,\tzeta,\sigma) = g\left[ z\equiv\tzeta + i\zeta, a \equiv - \sigma +
\frac{i}{2} (\zeta^2+\tzeta^2) \right]
\ee
It is straightforward to check that the quasi-conformal action reduces
on this invariant subspace to the action \eqref{su21za} induced from
the maximal parabolic at $\nu=-1$.

As far as the constraint $C_0=0$ itself is concerned, one may check
that it is invariant under the action of $G$ at the values
$k=1,\nu=0$ appropriate for the undeformed minimal representation, as well as
at $k=0,\nu=-1$ which we discussed above. Indeed, one may rewrite
\be
[F,C_0] = -2 \sigma C_0 + 2 (1-k) \left[ (\zeta^2+\tzeta^2)\pa_\sigma - J \right]
+2 i \nu ( \zeta \pa_\zeta + \tzeta \pa_{\tzeta} + k )
\ee
which is proportional to $C_0$ for $k=1,\nu=0$. From the point of view
of the magnetic problem, this corresponds to non-normalizable states
with energy below that of the lowest Landau level. In order to
find the eigenmodes explicitly, it is useful to Fourier transform
over $\tzeta,\sigma$,
\be
f(\zeta,\tzeta,\sigma) = \int dp\, dK
\, \exp\left( - i K \sigma -i p \tzeta \right) \
g(\zeta, p , K)
\ee
and redefine
\be
g(\zeta, p ,K) = \exp\left[ -\frac{(P-2K\zeta)^2}{4K} \right] \,h(\zeta, P, K)
\ee
where $P=p+K\zeta$.
The constraint $C_0=0$ becomes now an ordinary differential equation on $h$,
\be
\label{c0zetapk}
\left[ \pa_\zeta^2 + 2 (P -2 K \zeta) \pa_\zeta -2 K (\nu+1) \right]\,
h(\zeta,P,K)\equiv 0
\ee
For $\nu=-1$, the solutions are
\be
h(\zeta,P,K)=h_1(P,K) + h_2(P,K) \, K^{-1/2} \, e^{-\frac{P^2}{2K}}
\mbox{erfi}\left(  -\frac{P-2K\zeta}{\sqrt{2K}} \right)
\ee
Only the $\zeta$-independent part $h_1(P,K)$ obeys also
\eqref{defjoseph1}.  Changing variables again by setting $K=x^2/2=y/2$ and $P=- x u = - x_0$, it is easy
to check that the quasi-conformal action on $f(\zeta,\tzeta,\sigma)$ at $k=0$
gives precisely the deformed minimal representation \eqref{gknsu21def}
acting on $h_1(P,K)$ at $\nu=-1$.  We conclude that the deformed minimal
representation at $\nu=-1$ can be embedded
inside the principal series representation at $k=0$ by
\be
\label{mintoqc}
f(\zeta,\tzeta,\sigma) = \int dp\, dK
\, \exp\left( - i K \sigma - i p \tzeta -\frac{(p-K\zeta)^2}
{4K} \right) \ h(p+K\zeta, K)\ .
\ee
The formula \eqref{mintoqc} can also be viewed as the matrix element
\be
f(\zeta,\tzeta,\sigma) = \langle f_P \, | \,
e^{-\sigma E - \zeta E_q + \tzeta E_p} \, | \, h \rangle
\ee
where $f_P$ is the $P$-covariant vector in the deformed minimal
representation \eqref{pinvsu21}.

For $\nu=1$, similar arguments show that the
deformed minimal representation can be embedded inside
the principal series representation at $k=0$ via
\be
\label{mintoqc2}
f(\zeta,\tzeta,\sigma) = \int dp\, dK
\, \exp\left( - i K \sigma - i p \tzeta +\frac{(p-K\zeta)^2}
{4K} \right) \ h(p+K\zeta, K)\ .
\ee
For other values of $\nu$, the solution of \eqref{c0zetapk} involves
Hermite and hypergeometric functions,
\be
h=h_1(P,K)\,H_{-\frac{\nu+1}{2}}
\left( -\frac{P-2K\zeta}{\sqrt{2K}} \right)
+ h_2(P,K)\,\1F1\left( \frac{1+\nu}{4},\frac12 ;\,
\frac{(P-2K\zeta)^2}{2K}\right)
\ee

It is worth noting that the formula \eqref{mintoqc} admits a simple
generalization to all quaternionic groups, as we shall see
for $G_{2(2)}$ in \eqref{embmrg2} below.

\subsection{The minimal representation as a quantized quasi-conformal
action\label{minclasa2}}
We now explain how the minimal representation can be viewed as the
quantization of the quasi-conformal action of $G$, or equivalently how
the quaternionic discrete series arises as a semi-classical limit
of the minimal representation.

\subsubsection{Lifting the quasi-conformal action to the  hyperk\"ahler cone
\label{lifthkc}}
As a first step, it is useful to ``deprojectivize''
the quasi-conformal action, i.e. lift it to an action on the
hyperk\"ahler cone $\cS$. For this purpose, we introduce an extra variable
$t$ and interpret the $k$-dependent quasi-conformal action as an
action of functions of four variables $\hat f(t,\xi,\txi,\alpha)=
e^{-k t} f(\xi,\txi,\alpha)$. We then implement the change of variables
found in \cite{Neitzke:2007ke},
\be
\label{vtoxiQC}
v^\flat=e^{2t}\ ,\quad
v^0=\xi\, e^{2t}\ ,\quad
w_0=\frac{i}{2}\,\txi\ ,\quad
w_\flat=\frac{1}{4i}(\alpha+\xi \txi)
\ee
The coordinates $v^\flat,v^0,w_\flat,w_0$ are complex coordinates
on the Swann space $\cS$, such that the holomorphic symplectic form takes
the Darboux form
\be
\Omega =dw_\flat\wedge dv^\flat + dw_0\wedge dv^0
\ee
The quasi-conformal action on $\cO(-k)$ over $\cZ$ now corresponds to the
holomorphic action of $G$ on $\cS$, restricted to the
subspace of homogeneous functions of degree $-k$ under the rescaling
\be
v^I \to \mu^2 v^I\ ,\quad w_I \to w_I
\ee
The holomorphic vector fields generating the action of $G$ are given
by\footnote{Note that only real combinations $E+\bar E$ are actual
isometries of the hyperk\"ahler metric.}
\bea
E_{p} &=& \frac{i}{2}\pa_{w_0}\ ,\quad
E_{q} = w_0 \pa_{w_\flat} - v^\flat \pa_{v^0}\ ,\quad
E = -\frac{i}{4} \pa_{w_\flat}\\
H &=& 2 v^\flat \pa_{v^\flat} - 2 w_\flat \pa_{w_\flat}
+ v^0 \pa_{v^0} - w_0 \pa_{w_0}\ ,\nn\\
J&=&-i\left( w_0^2 + \frac{(v^0)^2}{v^\flat}\right)\pa_{w_\flat}
+2 i v^\flat w_0 \pa_{v^0} + \frac{i v^0}{v^\flat} \pa_{w_0}\nn\\
F_{p}&=&-4 i {v^\flat} {w_0} \pa_{v^\flat} +
\left( 4 i {w_\flat} {w_0}-\frac{i {v^0}^3}{2 {v^\flat}^3}\right)
\pa_{w_\flat}-2
   i (2 {v^\flat} {w_\flat}+{v^0} {w_0})\pa_{v^0}+
\left( \frac{3 i {v^0}^2}{4 {v^\flat}^2}+i
   {w_0}^2 \right) \pa_{w_0}\nn\\
F_{q}&=&-2 {v^0}\pa_{v^\flat}-\left(2 {w_0}^3+\frac{3 (v^0)^2 {w_0}}
{2 {(v^\flat})^2}\right) \pa_{w_\flat} +\left( 6 {v^\flat}
   {w_0}^2-\frac{3 (v^0)^2}{2 {v^\flat}}\right) \pa_{v^0} +
\left(2 {w_\flat}+\frac{3 {v^0}
   {w_0}}{{v^\flat}}\right)\pa_{w_0}\nn
\\
F&=&4 i (2 {v^\flat} {w_\flat}+{v^0} {w_0})\pa_{v^\flat}+\frac{i \left(16
\left({w_0}^4-4
   {w_\flat}^2\right) (v^\flat)^4+24 {v^0}^2 {w_0}^2 {v^\flat}^2-3
   {v^0}^4\right)}{16 (v^\flat)^4}\pa_{w_\flat} \nn\\
&& + i \left(-4 {v^\flat} {w_0}^3+\frac{3 (v^0)^2
   {w_0}}{{v^\flat}}+4 {v^0} {w_\flat}\right)\pa_{v^0}
-\frac{i \left(16 {w_\flat} {w_0}
   {v^\flat}^3+12 {v^0} {w_0}^2 (v^\flat)^2+(v^0)^3\right)}{4 {(v^\flat})^3}\pa_{w_0}\nn
\eea
Being tri-holomorphic isometries, these vector fields in particular
preserve the holomorphic symplectic form $\Omega$. They
can be represented by holomorphic moment maps $\moment{X}$,
such that the contraction $\iota_X \Omega = d \moment{X}$:
\bea
\label{holmomsu21}
\moment{E}_{p} &=& \frac{i}{2} v^0\ ,\quad
\moment{E}_{q}=v^\flat w_0\ ,\quad
\moment{E}=-\frac{i}{4} v^\flat\ ,\quad
\moment{H}=-2 v^\flat w_\flat - v^0 w_0\ ,\quad
\moment{J} =- w_0^2 -\frac{i}{4} \frac{(v^0)^2}{(v^\flat)^2} \nn \\
\moment{F}_{p}
&=&\frac{i}{4} \left(\frac{(v^0)^3}{(v^\flat)^2}+4 (w_0)^2 {v^0}
+16 {v^\flat}  {w_\flat} {w_0}\right)\ ,\quad
\moment{F}_{q}
=-2 {v^\flat} (w_0)^3+\frac{3 (v^0)^2 {w_0}}{2 {v^\flat}}+2 {v^0}
{w_\flat}\nn\\
\moment{F}
&=&\frac{i}{16   {(v^\flat)^3}}
 \left(16 \left(w_0^4-4 {w_\flat}^2\right) (v^\flat)^4-64 {v^0}
{w_\flat}   {w_0} ({v^\flat})^3-24 (v^0 w_0 v^\flat)^2+(v^0)^4\right)
\eea
Returning to the variables $t,\xi,\txi,\alpha$, the holomorphic moment maps
become,
\bea
\moment{E}_p&=&\frac{i}{2} e^{2t}\, \xi\ ,\quad
\moment{E}_q=\frac{i}{2} e^{2t}\, \txi\ ,\quad
\moment{E}=-\frac{i}{4} e^{2t}\ ,\quad
\moment{H}=\frac{i}{2} e^{2t}\, \alpha\ ,\nn\\
\moment{F}_p&=&\frac{i}{4} e^{2t}\,
\left[\xi(\xi^2+\txi^2)+2\alpha \txi\right]\ ,\quad
\moment{F}_q=\frac{i}{4} e^{2t}\,
\left[\txi(\xi^2+\txi^2)-2\alpha \xi\right]\ ,\quad\\
\moment{F}&=&\frac{i}{16} e^{2t}\, \left[ 4\alpha^2 + (\xi^2+\txi^2)^2
\right]\ ,\quad
\moment{J}=\frac{i}{4} e^{2t}\, (\xi^2+\txi^2)\nn
\eea
Arranging these holomorphic moment maps into an element $Q$ of $\fg^*$, one may
check that $Q^2=0$ in our matrix representation, i.e. that $Q$ is valued in the minimal co-adjoint orbit.
In particular, the holomorphic moment maps satisfy classical versions
of the identities \eqref{opadsu21},
\bse
\label{opadsu21clas}
\bea
\moment{H}^2 + 4 \moment{E} \moment{F} + \moment{J}^2  &=& 0 \\
\moment{E}_p^2+\moment{E}_q^2+4 \moment{J} \moment{E} &=& 0\\
C_2(\moment{J}) + \frac19 \moment{S}^2 &=& 0 \\
\moment{C}_2 = \moment{C}_3 &=& 0\\
\moment{J}_{\frac12,\frac32}\,\moment{J}_{\frac12,-\frac32}
-\frac{4i}{3} \moment{S} \moment{J}_+ &=&0
\eea
\ese
The holomorphic moment maps of the compact generators are given by
\bse
\bea
\moment{S}&=&\frac{3i}{16} e^{2t}\, \left( 1 + \xi^2 + \txi^2 + \alpha^2
+ \frac14 (\xi^2+\txi)^2 \right)\\
\moment{J}_3&=&\frac{i}{16} e^{2t}\,\left(1-3 \xi^3-3 \txi^2+\alpha^2
+\frac14(\xi^2+\txi^2)^2\right)\\
\moment{J}_\pm&=&\frac{i}{8\sqrt{2}} e^{2t}\,
\,\left(\xi\mp i \txi\right)\,
\left(\xi^2+\txi^2\pm 2 i \alpha-2\right)\\
\moment{J}_{\pm\frac12,\pm\frac32}
&=&
\frac{1}{16} e^{2t}
\left[ (\xi^2+\txi^2)^2+4\alpha(\alpha\pm 2i)-4 \right]\\
\moment{J}_{\pm\frac12,\mp\frac32}
&=&-\frac{i}{8\sqrt{2}} e^{2t} \,
\left(\xi\mp i \txi\right)\,\left(\xi^2+\txi^2\pm 2 i \alpha+2\right)
\eea
\ese
It is interesting to note that the $K$-type \eqref{su21ktype}
may be rewritten in terms of these moment maps, up to an overall
numerical factor, as
\be \label{ktypemoment}
f_{j,s}=\left( \frac{\moment{J}_{\frac12,\frac32}}{\moment{S}}
\right)^{j+\frac13 s} \,
\left( \frac{\moment{J}_{\frac12,-\frac32}}{\moment{S}}
\right)^{j-\frac13 s}\,
\moment{S}^{-\frac12 k}
\ee
The covariance of \eqref{ktypemoment} under $K$ is then easy to
see, using the identities \eqref{opadsu21clas} obeyed by the holomorphic
moment maps, and the fact that $G$ acts by Poisson brackets,
$X f = \{ \moment{X}, f \}$.

Moreover, the vanishing locus of the holomorphic moment maps associated
to the compact generators $S,J_3,J_\pm$ consists of two branches
\be
\xi = \pm i \txi\ ,\quad \alpha=\pm i
\ee
or equivalently, in terms of the variables on the hyperk\"ahler cone,
\be
w_\flat = \pm \frac14 \left( 1 - \frac{(v^0)^2}{(v^\flat)^2}\right)\ ,\quad
w_0 = \pm \frac{v^0}{2v^\flat}
\ee
These relations define two lagrangian cones $\cC_\pm$,
with generating functions $S_\pm$:  they may be rewritten as
\be
\label{slag}
w_\flat = \pa_{v^\flat} S_\pm \ ,\quad
w_0 = \pa_{v^0} S_\pm\ ,\quad
S_\pm (v^\flat,v^0)= \pm \frac{1}{4} \left( v^\flat + \frac{(v^0)^2}{v^\flat}
\right)
\ee
By construction, $\cC_\pm$ are invariant under
the holomorphic action of $K$,
since the Poisson brackets with the constraints vanish on the
constraint locus. As we shall see momentarily,
$S_-$ describes the semi-classical limit
of the lowest $K$-type \eqref{su21mink} of the minimal representation.

\subsubsection{The classical limit of the minimal representation\label{claslag}}
We now return to the presentation \eqref{kpwsu21} of the minimal
representation acting on functions of $y,x_0$. The form \eqref{su21mink}
of the spherical vector suggests that a semi-classical limit exists
as $y,x_0$ are scaled simultaneously to infinity, if one restricts
to wave functions of the form
\be
\label{semiclas}
f(y,x_0) = \exp\left[ S(y,x_0) \right]\ ,\quad
S(y,x_0) = y \,\hat S(\hat x_0) + {\cal O}(1)
\ee
where $\hat x_0=x_0/y$ is kept fixed in the limit $y\to \infty$.
Indeed, it is easy to check that, to leading order in this limit,
the action of the infinitesimal generators on $f$ produces
\be
X \cdot f = y \,\moment{X}
\left( \hat S, \pa_{\hat 0}\hat S, \hat x^0 \right)\, f
+ {\cal O}(y),
\ee
for some (so far unspecified) function $\moment{X}$.
Changing variables to
\be
p_0 = \pa_{x_0}S(y,x_0) = \pa_{\hat x_0} \hat S\ ,\quad
p_y = \pa_{y} S(y,x_0) = \hat S - \hat x_0 \pa_{\hat x_0} \hat S\ ,
\ee
identifies the function $\moment{X}(y,x_0,p_y,p_0)$
as the leading differential
symbol of the differential operator $X$ in the semi-classical limit
\eqref{semiclas}. Further setting
\be
v^\flat = -2 y \ ,\quad v^0=2 x_0\ ,\quad
w_\flat = \frac12 p_y \ ,\quad w_0=-\frac12 p_0\ ,\quad
\ee
identifies $\moment{X}$ as the holomorphic moment map \eqref{holmomsu21}
associated to the tri-holomorphic action of $G$ on its hyperk\"ahler
cone $\cS$. Thus, we conclude that the minimal representation can be viewed
as the quantization of the holomorphic symplectic manifold $\cS$. Furthermore, the semi-classical
limit of the spherical vector \eqref{su21mink} is given by the generating
function $S_-$ of the Lagrangian cone $\cS_-$ defined in
\eqref{slag}.

One could also have given a real version of this construction, lifting the
action of $G$ on $P \bas G$ to a real symplectic manifold by adding a single real coordinate.
In that case we would say that the minimal representation arises by quantizing the
real symplectic structure.  Indeed, the real manifold so obtained is at least locally
isomorphic to the minimal coadjoint orbit of $G$, so this makes contact with one of
the standard ways of thinking about the minimal representation.

\section{$G_{2(2)}$}

In this section, we describe the geometry of the quaternionic-K\"ahler
space $SO(4) \bas G_{2(2)}$, and various associated unitary representations
of $G = G_{2(2)}$.  We will be somewhat briefer in this section since
many of the constructions are parallel to ones we described for $G =
SU(2,1)$ above.

\subsection{Some group theory}

It is convenient to represent the Lie algebra $\fg$ of $G=G_{2(2)}$
by the 7-dimensional matrix representation described in
\cite{Bodner:1989cg} (after some relabelings and change of normalization)
\be
\label{g2X}
\begin{pmatrix}
\underline{Y}_0 & -\underline{Y}_+ & 0 & -\sqrt{\frac{2}{3}}
\underline{F}_{q_1} & \sqrt{\frac{2}{3}} \underline{E}_{p^1} &
-\sqrt{2} \underline{F}_{q_0} & \sqrt{2}   \underline{E}_{p^0} \\
\underline{Y}_- & 0 & -\underline{Y}_+ &
\frac{2}{\sqrt{3}} \underline{F}_{p^1}&
\frac{2}{\sqrt{3}}\underline{E}_{q_1} &
\frac{2}{\sqrt{3}} \underline{F}_{q_1} &
 -\frac{2}{\sqrt{3}}\underline{E}_{p^1} \\
 0 & \underline{Y}_- & -\underline{Y}_0 & \sqrt{2} \underline{F}_{p^0} &
\sqrt{2} \underline{E}_{q_0} & -\sqrt{\frac{2}{3}} \underline{F}_{p^1} &
-\sqrt{\frac{2}{3}} \underline{E}_{q_1} \\
 -\sqrt{\frac{2}{3}} \underline{E}_{q_1} &
\frac{2}{\sqrt{3}} \underline{E}_{p^1} &
\sqrt{2} \underline{E}_{p^0} &
\underline{H}+\frac12\underline{Y}_0 & -\underline{E} &
   -\frac{1}{\sqrt{2}}\underline{Y}_+ & 0 \\
 \sqrt{\frac{2}{3}} \underline{F}_{p^1} &
\frac{2}{\sqrt{3}} \underline{F}_{q_1} &
   \sqrt{2} \underline{F}_{q_0} &
-\underline{F} & -\underline{H}+\frac12\underline{Y}_0 & 0 &
-\frac1{\sqrt{2}}\underline{Y}_+ \\
-\sqrt{2} \underline{E}_{q_0} & \frac{2}{\sqrt{3}} \underline{E}_{q_1} &
-\sqrt{\frac{2}{3}} \underline{E}_{p^1} &
\frac1{\sqrt{2}}\underline{Y_-} & 0 &
\underline{H}-\frac12\underline{Y}_0 & -\underline{E} \\
\sqrt{2} \underline{F}_{p^0} & -\frac{2}{\sqrt{3}}
\underline{F}_{p^1} & -\sqrt{\frac{2}{3}} \underline{F}_{q_1} & 0 &
\frac1{\sqrt{2}}\underline{Y}_- &
   -\underline{F} & -\frac12\underline{Y}_0-\underline{H}
\end{pmatrix}
\equiv \underline{X}_i\, X_i
\ee
where, as in \eqref{su21X}, $\underline{X}_i$ are real coordinates dual to the
generators, to be represented by anti-hermitian operators in a
given unitary representation.
These matrices preserve the signature $(4^+,3^-)$ metric
\be
\eta_{ij} dx_i dx_j = -2 dx_1 dx_3 + dx_2^2 + 2 dx_4 dx_7 - 2 dx_5 dx_6
\ee
and the three-form
\be
dx_{123} -dx_{247} + dx_{256} + \sqrt2 dx_{167} + \sqrt2 dx_{345}\ ,
\ee
thus providing an embedding of $G_{2(2)}$ inside $SO(3,4)$.
In addition to the universal commutation relations
\be
\begin{array}{l!{\hspace{7mm}}l!{\hspace{7mm}}l!{\hspace{7mm}}l}
\multicolumn{4}{c}{\left[E,F\right] = H, \quad
\left[ H,E \right]= 2E, \quad
\left[ H,F \right]= -2F,}  \\
\multicolumn{2}{c}{\left[ E_{p^I}, E_{q_J} \right] = - 2 \delta^I_J E\ ,}
&\multicolumn{2}{c}{\left[ F_{p^I}, F_{q_J} \right] = 2 \delta^I_J F\ ,} \\
\left[ E_{p^I}, E_{p^J} \right] = 0, &
\left[ F_{p^I}, F_{p^J} \right] = 0, &
\left[ E_{q_I}, E_{q_J} \right] = 0, &
\left[ F_{q_I}, F_{q_J} \right] = 0, \\
\left[ H, E_{p^I} \right] = E_{p^I},&
\left[ H, F_{p^I} \right] = -F_{p^I},&
\left[ H, E_{q_I} \right] = E_{q_I},&
\left[ H, F_{q_I} \right] = -F_{q_I},\\
\left[ F, E_{p^I} \right] = -F_{q_I},&
\left[ E, F_{q_I} \right] = -F_{p^I},&
\left[ F, E_{q_I} \right] = F_{p^I},&
\left[ E, F_{p^I} \right] = E_{q_I},
\end{array}
\ee
we have the $SL(2,\IR)$ algebra
\be
[Y_0,Y_{\pm}]=\pm Y_{\pm}\ ,\quad [Y_-,Y_+]=Y_0
\ee
under which $E,F$ are singlets
\be
[Y_0,E]=[Y_{\pm},E]=[Y_0,F]=[Y_{\pm},F]=0
\ee
and $E_{p,q}$ and $F_{p,q}$ transform as a spin 3/2,
\bse
\bea
\left[Y_0,
\begin{pmatrix}E_{p^0} \\E_{p^1} \\E_{q_1} \\E_{q_0}
\end{pmatrix}
\right] = \frac12
\begin{pmatrix} 3 E_{p^0} \\E_{p^1} \\ - E_{q_1} \\ -3 E_{q_0}
\end{pmatrix} \ ,&\quad&
\left[Y_0,
\begin{pmatrix}F_{p^0} \\F_{p^1} \\F_{q_1} \\F_{q_0}
\end{pmatrix}
\right] = \frac12
\begin{pmatrix}-3 F_{p^0} \\- F_{p^1} \\F_{q_1} \\3 F_{q_0}
\end{pmatrix} \\
\left[Y_+,
\begin{pmatrix}E_{p^0} \\E_{p^1} \\E_{q_1} \\E_{q_0}
\end{pmatrix}
\right] =
\begin{pmatrix}
0 \\
\sqrt{\frac{3}{2}} E_{p^0} \\
- \sqrt{2} E_{p^1} \\
-\sqrt{\frac{3}{2}} E_{q_1}
\end{pmatrix} \ ,&\quad&
\left[Y_+,
\begin{pmatrix}F_{p^0} \\F_{p^1} \\F_{q_1} \\F_{q_0}
\end{pmatrix}
\right] =
\begin{pmatrix}
-\sqrt{\frac{3}{2}} F_{p^1} \\
\sqrt{2} F_{q_1} \\
\sqrt{\frac{3}{2}} F_{q_0} \\
0
\end{pmatrix} \\
\left[Y_-,
\begin{pmatrix}E_{p^0} \\E_{p^1} \\E_{q_1} \\E_{q_0}
\end{pmatrix}
\right] =
\begin{pmatrix}
-\sqrt{\frac{3}{2}} E_{p^1} \\
\sqrt{2} E_{q_1} \\
\sqrt{\frac{3}{2}} E_{q_0} \\
0
\end{pmatrix}
\ , &\quad&
\left[Y_-,
\begin{pmatrix}F_{p^0} \\F_{p^1} \\F_{q_1} \\F_{q_0}
\end{pmatrix}
\right] =
\begin{pmatrix}
0 \\
\sqrt{\frac{3}{2}} F_{p^0} \\
- \sqrt{2} F_{p^1} \\
-\sqrt{\frac{3}{2}} F_{q_1}
\end{pmatrix}
\eea
\ese
Moreover, the commutation between positive and negative roots give
\bse
\bea
\left[E_{p^0}, F_{p^0}\right] = H+2 Y_0 \ ,&\quad&
\left[E_{q_0}, F_{q_0}\right] = H-2 Y_0\\
\left[E_{p^1}, F_{p^1}\right] = \frac13 (3H+2 Y_0)  \ ,&\quad&
\left[E_{q_1}, F_{q_1}\right] = \frac13 (3H-2 Y_0), \\
\left[E_{p^1}, F_{q_1}\right] =- \frac{4\sqrt{2}}{3} Y_+  \ ,&\quad&
\left[E_{q_1}, F_{p^1}\right] = \frac{4\sqrt{2}}{3} Y_-
\eea
\ese
The quadratic Casimir is
\be
\begin{split}
C_2 =& \frac14 \left( H^2 +  2 EF + 2 FE \right) +
\frac13 (Y_0^2- Y_+ Y_- - Y_- Y_+)\\
&+ \frac14 \sum_{I=0,1}
( E_{p^I} F_{p^I} + F_{p^I} E_{p^I} + E_{q_I} F_{q_I} + F_{q_I} E_{q_I} )
\end{split}
\ee
normalized so that
\be
C_2(adj) = 4 \ ,\quad C_2({\bf 7}) = 2
\ee
There is also a degree 6 Casimir, corresponding to the trace of the
sixth power of the matrix \eqref{g2X}, which we shall not attempt to write.

\FIGURE{\begin{picture}(0,0)%
\includegraphics{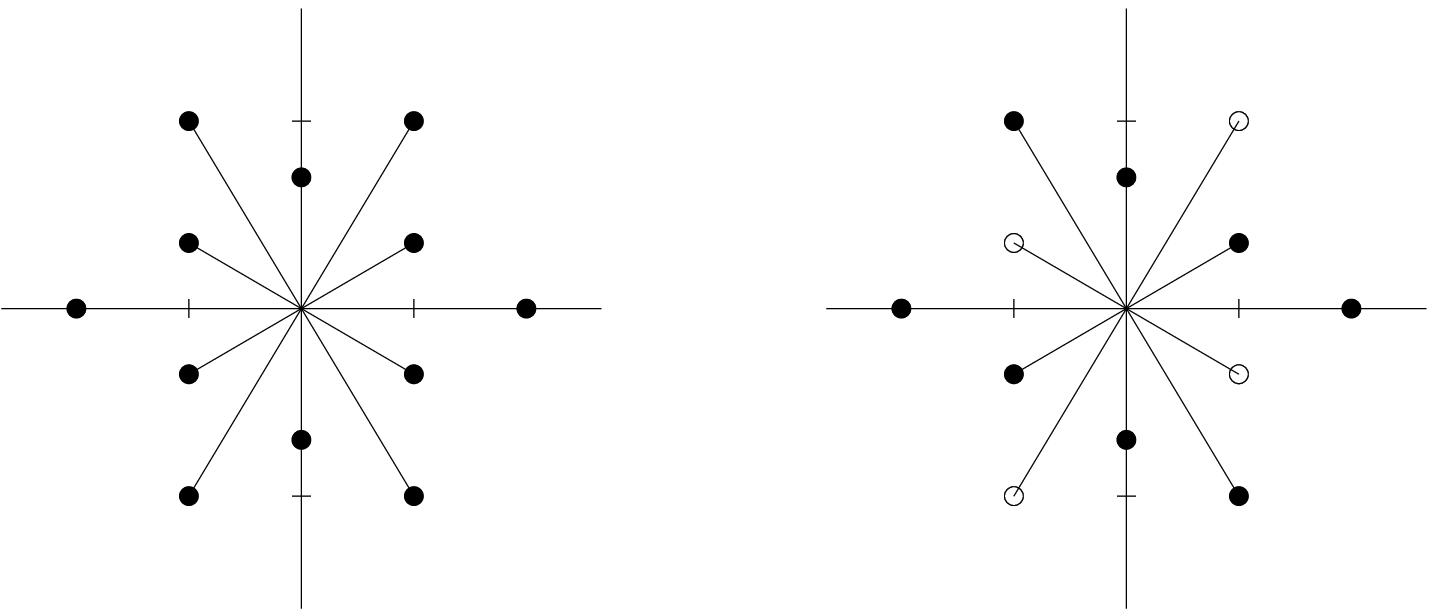}%
\end{picture}%
\setlength{\unitlength}{2368sp}%
\begingroup\makeatletter\ifx\SetFigFont\undefined%
\gdef\SetFigFont#1#2#3#4#5{%
  \reset@font\fontsize{#1}{#2pt}%
  \fontfamily{#3}\fontseries{#4}\fontshape{#5}%
  \selectfont}%
\fi\endgroup%
\begin{picture}(11424,4824)(1189,-6373)
\put(4801,-5761){\makebox(0,0)[lb]{\smash{{\SetFigFont{11}{13.2}{\rmdefault}{\mddefault}{\updefault}{\color[rgb]{0,0,0}$E_{q^0}$}%
}}}}
\put(12451,-3661){\makebox(0,0)[lb]{\smash{{\SetFigFont{11}{13.2}{\rmdefault}{\mddefault}{\updefault}{\color[rgb]{0,0,0}$L_0$}%
}}}}
\put(8326,-3811){\makebox(0,0)[lb]{\smash{{\SetFigFont{7}{8.4}{\rmdefault}{\mddefault}{\updefault}{\color[rgb]{0,0,0}$-i$}%
}}}}
\put(11926,-3811){\makebox(0,0)[lb]{\smash{{\SetFigFont{7}{8.4}{\rmdefault}{\mddefault}{\updefault}{\color[rgb]{0,0,0}$+i$}%
}}}}
\put(9226,-3811){\makebox(0,0)[lb]{\smash{{\SetFigFont{7}{8.4}{\rmdefault}{\mddefault}{\updefault}{\color[rgb]{0,0,0}$-i/2$}%
}}}}
\put(11026,-3811){\makebox(0,0)[lb]{\smash{{\SetFigFont{7}{8.4}{\rmdefault}{\mddefault}{\updefault}{\color[rgb]{0,0,0}$+i/2$}%
}}}}
\put(10426,-1861){\makebox(0,0)[lb]{\smash{{\SetFigFont{11}{13.2}{\rmdefault}{\mddefault}{\updefault}{\color[rgb]{0,0,0}$R_0$}%
}}}}
\put(10351,-2536){\makebox(0,0)[lb]{\smash{{\SetFigFont{7}{8.4}{\rmdefault}{\mddefault}{\updefault}{\color[rgb]{0,0,0}$+3/2i$}%
}}}}
\put(10351,-5536){\makebox(0,0)[lb]{\smash{{\SetFigFont{7}{8.4}{\rmdefault}{\mddefault}{\updefault}{\color[rgb]{0,0,0}$-3i/2$}%
}}}}
\put(11401,-2461){\makebox(0,0)[lb]{\smash{{\SetFigFont{11}{13.2}{\rmdefault}{\mddefault}{\updefault}{\color[rgb]{0,0,0}$J_+$}%
}}}}
\put(11401,-5761){\makebox(0,0)[lb]{\smash{{\SetFigFont{11}{13.2}{\rmdefault}{\mddefault}{\updefault}{\color[rgb]{0,0,0}$T_-$}%
}}}}
\put(11401,-4711){\makebox(0,0)[lb]{\smash{{\SetFigFont{11}{13.2}{\rmdefault}{\mddefault}{\updefault}{\color[rgb]{0,0,0}$S_+$}%
}}}}
\put(11401,-3436){\makebox(0,0)[lb]{\smash{{\SetFigFont{11}{13.2}{\rmdefault}{\mddefault}{\updefault}{\color[rgb]{0,0,0}$U_+$}%
}}}}
\put(8626,-3436){\makebox(0,0)[lb]{\smash{{\SetFigFont{11}{13.2}{\rmdefault}{\mddefault}{\updefault}{\color[rgb]{0,0,0}$S_-$}%
}}}}
\put(8626,-4711){\makebox(0,0)[lb]{\smash{{\SetFigFont{11}{13.2}{\rmdefault}{\mddefault}{\updefault}{\color[rgb]{0,0,0}$U_-$}%
}}}}
\put(8626,-5761){\makebox(0,0)[lb]{\smash{{\SetFigFont{11}{13.2}{\rmdefault}{\mddefault}{\updefault}{\color[rgb]{0,0,0}$J_-$}%
}}}}
\put(8626,-2461){\makebox(0,0)[lb]{\smash{{\SetFigFont{11}{13.2}{\rmdefault}{\mddefault}{\updefault}{\color[rgb]{0,0,0}$T_+$}%
}}}}
\put(12151,-4261){\makebox(0,0)[lb]{\smash{{\SetFigFont{11}{13.2}{\rmdefault}{\mddefault}{\updefault}{\color[rgb]{0,0,0}$L_+$}%
}}}}
\put(8101,-4261){\makebox(0,0)[lb]{\smash{{\SetFigFont{11}{13.2}{\rmdefault}{\mddefault}{\updefault}{\color[rgb]{0,0,0}$L_-$}%
}}}}
\put(9751,-2986){\makebox(0,0)[lb]{\smash{{\SetFigFont{11}{13.2}{\rmdefault}{\mddefault}{\updefault}{\color[rgb]{0,0,0}$R_+$}%
}}}}
\put(9676,-5161){\makebox(0,0)[lb]{\smash{{\SetFigFont{11}{13.2}{\rmdefault}{\mddefault}{\updefault}{\color[rgb]{0,0,0}$R_-$}%
}}}}
\put(5851,-3661){\makebox(0,0)[lb]{\smash{{\SetFigFont{11}{13.2}{\rmdefault}{\mddefault}{\updefault}{\color[rgb]{0,0,0}$H$}%
}}}}
\put(1726,-3811){\makebox(0,0)[lb]{\smash{{\SetFigFont{7}{8.4}{\rmdefault}{\mddefault}{\updefault}{\color[rgb]{0,0,0}$-2$}%
}}}}
\put(5326,-3811){\makebox(0,0)[lb]{\smash{{\SetFigFont{7}{8.4}{\rmdefault}{\mddefault}{\updefault}{\color[rgb]{0,0,0}$+2$}%
}}}}
\put(2626,-3811){\makebox(0,0)[lb]{\smash{{\SetFigFont{7}{8.4}{\rmdefault}{\mddefault}{\updefault}{\color[rgb]{0,0,0}$-1$}%
}}}}
\put(4426,-3811){\makebox(0,0)[lb]{\smash{{\SetFigFont{7}{8.4}{\rmdefault}{\mddefault}{\updefault}{\color[rgb]{0,0,0}$+1$}%
}}}}
\put(3751,-2536){\makebox(0,0)[lb]{\smash{{\SetFigFont{7}{8.4}{\rmdefault}{\mddefault}{\updefault}{\color[rgb]{0,0,0}$+3/2$}%
}}}}
\put(3751,-5536){\makebox(0,0)[lb]{\smash{{\SetFigFont{7}{8.4}{\rmdefault}{\mddefault}{\updefault}{\color[rgb]{0,0,0}$-3/2$}%
}}}}
\put(4801,-2461){\makebox(0,0)[lb]{\smash{{\SetFigFont{11}{13.2}{\rmdefault}{\mddefault}{\updefault}{\color[rgb]{0,0,0}$E_{p^0}$}%
}}}}
\put(4801,-4711){\makebox(0,0)[lb]{\smash{{\SetFigFont{11}{13.2}{\rmdefault}{\mddefault}{\updefault}{\color[rgb]{0,0,0}$E_{q^1}$}%
}}}}
\put(4801,-3436){\makebox(0,0)[lb]{\smash{{\SetFigFont{11}{13.2}{\rmdefault}{\mddefault}{\updefault}{\color[rgb]{0,0,0}$E_{p^1}$}%
}}}}
\put(2026,-3436){\makebox(0,0)[lb]{\smash{{\SetFigFont{11}{13.2}{\rmdefault}{\mddefault}{\updefault}{\color[rgb]{0,0,0}$F_{q_1}$}%
}}}}
\put(2026,-4711){\makebox(0,0)[lb]{\smash{{\SetFigFont{11}{13.2}{\rmdefault}{\mddefault}{\updefault}{\color[rgb]{0,0,0}$F_{p^1}$}%
}}}}
\put(2026,-5761){\makebox(0,0)[lb]{\smash{{\SetFigFont{11}{13.2}{\rmdefault}{\mddefault}{\updefault}{\color[rgb]{0,0,0}$F_{p^0}$}%
}}}}
\put(5551,-4261){\makebox(0,0)[lb]{\smash{{\SetFigFont{11}{13.2}{\rmdefault}{\mddefault}{\updefault}{\color[rgb]{0,0,0}$E$}%
}}}}
\put(1501,-4261){\makebox(0,0)[lb]{\smash{{\SetFigFont{11}{13.2}{\rmdefault}{\mddefault}{\updefault}{\color[rgb]{0,0,0}$F$}%
}}}}
\put(3151,-5086){\makebox(0,0)[lb]{\smash{{\SetFigFont{11}{13.2}{\rmdefault}{\mddefault}{\updefault}{\color[rgb]{0,0,0}$Y_-$}%
}}}}
\put(3151,-2986){\makebox(0,0)[lb]{\smash{{\SetFigFont{11}{13.2}{\rmdefault}{\mddefault}{\updefault}{\color[rgb]{0,0,0}$Y_+$}%
}}}}
\put(3826,-1861){\makebox(0,0)[lb]{\smash{{\SetFigFont{11}{13.2}{\rmdefault}{\mddefault}{\updefault}{\color[rgb]{0,0,0}$Y_0$}%
}}}}
\put(2026,-2461){\makebox(0,0)[lb]{\smash{{\SetFigFont{11}{13.2}{\rmdefault}{\mddefault}{\updefault}{\color[rgb]{0,0,0}$F_{q_0}$}%
}}}}
\end{picture}%

\caption{Root diagram of $G_{2(2)}$ with respect to the split
Cartan torus $H,Y_0$ (left) and the compact Cartan torus $L_0,R_0$ (right).
The compact (resp. non-compact) roots are indicated by a white
(resp. black) dot. The long roots generate $SL(3,\IR)$ (left)
and $SU(2,1)$ (right) subgroups, respectively.\label{g2root}}}

%
%\EPSFIGURE{g2root,height=6cm}{
%Root diagram of $G_{2(2)}$ with respect to the split
%Cartan torus $H,Y_0$ (left) and the compact Cartan torus $L_0,R_0$ (right).
%The compact (resp. non-compact) roots are indicated by a white
%(resp. black) dot. The long roots generate $SL(3,\IR)$ (left)
%and $SU(2,1)$ (right) subgroups, respectively.\label{g2root}}

This basis is adapted to the maximal subgroup
$SL(2,\IR)_{\rm short} \times SL(2,\IR)_{\rm long}$,
where the first factor is generated by $\{H,E,F\}$ while the
second is generated by $\{Y_-,Y_0,Y_+\}$.
The Cartan generators $H, Y_0$ are non-compact, with spectrum
\be
\mbox{Spec}(Y_0)=\{0,\pm 1/2,\pm 1,\pm 3/2\}\ ,
\quad \mbox{Spec}(H)=\{0,\pm 1,\pm2\}
\ee
The non-compact generator $H$ gives rise to the ``real non-compact 5-grading''
\be
\label{5gradg2}
{F}\vert_{-2} \oplus
\{ {F_{p^I}}, {F_{q_I}} \}\vert_{-1} \oplus
\{ H, Y_0, Y_{\pm} \}\vert_0  \oplus
\{ {E_{p^I}}, {E_{q_I}} \}\vert_{1} \oplus
{E}\vert_{2}
\ee
Define a parabolic subgroup $P=L N$ with Levi $L=\IR \times SL(2,\IR)$ generated by
$\{H,Y_0,Y_+,Y_-\}$ and unipotent radical $N$ generated by $\{F_{q_0},F_{q_1},F_{p^0},F_{p^1},
F\}$, corresponding to the spaces with zero and negative grade in the
decomposition \eqref{5gradg2}.  We call $P$ the Heisenberg
parabolic subgroup. In section \ref{3stepsec}, we will also be interested
in the parabolic subgroup $P_3=L_3 N_3$ associated to the 7-grading
induced by $Y_0$, with Levi $L_3=\IR \times SL(2,\IR)=
\{Y_0, E,F,H\}$ and unipotent radical $N_3=\{E_{q_0},E_{q_1},F_{p^0},F_{p^1},
Y_-\}$, nilpotent of degree 3.

Now we introduce a different basis adapted to the maximal compact
subgroup $SU(2)\times SU(2)$.  We first go to a compact basis for
the $SL(2,\R)\times SL(2,\R)$ group:
\bse
\bea
L_{\pm} = -\frac{1}{2\sqrt{2}}
\left(E+F \pm i H\right)\ ,&\quad& L_0 = \frac12 ( F-E) \\
R_0 = \frac{1}{\sqrt{2}}(Y_+ + Y_-) \ ,&\quad&
R_{\pm} = \frac{1}{2} \left( Y_+ - Y_- \mp i \sqrt{2} Y_0 \right)
\eea
\ese
such that
\bse
\bea
\left[ L_0, L_\pm \right] = \pm i L_\pm \ ,&\quad& [L_+,L_-]= - i L_0 \\
\left[ R_0, R_\pm \right] = \pm i R_\pm \ ,&\quad& [R_+,R_-]= - i R_0
\eea
\ese
Moreover, we define the eigenmodes
\begin{eqnarray}
K_{\frac12,\frac32}
&=& \frac{\sqrt{2}}{4}
\left[ -(E_{p^0}-i E_{q_0}) - i \sqrt{3} (E_{p^1}-i E_{q_1})
+ (F_{p^0}-i F_{q_0}) + i \sqrt{3} (F_{p^1}-i F_{q_1})\right]\nn\\
K_{\frac12,\frac12}
&=& \frac{\sqrt{2}}{4}
\left[ \sqrt{3}(E_{p^0}+i E_{q_0}) +i (E_{p^1}+i E_{q_1})
+ \sqrt{3} (F_{p^0}+i F_{q_0}) + i (F_{p^1}+i F_{q_1})\right]\nn\\
K_{\frac12,-\frac12}
&=& \frac{\sqrt{2}}{4}
\left[ \sqrt{3}(E_{p^0}-i E_{q_0}) -i (E_{p^1}-i E_{q_1})
- \sqrt{3} (F_{p^0}-i F_{q_0}) + i (F_{p^1}-i F_{q_1})\right]\nn\\
K_{\frac12,-\frac32}
&=& \frac{\sqrt{2}}{4}
\left[ -(E_{p^0}+i E_{q_0}) + i \sqrt{3} (E_{p^1}+i E_{q_1})
- (F_{p^0}-i F_{q_0}) + i \sqrt{3} (F_{p^1}+i F_{q_1})\right]\nn\\
\end{eqnarray}
where the eigenvalues under of $(-iL_0,-iR_0)$ are indicated in subscript.
Note that the hermiticity conditions are now
\be
L_\pm^\dagger = - L_\mp\ ,\quad
R_\pm^\dagger = - R_\mp\ ,\quad
K_{-l_0,-r_0}=- K^{\dagger}_{l_0,r_0}
\ee
In this basis, the quadratic Casimir becomes
\bea
C_2 &=& -\frac13(R_0^2-R_-R_+-R_+R_-) - (L_0^2-L_-L_+-L_+L_-) \\
&&
-\frac18 \left( K_{\frac12,\frac32}K_{-\frac12,-\frac32}
+K_{-\frac12,-\frac32}K_{\frac12,\frac32}\right)
+\frac18 \left( K_{-\frac12,\frac32}K_{\frac12,-\frac32}
+K_{\frac12,-\frac32}K_{-\frac12,\frac32}\right) \nn \\
&&-\frac18 \left( K_{\frac12,-\frac12}K_{-\frac12,\frac12}
+K_{-\frac12,\frac12}K_{\frac12,-\frac12}\right)
+\frac18 \left( K_{\frac12,\frac12}K_{-\frac12,-\frac12}
+K_{-\frac12,-\frac12}K_{\frac12,\frac12}\right) \nn
\eea
This shows in particular that the roots $K_{\pm\frac12,\pm\frac32}$
and  $K_{\pm\frac12,\mp\frac12}$ are compact, as indicated on the
diagram.

The compact Cartan generator $L_0$
gives rise to the ``non-compact holomorphic 5-grading''
\be
L_- \vert_{-2i} \oplus
\{ K_{-\frac12,\pm \frac12}, K_{-\frac12,\pm\frac32}  \}\vert_{-i} \oplus
\{ L_0, R_0, R_\pm \}\vert_0  \oplus
\{ K_{\frac12,\pm \frac12}, K_{\frac12,\pm\frac32}  \}\vert_{i} \oplus
L_+ \vert_{2i}
\ee

Now, we perform a $\pi/3$ rotation of the root diagram, and define
\bse
\bea
J_3 &=&
\frac12 ( L_0 + R_0 ) = \frac14 (F-E) + \frac{1}{2\sqrt{2}}( Y_+ + Y_-)\\
S_3 &=&
\frac12 ( 3L_0 - R_0 ) = \frac34 (F-E) - \frac{1}{2\sqrt{2}}( Y_+ + Y_-)
\eea
\ese
as the new Cartan algebra. The new eigenmodes are now
\be
J_- = \frac12 K_{-\frac12,-\frac32} \ ,\quad
J_+ = \frac12 K_{\frac12,\frac32} \ ,\quad
S_-=\frac{3}{2} K_{-\frac12,\frac12} \ ,\quad
S_+=\frac{3}{2} K_{\frac12,-\frac12} \nn
\ee
\be
J_{\frac12,-\frac32} = K_{-\frac12,\frac32}\ ,\quad
J_{\frac12,\frac32}=2\sqrt{2} L_+\, \quad
J_{\frac12,-\frac12} =  2\sqrt{\frac{2}{3}} R_+\ ,\quad
J_{\frac12,\frac12}=K_{\frac12,\frac12}
\ee
together with their hermitian conjugates. They satisfy the $SU(2)\times
SU(2)$ algebra
\bse
\bea
\left[ J_3, J_\pm \right] = \pm i J_\pm \ ,&\quad& [J_+,J_-]= 2 i J_3 \\
\left[ S_3, S_\pm \right] = \pm i S_\pm \ ,&\quad& [S_+,S_-]= 2 i S_3
\eea
\ese
The subscript on $J_{}$ now denotes the eigenvalues under $(-iJ_3,-iS_3)$.
In terms of the non-compact basis, the compact generators are
as usual differences between positive and negative roots,
\bse
\bea
F_{p^0}-E_{p^0} &=& \frac12 \left( S_+ + S_- + J_+ + J_- \right)\\
F_{p^1}-E_{p^1} &=& \frac{i}{2\sqrt{3}}
\left( S_+ - S_- -3 J_+ + 3 J_- \right)\\
F_{q_0}-E_{q_0} &=& \frac{i}{2}
\left( S_+ - S_- + J_+ - J_- \right)\\
F_{q_1}-E_{q_1} &=& \frac{1}{2\sqrt{3}}
\left( -S_+ - S_- +3 J_+ + 3 J_- \right)\\
F - E &=& S_3 + J_3 \\
Y_+ + Y_- &=& \frac{1}{\sqrt{2}}(3J_3 - S_3)
\eea
\ese
The matrix representation adapted to this compact basis is obtained
from \eqref{g2X} by a Cayley rotation
\be
\label{cayleyg2}
C\eta C^t= \begin{pmatrix}
 0 & 0 & 0 & 0 & 0 & 0 & 1 \\
 0 & 0 & 0 & 0 & 0 & 1 & 0 \\
 0 & 0 & 0 & 0 & -1 & 0 & 0 \\
 0 & 0 & 0 & -1 & 0 & 0 & 0 \\
 0 & 0 & -1 & 0 & 0 & 0 & 0 \\
 0 & 1 & 0 & 0 & 0 & 0 & 0 \\
 1 & 0 & 0 & 0 & 0 & 0 & 0
\end{pmatrix} , \quad
C=e^{\frac{\pi \sqrt{2}}{4}(R_+ + R_-)} = \begin{pmatrix}
 0 & 0 & 0 & \frac{i}{2} & \frac{1}{2} & -\frac{1}{2} & \frac{i}{2} \\
 -\frac{i}{2} & \frac{1}{\sqrt{2}} & \frac{i}{2} & 0 & 0 & 0 & 0 \\
 0 & 0 & 0 & \frac{i}{2} & \frac{1}{2} & \frac{1}{2} & -\frac{i}{2} \\
 \frac{1}{\sqrt{2}} & 0 & \frac{1}{\sqrt{2}} & 0 & 0 & 0 & 0 \\
 0 & 0 & 0 & -\frac{i}{2} & \frac{1}{2} & \frac{1}{2} & \frac{i}{2} \\
 \frac{i}{2} & \frac{1}{\sqrt{2}} & -\frac{i}{2} & 0 & 0 & 0 & 0 \\
 0 & 0 & 0 & -\frac{i}{2} & \frac{1}{2} & -\frac{1}{2} & -\frac{i}{2}
\end{pmatrix}
\ee
The generators in the compact basis then have the matrix representation
\scriptsize
\begin{equation}
\label{Xcayleyg2}
\begin{pmatrix}
 \frac{1}{2} i (\underline{J}_3+\underline{S}_3) & -\underline{S}_+ &
-\frac{2 i \underline{J}_{\frac{1}{2},-\frac{1}{2}}}{\sqrt{3}} & 2 i \sqrt{\frac{2}{3}}
   \underline{J}_{\frac{1}{2},\frac{1}{2}} & -2 \underline{J}_{\frac{1}{2},\frac{3}{2}} & \underline{J}_{+} & 0 \\
 \underline{S}_{-} & \frac{1}{2} i (\underline{J}_3-\underline{S}_3) & -2 \underline{J}_{\frac{1}{2},-\frac{3}{2}} & -2 \sqrt{\frac{2}{3}} \underline{J}_{\frac{1}{2},-\frac{1}{2}} & -\frac{2
   \underline{J}_{\frac{1}{2},\frac{1}{2}}}{\sqrt{3}} & 0 & -\underline{J}_{+} \\
 \frac{2 i \underline{J}_{-\frac{1}{2},\frac{1}{2}}}{\sqrt{3}} & -2 \underline{J}_{-\frac{1}{2},\frac{3}{2}} & i \underline{S}_3 & -i \sqrt{2} \underline{S}_+ & 0 & -\frac{2
   \underline{J}_{\frac{1}{2},\frac{1}{2}}}{\sqrt{3}} & -2 \underline{J}_{\frac{1}{2},\frac{3}{2}} \\
 -2 i \sqrt{\frac{2}{3}} \underline{J}_{-\frac{1}{2},-\frac{1}{2}} & -2 \sqrt{\frac{2}{3}} \underline{J}_{-\frac{1}{2},\frac{1}{2}} & -i \sqrt{2} \underline{S}_{-} & 0 & i \sqrt{2}
   \underline{S}_+ & -2 \sqrt{\frac{2}{3}} \underline{J}_{\frac{1}{2},-\frac{1}{2}} & 2 i \sqrt{\frac{2}{3}} \underline{J}_{\frac{1}{2},\frac{1}{2}} \\
 -2 \underline{J}_{-\frac{1}{2},-\frac{3}{2}} & -\frac{2 \underline{J}_{-\frac{1}{2},-\frac{1}{2}}}{\sqrt{3}} & 0 & i \sqrt{2} \underline{S}_{-} & -i \underline{S}_3 & -2
   \underline{J}_{\frac{1}{2},-\frac{3}{2}} & -\frac{2 i \underline{J}_{\frac{1}{2},-\frac{1}{2}}}{\sqrt{3}} \\
 -\underline{J}_{-} & 0 & -\frac{2 \underline{J}_{-\frac{1}{2},-\frac{1}{2}}}{\sqrt{3}} & -2 \sqrt{\frac{2}{3}} \underline{J}_{-\frac{1}{2},\frac{1}{2}} & -2
   \underline{J}_{-\frac{1}{2},\frac{3}{2}} & -\frac{1}{2} i (\underline{J}_3-\underline{S}_3) & \underline{S}_+ \\
 0 & \underline{J}_{-} & -2 \underline{J}_{-\frac{1}{2},-\frac{3}{2}} & -2 i \sqrt{\frac{2}{3}} \underline{J}_{-\frac{1}{2},-\frac{1}{2}} & \frac{2 i
   \underline{J}_{-\frac{1}{2},\frac{1}{2}}}{\sqrt{3}} & -\underline{S}_{-} & -\frac{1}{2} i (\underline{J}_3+\underline{S}_3)
\end{pmatrix}
\end{equation}
\normalsize
The advantage of this description is that the Harish-Chandra decomposition
with respect to $J_3$ is simply a generalized $LU$ decomposition in 2+3+2 blocks.

The quadratic Casimir in the compact basis becomes
\bea
C_2 &=& -\frac13(S_3^2+ \frac12 S_-S_+ + \frac12 S_+S_-)
- (J_3^2+\frac12 J_-J_++\frac12 J_+J_-) \\
&&
+\frac18 \left( J_{\frac12,\frac32}J_{-\frac12,-\frac32}
+J_{-\frac12,-\frac32}J_{\frac12,\frac32}\right)
+\frac18 \left( J_{-\frac12,\frac32}J_{\frac12,-\frac32}
+J_{\frac12,-\frac32}J_{-\frac12,\frac32}\right) \nn \\
&&+\frac18 \left( J_{\frac12,-\frac12}J_{-\frac12,\frac12}
+J_{-\frac12,\frac12}J_{\frac12,-\frac12}\right)
+\frac18 \left( J_{\frac12,\frac12}J_{-\frac12,-\frac12}
+J_{-\frac12,-\frac12}J_{\frac12,\frac12}\right) \nn
\eea
which makes it clear that the compact roots are $J_\pm, S_\pm$.

The generator $J_3$
now gives rise to the  ``compact 5-grading''
\be
J_- \vert_{-2i} \oplus
\{ T_-,R_-,U_-,L_- \}\vert_{-i} \oplus
\{ J_3, S_3, S_+, S_- \}\vert_0  \oplus
\{ T_+,R_+,U_+,L_+ \}\vert_{i} \oplus
J_+ \vert_{2i}
\ee
where we denoted $T_+=J_{\frac12,-\frac32}, U_+=J_{\frac12,\frac12}$.

We now consider the $SL(3,\IR)$ and $SU(2,1)$ subalgebras of $G_{2(2)}$.
The $SL(3,\IR)$ subalgebra is generated by the long roots in the non-compact basis.
Its quadratic
Casimir reads
\be
C_2[SL(3,\R)]= \frac14 H^2 + \frac13 Y_0^2
+ \frac14
( E_{p^0} F_{p^0} + F_{p^0} E_{p^0} + E_{q_0} F_{q_0} + F_{q_0} E_{q_0} )
+ \frac12 (EF+FE)
\ee
The $SU(2,1)$ subalgebra is generated by the long roots in the compact basis.
Its quadratic Casimir reads
\be
\begin{split}
C_2[SU(2,1)] =& -\frac13 S_3^2 - (J_3^2+J_-J_++J_+J_-) \\
&
+\frac18 \left( J_{\frac12,\frac32}J_{-\frac12,-\frac32}
+J_{-\frac12,-\frac32}J_{\frac12,\frac32}\right)
+\frac18 \left( J_{-\frac12,\frac32}J_{\frac12,-\frac32}
+J_{\frac12,-\frac32}J_{-\frac12,\frac32}\right)
\end{split}
\ee

\subsection{Quaternionic symmetric space}

The long $SU(2)_{J_\pm,J_3}$ endows $K \bas G = (SU(2)\times SU(2)) \backslash G_{2(2)}$ with a
quaternionic-K\"ahler geometry. In order to describe its geometry, we perform the Iwasawa
decomposition $g=k\cdot e_{QK}$ (slightly adapted from \cite{Mizoguchi:1999fu}) 
\be 
e_{QK} =
\tau_2^{-Y_0} \cdot e^{\sqrt{2} \tau_1 Y_+} \cdot e^{-U H} \cdot e^{-\zeta^0 {E_{q_0}} + \tzeta_0
{E_{p^0}}} \cdot e^{-\sqrt{3} \zeta^1 {E_{q_1}} + \frac{\sqrt{3}}{3} \tzeta_1 {E_{p^1}}} \cdot
e^{\sigma E} 
\ee 
where $k$ is an element of the maximal compact subgroup.  This decomposition defines coordinates 
$(\tau_1, \tau_2, \zeta^0, \zeta^1, \tzeta_0, \tzeta_1, U, \sigma)$ on $K \bas G$, where
$\tau = \tau_1 + i \tau_2$ is an element of the upper half-plane.

The invariant $\fg$-valued one-form $\theta=de_{QK} \cdot e_{QK}^{-1}$ may be expanded on the
compact basis, leading to the quaternionic viel-bein \be
\begin{pmatrix}
\underline{J}_{-\frac12,\frac32}  & \underline{J}_{\frac12,\frac32} \\
i \underline{J}_{-\frac12,\frac12}  & \underline{J}_{\frac12,\frac12} \\
\underline{J}_{-\frac12,-\frac12} & i \underline{J}_{\frac12,-\frac12} \\
-\underline{J}_{-\frac12,-\frac32} & \underline{J}_{\frac12,-\frac32}
\end{pmatrix}
= -
\begin{pmatrix}
\bar u &  v \\ -\bar  e^1 & E_1 \\ \bar E_1 & e^1 \\ - \bar  v & u
\end{pmatrix}\ ,\quad
\ee
and the $SU(2)\times SU(2)$ spin connection
\be
\begin{pmatrix}
\underline{J}_- \\ \underline{J}_3 \\ \underline{J}_+
\end{pmatrix}
=-\frac12
\begin{pmatrix}
\bar u \\
\frac{1}{4i}(v-\bar v) + \frac{i\sqrt{3}}{4}(e^1-\bar e^1) \\
u
\end{pmatrix}\ ,\quad
\begin{pmatrix}
\underline{S}_- \\ \underline{S}_3 \\ \underline{S}_+
\end{pmatrix}
=\frac{\sqrt{3}}{2}
\begin{pmatrix}
E_1 \\
\frac{i\sqrt{3}}{4}(v-\bar v) +\frac{i}{4}(e^1-\bar e^1) \\
\bar E_1
\end{pmatrix}\ ,\quad
\ee
where
\bse
\bea
u &=& \frac{e^{-U}}{ 2\sqrt{2}\, \tau_2^{3/2}}
\left( d\tzeta_0 + \tau d\tzeta_1
+ 3 \tau^2 d\zeta^1 - \tau^3 d\zeta^0 \right)\\
v &=& dU + \frac{i}{2} e^{-2U} (
d\sigma - \zeta^0 d\tzeta_0 -\zeta^1 d\tzeta_1 
+  \tzeta_0 d\zeta^0 +\tzeta_1 d\zeta^1)\\
e^1 &=&  \frac{i \sqrt{3}}{2\tau_2} d\tau  \\
E_1 &=&
-\frac{e^{-U}}{  2\sqrt{6}\,\tau_2^{3/2}}
\left( 3 d\tzeta_0 + d\tzeta_1 (\bar\tau+2\tau)
+3 \tau (2\bar\tau+\tau) d\zeta^1 -3 \bar\tau \tau^2 d\zeta^0 \right)
\eea
\ese
This form of the vielbein agrees with that of the $c$-map space with
prepotential $F =-(X^1)^3/X^0$ \cite{Ferrara:1989ik}, as expected.
The metric is then
\be
ds^2 = 2 \left(
u \,\bar u +
v \, \bar v +
e^1 \, \bar e^1 +
E_1 \, \bar E_1 \right)
\ee
The right action of $G$ on $K\backslash G$ is given by the vector fields
\bse
\bea
E&=&\pa_{\sigma}\\
E_{p^0}&=&\pa_{\tzeta_0} - \zeta^0 \pa_{\sigma} \quad,\quad
E_{q_0}= -\pa_{\zeta^0} - \tzeta_0 \pa_{\sigma} \\
E_{p^1}&=& \sqrt{3}(\pa_{\tzeta_1} -  \zeta^1 \pa_{\sigma}) \quad,\quad
E_{q_1}= \frac{1}{\sqrt{3}}(-\pa_{\zeta^1} -  \tzeta_1 \pa_{\sigma})
\\
%\eea
%\bea
H &=& -\pa_U -2 \sigma \pa_{\sigma}
- \zeta^0 \pa_{\zeta^0} - \zeta^1 \pa_{\zeta^1}
-   \tzeta_0 \pa_{\tzeta_0}- \tzeta_1 \pa_{\tzeta_1}\\
Y_0&=&-\frac12(2\tau_1 \pa_{\tau_1} + 2\tau_2 \pa_{\tau_2}
- 3 \zeta^0 \pa_{\zeta^0}+ 3 \tzeta_0 \pa_{\tzeta_0}
- \zeta^1 \pa_{\zeta^1}+ \tzeta_1 \pa_{\tzeta_1})\\
Y_+&=& \frac{1}{\sqrt{2}}(\pa_{\tau_1} + \zeta^0 \pa_{\zeta^1}
- 6\zeta^1\pa_{\tzeta_1} - \tzeta_1\pa_{\tzeta_0})\\
Y_-&=&  \frac{1}{3\sqrt{2}}
(6\tau_1\tau_2 \pa_{\tau_2} + 3 (\tau_1^2 - \tau_2^2) \pa_{\tau_1}
+ 9\tzeta_0 \pa_{\tzeta_1}- 9\zeta^1 \pa_{\zeta^0}
+ 2\tzeta_1\pa_{\zeta^1})
\eea
\ese
The other negative roots are too bulky to be displayed.

\subsection{Twistor space}
The twistor space
\be
\cZ =  (SU(2)_{S_\pm,S_3} \times U(1)_{J_3}) \bas G_{2(2)}
\ee
can be parameterized by the coset representative
\be
\label{ezsu2}
e_{\cZ} = e^{-{\bar z} J_+}\ (1+z\bar z)^{-i J_3} \ e^{-z J_-} \ e_{QK}
\ee
As in Section \ref{hchandsu21}, we can construct
complex coordinates on $\cZ$ using the Borel embedding:
\be
\label{hcg2}
C e_{\cZ} C^{-1} =
\begin{pmatrix}
1&&&&&&\\
&1&&&&&\\
*&*&1&&&&\\
*&*&&1&&&\\
*&*&&&1&&\\
*&*&*&*&*&1&\\
*&*&*&*&*&&1
\end{pmatrix}
\cdot
\begin{pmatrix}
*&*&&&&&\\
*&*&&&&&\\
&&*&*&*&&\\
&&*&*&*&&\\
&&*&*&*&&\\
&&&&&*&*\\
&&&&&*&*\\
\end{pmatrix}
\cdot
\begin{pmatrix}
1&&*&*&*&*&*\\
&1&*&*&*&*&*\\
&&1&&&*&*\\
&&&1&&*&*\\
&&&&1&*&*\\
&&&&&1&\\
&&&&&&1
\end{pmatrix}
\ee The entries of the upper triangular matrix then provide a complex coordinate system such that
the ``compact Heisenberg algebra'' $J_{\frac12,\pm\frac12/\pm\frac32},J_+$ acts in a simple way,
essentially by shifts.
The K\"ahler potential in these coordinates (analogous to the bounded domain coordinates of
$SL(2,\R)/U(1)$) appears to be complicated; we do not consider them further here.

To get a simpler form for the K\"ahler potential we construct
complex coordinates adapted to the ``real'' Heisenberg algebra, by performing a decomposition
analogous to \eqref{hcg2} but using $e_\cZ$ directly rather than $C e_\cZ C^{-1}$, and then
substituting $\bar z \to - z^{-1}$ at the end, as we did at the end of Section \ref{sectwimapsu21}.
The resulting complex coordinates $(\xi^0, \xi^1, \txi_0, \txi_1, \alpha)$ are related to the
coordinates on the base and the stereographic coordinate on the fiber by \bea \label{g2twi} \xi^0
&=& \zeta^0 + \frac{i}{2\sqrt{2}} \frac{e^U}{\tau_2^{3/2}} \left( z + z^{-1} \right) \ ,\quad \xi^1
= \zeta^1 + \frac{i}{2\sqrt{2}} \frac{e^U}{\tau_2^{3/2}}
\left( \bar \tau \ z + \tau \ z^{-1} \right) \\
\txi_1 &=& \tzeta_1 - \frac{i}{2\sqrt{2}} \frac{e^U}{\tau_2^{3/2}}
\left( 3\bar\tau^2 \ z + 3\tau^2 \ z^{-1} \right) \ ,\quad
\txi_0 = \tzeta_0 + \frac{i}{2\sqrt{2}} \frac{e^U}{\tau_2^{3/2}}
\left( \bar\tau^3 \ z + \tau^3 \ z^{-1} \right) \nn\\
\alpha&=& \sigma +  \frac{i}{2\sqrt{2}} \frac{e^U}{\tau_2^{3/2}}
\left[ \left( \bar\tau^3 \zeta^0-3\bar\tau^2 \zeta^1
-\bar\tau\tzeta_1-\tzeta_0\right)
z + \left( \tau^3 \zeta^0-3\tau^2 \zeta^1-\tau\tzeta_1-
\tzeta_0\right) z^{-1} \right]\nn
\eea
These coordinates are an example of the ``canonical'' coordinates of
quasi-conformal geometries defined by Jordan algebras
\cite{Gunaydin:2000xr,Gunaydin:2005zz}.  Such coordinates in fact
exist for all $c$-map spaces \cite{Neitzke:2007ke}, and \eqref{g2twi}
agrees with the ``twistor map'' derived in that context in
\cite{Neitzke:2007ke}.  To compare the two, recall that the K\"ahler
potential on the special K\"ahler base (the upper half-plane in this
case) is $e^{-K}=8\tau_2^3$.

The K\"ahler potential on $\cZ$ is
\be
K_\cZ =\frac{1}{2} \log N_4 =  \frac12 \log\left[ I_4(\xi^I-\bar\xi^I,
\txi_I-\bar\txi_I) + (\alpha-\bar \alpha+ \xi^I \bar\txi_I 
- \bar\xi^I \txi_I)^2
\right]
\ee
where
\be
\label{i4g2}
I_4(\xi,\txi) =
(\xi^0)^2 \txi_0^2
+ 4(\xi^1)^3 \txi_0
+2 \xi^0 \txi_0 \xi^1 \txi_1
- \frac13 (\xi^1)^2 \txi_1^2
- \frac{4}{27} \txi_1^3 \xi^0
\ee
is the quartic invariant of $SL(2,\IR)$ in its spin-3/2
representation, and $N_4$ is the
quartic distance function of quasi-conformal geometry, which defines
the ``quartic light-cone''.

In parallel to the discussion of Section \ref{secswann}, one can also define the homogeneous
line bundles $\cO(k)$ over $\cZ$, and the total space of $\cO(-2)$ gives the Swann space $\cS$.

\subsection{Quasi-conformal representations}

We construct a principal series representation of $G$ by induction from the parabolic $P$,
with the character $\chi_k = e^{-k \underline{H}/2}$ of $P$.
The infinitesimal action of $G$ in this representation is determined as in Section \ref{secprincipal},
this time using the decomposition of $G$ as
\bea
\label{gpg2}
g &=& p\cdot e^{-\zeta^0 {E_{q_0}} + \tzeta_0 {E_{p^0}}}
\cdot e^{-\sqrt{3} \zeta^1 {E_{q_1}} + \frac{\sqrt{3}}{3} \tzeta_1 {E_{p^1}}}
\cdot e^{\sigma E}\\
&=& p\cdot \begin{pmatrix}
 1 & 0 & 0 & 0 & \frac{\sqrt{2}}{3} {\tzeta_1} & 0 & \sqrt{2} {\tzeta_0} \\
 0 & 1 & 0 & 0 & -2 {\zeta^1} & 0 & -\frac{2}{3} {\tzeta_1} \\
 0 & 0 & 1 & 0 & -\sqrt{2} {\zeta^0} & 0 & \sqrt{2} {\zeta^1} \\
 \sqrt{2} {\zeta^1} & \frac{2}{3} {\tzeta_1} &
\sqrt{2} {\tzeta_0} & 1 &
\sigma-{\zeta^0}
   {\tzeta_0}-\frac13 \zeta^1 \tzeta_1 & 0 & 2 {\tzeta_0} {\zeta^1}
-\frac{2}{9}
   {\tzeta_1}^2 \\
 0 & 0 & 0 & 0 & 1 & 0 & 0 \\
 \sqrt{2} {\zeta^0} & -2 {\zeta^1} & -\frac{\sqrt{2}}{3} \tzeta_1 & 0 & 2
   (\zeta^1)^2+\frac{2}{3} \zeta^0 {\tzeta_1} & 1 & \sigma+{\zeta^0}
   {\tzeta_0}+\frac13 \zeta^1 \tzeta_1 \\
 0 & 0 & 0 & 0 & 0 & 0 & 1
\end{pmatrix}
\eea
where $p \in P$.  The generators act by
\bea
H^{QC} &=& -2 \sigma {\pa_\sigma}
- {\zeta^0}{\pa_{\zeta^0}}
-{\zeta^1}{\pa_{\zeta^1}}
-{\tzeta_0}{\pa_{\tzeta_0}}
-{\tzeta_1} {\pa_{\tzeta_1}}-k  \ ,\quad
E^{QC} = -\pa_\sigma \nn\\
E_{p^0}^{QC} &=& \pa_{\tzeta_0} + \zeta^0 \pa_\sigma \ ,\quad
E_{p^1}^{QC} = \sqrt{3} \left( \pa_{\tzeta_1} + \zeta^1 \pa_\sigma \right) \nn\\
E_{q_0}^{QC} &=& -\pa_{\zeta^0} + \tzeta_0 \pa_\sigma \ ,\quad
E_{q_1}^{QC} = \frac{1}{\sqrt{3}}
\left( -\pa_{\zeta^1} + \tzeta_1 \pa_\sigma \right) \nn\\
{Y_+}^{QC} &=& \frac{1}{\sqrt{2}} {\zeta^0} {\pa_{\zeta^1}}
   -3 \sqrt{2} {\zeta^1} {\pa_{\tzeta_1}}
   -\frac{1}{\sqrt{2}}  {\tzeta_1}{\pa_{\tzeta_0}}\ ,\quad
{Y_-}^{QC} = \frac{3}{\sqrt{2}} \left( {\tzeta_0}\pa_{\tzeta_1}
  -{\zeta^1}{\pa_{\zeta^0}} \right)
   +\frac{\sqrt{2}}{3}  {\tzeta_1} {\pa_{\zeta^1}}\nn\\
{Y_0}^{QC} &=& \frac12 \left(
3 {\zeta^0} {\pa_{\zeta^0}}
+ {\zeta^1} {\pa_{\zeta^1}}
-3 {\tzeta_0} {\pa_{\tzeta_0}}
- {\tzeta_1} {\pa_{\tzeta_1}} \right)\nn\\
F^{QC} &=&
\left(2(\zeta^1)^3 +
{\zeta^0}{\tzeta_1}{\zeta^1}
+ (\zeta^0)^2{\tzeta_0}
+ \sigma {\zeta^0}  \right)
{\pa_{\zeta^0}}
-\left( \frac13 {\tzeta_1}(\zeta^1)^2
-{\zeta^0}{\tzeta_0}{\zeta^1} + \frac29 {\zeta^0}{\tzeta_1}^2
- \sigma {\zeta^1}  \right)
{\pa_{\zeta^1}}\nn\\
&&+
\left(-6{\tzeta_0}(\zeta^1)^2
+\frac13 {\tzeta_1}^2{\zeta^1}
-{\zeta^0}{\tzeta_0} {\tzeta_1}
+\sigma {\tzeta_1}
\right)
{\pa_{\tzeta_1}}
+
\left( \frac{2}{27} {\tzeta_1^3} - {\tzeta_0} {\zeta^1}{\tzeta_1} -
{\zeta^0}{\tzeta_0}^2 + \sigma {\tzeta_0}\right) {\pa_{\tzeta_0}} \nn \\
&&- [I_4(\zeta,\tzeta)+\sigma^2] \pa_\sigma  - k \sigma
\label{g2qcg}
\eea
where $I_4(\zeta,\tzeta)$ is the quartic polynomial defined in \eqref{i4g2}.
The quadratic Casimir evaluates to
\be
C_2 = -\frac14 k(6-k)
\ee
This representation is unitary for $k\in 3+i\IR$, with respect to the
inner product
\be
\label{quatdisckg2c}
\langle f_1 | f_2 \rangle =
\int d\ze^\Lambda\, d\tze_\Lambda\, d\sigma\,
f_1^*(\ze^\Lambda,\tze_\Lambda,\sigma)\,
f_2(\ze^\Lambda,\tze_\Lambda,\sigma)\ .
\ee

Complexifying, we can also consider the holomorphic
right action of $G$ on sections of the line bundle $\cO(-k)$ on the
twistor space $\cZ=P_\IC \backslash G_{\IC}$.
The infinitesimal action is the complexification of the one above,
\bea
H^{QC} &=& -2 \alpha {\pa_\alpha}
- {\xi^0}{\pa_{\xi^0}}
-{\xi^1}{\pa_{\xi^1}}
-{\txi_0}{\pa_{\txi0}}
-{\txi_1} {\pa_{\txi_1}}-k  \ ,\quad
E^{QC} = -\pa_\alpha \nn\\
E_{p^0}^{QC} &=& \pa_{\txi_0} + \xi^0 \pa_\alpha \ ,\quad
E_{p^1}^{QC} = \sqrt{3} \left( \pa_{\txi_1} + \xi^1 \pa_\alpha \right) \nn\\
E_{q_0}^{QC} &=& -\pa_{\xi^0} + \txi_0 \pa_\alpha \ ,\quad
E_{q_1}^{QC} = \frac{1}{\sqrt{3}}
\left( -\pa_{\xi^1} + \txi_1 \pa_\alpha \right) \nn\\
{Y_+}^{QC} &=& \frac{1}{\sqrt{2}} {\xi^0} {\pa_{\xi^1}}
   -3 \sqrt{2} {\xi^1} {\pa_{\txi_1}}
   -\frac{1}{\sqrt{2}}  {\txi_1}{\pa_{\txi0}}\ ,\quad
{Y_-}^{QC} = \frac{3}{\sqrt{2}} \left( {\txi_0}\pa_{\txi_1}
  -{\xi^1}{\pa_{\xi^0}} \right)
   +\frac{\sqrt{2}}{3}  {\txi_1} {\pa_{\xi^1}}\nn\\
{Y_0}^{QC} &=& \frac12 \left(
3 {\xi^0} {\pa_{\xi^0}}
+ {\xi^1} {\pa_{\xi^1}}
-3 {\txi_0} {\pa_{\txi_0}}
- {\txi_1} {\pa_{\txi_1}} \right)\nn\\
F^{QC} &=&
\left(2(\xi^1)^3 +
{\xi^0}{\txi_1}{\xi^1}
+ (\xi^0)^2{\txi_0}
+ \alpha {\xi^0}  \right)
{\pa_{\xi^0}}
-\left( \frac13 {\txi_1}(\xi^1)^2
-{\xi^0}{\txi_0}{\xi^1} + \frac29 {\xi^0}{\txi_1}^2
- \alpha {\xi^1}  \right)
{\pa_{\xi^1}}\nn\\
&&+
\left(-6{\txi_0}(\xi^1)^2
+\frac13 {\txi_1}^2{\xi^1}
-{\xi^0}{\txi_0} {\txi_1}
+\alpha {\txi_1}
\right)
{\pa_{\txi_1}}
+
\left( \frac{2}{27} {\txi_1^3} - {\txi_0} {\xi^1}{\txi_1} -
{\xi^0}{\txi_0}^2 + \alpha {\txi_0}\right) {\pa_{\txi0}} \nn \\
&&- [I_4(\xi,\txi)+\alpha^2] \pa_\alpha  - k \alpha
\eea
Naively this construction would require $k \in \Z$, but
in what follows, we will sometimes consider this action not only when $k$ is integral 
but even when $k \in \frac{1}{3}\Z$.  Presumably this should be understood in terms of
a triple cover of $\cZ$.

This representation is formally unitary under the inner product
\be
\label{quatdisckg2}
\langle f_1 | f_2 \rangle =
\int d\xi^\Lambda\, d\txi_\Lambda\, d\alpha\, d\bar\xi^\Lambda\, d\bar\txi_\Lambda\, d\bar\alpha\,
e^{(k-6)K_\cZ}\,
f_1^*(\bar\xi,\bar\txi,\bar\alpha)\,
f_2(\xi,\txi,\alpha)
\ee
As discussed in Section \ref{secqdisc} for $G = SU(2,1)$, we expect this formally
unitary action on sections of $\cO(-k)$ to yield a genuine unitary
action on $H^1(\cZ, \cO(-k))$ at least for $k \ge 3$; for
$k \ge 5$ it should give the quaternionic discrete
series of $G_{2(2)}$.  Moreover, it should occur as a subquotient of the principal
series for some $k$.  We will discuss this further at the level of the
$K$-finite vectors below.

\subsubsection{Lift to hyperkahler cone \label{hkcliftg2}}

Similar to the discussion of Section \ref{lifthkc} for $G = SU(2,1)$, the
action of $G = G_{2(2)}$ on holomorphic sections of $\cO(-k)$ over $\cZ$ is equivalent to an action on
holomorphic functions of homogeneity degree $-k$ on the Swann space $\cS$.
Introducing the complex coordinates on $\cS$
\bea
\label{vtoxi}
v^\flat&=&e^{2t}\ ,\quad
w_\flat=-\frac{1}{4i}(\alpha+\xi^0 \txi_0+\xi^1 \txi_1)\\
v^0&=& 3\sqrt{3}\xi^0\, e^{2t}\ ,\quad
w_0= -\frac{i}{6\sqrt{3}}\,\txi_0\ ,\quad
v^1= - \frac{1}{\sqrt{3}}\xi^1\, e^{2t}\ ,\quad
w_1=\frac{i \sqrt{3}}{2}\,\txi_1\ ,\quad
\eea
it is straightforward to compute the holomorphic vector fields
corresponding to the action of $G_{\IC}$ on $\cS$ and determine their
holomorphic moment maps. Expressing the result in
terms of the coordinates $\xi^I,\txi_I,\alpha$ on $\cZ$
and $t$ in the $\IC^*$ fiber, we find the holomorphic moment maps
\bea
\label{ehamg2}
\moment{H}&=&\frac{i}{2} e^{2t}\,\alpha\ ,\quad
\moment{Y}_0=\frac{i}{4} e^{2t} (3 {\xi^0} {\txi_0}+{\xi^1} {\txi_1})\ ,\quad
\moment{E}=-\frac{i}{4} e^{2t}\ ,\nn\\
\moment{Y}_+&=&-\frac{i}{2\sqrt{2}} e^{2t} \left( {\txi_1}^2+9 {\txi_0} {\xi^1}
\right)\ ,\quad
\moment{Y}_-=\frac{i}{6 \sqrt{2}} e^{2t}
\left(\xi^0 \txi_1-3 (\xi^1)^2\right) \nn\\
\moment{E}_{p^0}&=&- \frac{3 i\sqrt{3}}{2} e^{2t} {\txi_0}\ ,\quad
\moment{E}_{p^1}=\frac{i}{2} e^{2t} {\txi_1}\ ,\quad
\moment{E}_{q_0}=\frac{i}{6 \sqrt{3}} e^{2t} \,\xi^0\ ,\quad
\moment{E}_{q_1}=-\frac{i}{2} e^{2t}\,{\xi^1}\nn\\
\moment{F}_{p^0}&=&\frac{i}{6\sqrt{3}} e^{2t}
\left( -2(\xi^1)^3+{\xi^0} {\txi_1} {\xi^1}
+\alpha {\xi^0}+(\xi^0)^2 {\txi_0}\right) \nn\\
\moment{F}_{p^1}&=&\frac{i}{18} e^{2t}
\left(3 {\txi_1} (\xi^1)^2-9 \alpha {\xi^1}-9 {\xi^0}
{\txi_0}   {\xi^1}-2 {\xi^0} {\txi_1}^2\right)\nn \\
\moment{F}_{q_0}&=&- \frac{i}{6 \sqrt{3}}
e^{2t} \left( 2 {\txi_1}^3+27 {\txi_0} {\xi^1}
   {\txi_1}+27 {\xi^0} {\txi_0}^2-27 \alpha {\txi_0}\right) \nn \\
\moment{F}_{q_1}&=&\frac{i}{2} e^{2t} \left[ {\txi_0} \left({\xi^0}
   {\txi_1}-6 (\xi^1)^2\right)-\frac{1}{3}
{\txi_1} (3 \alpha+{\xi^1} {\txi_1})\right] \nn \\
\moment{F}&=&\frac{i}{2} e^{2t} \left[
2{\txi_0} {\xi^1}^3+\frac{{\txi_1}^2 (\xi^1)^2}{6}-{\xi^0} {\txi_0} {\txi_1}
   {\xi^1}-\frac{2 {\xi^0} {\txi_1}^3}{27}+\frac{\alpha^2}{2}
-\frac{(\xi^0)^2 {\txi_0}^2}{2}\right]
\eea
In order to compute the moment maps in the compact basis, which will
be relevant in the next section, it is convenient to change variables
to $(a,b,\bar a,\bar b)$ transforming
homogeneously under the compact generator $R_0$:
\bea
\label{xitoab}
\xi^0 &=& -\frac{1}{8} (a+{\bar a}-b-{\bar b})\ ,\quad
\xi^1 = \frac{1}{24} i (a-{\bar a}+3 b-3{\bar b})\\
\txi_1&=& \frac{1}{8} (a+{\bar a}+3 (b+{\bar b}))\ ,\quad
\txi_0= \frac{1}{8} i(a-{\bar a}-b+{\bar b})\\
R_0 &=& \frac{i}{2} \left( a\pa_a-\bar a\pa_{\bar a} -3 b {\pa_b}
+ 3 \bar b\pa_{\bar b} \right)
\eea
The holomorphic moment maps for the generators in the compact basis
then read
\begin{eqnarray}
\label{jhamg2}
\moment{J}_3&=&\frac{i\,e^{2t}}{27648}
\left[ 1728 (1+\alpha^2)-a^2 {\bar a}^2-4 ( a^3 b+  {\bar a}^3
{\bar b}) -1296 b {\bar b}+27 b^2 {\bar b}^2-18 a {\bar a} (b {\bar b}-8)
\right] \nn\\
\moment{S}_3&=&\frac{i\,e^{2t}}{9216}
\left[ 1728(1+\alpha^2)-a^2 {\bar a}^2-4 (a^3 b+
{\bar a}^3 {\bar b})+432 b {\bar b}+27 b^2 {\bar b}^2-6 a {\bar a}
(3 b {\bar b}+8)\right]\nn\\
\moment{J}_+&=&-\frac{i}{1728\sqrt{2}} e^{2t}
\left[2a^3+9 a {\bar a} {\bar b}+27 {\bar b}
(8+8 i \alpha-b {\bar b})\right]\ ,\quad
\nn\\
\moment{S}_+&=&\frac{i}{576\sqrt2} e^{2t}  \left[ a {\bar a}^2+6 a^2
   b+9 {\bar a} (8+8 i \alpha+b {\bar b})\right]\ ,\quad
\nn
\end{eqnarray}
\begin{eqnarray}
\moment{J}_{\frac12,\frac32}&=&
-\frac{e^{2t}}{6912}  \left[1728( 1- \alpha^2) +3456 i \alpha+a^2
   {\bar a}^2+4 a^3 b+4 {\bar a}^3 {\bar b}+18 a {\bar a} b {\bar b}-27 b^2
   {\bar b}^2\right]\ ,\nn\\
\moment{J}_{\frac12,\frac12}&=&\frac{i\,e^{2t}}{288\sqrt6}
\left( a^2 {\bar a}+6 {\bar a}^2 {\bar b}
+9 a (-8-8 i \alpha+b   {\bar b})\right)\ ,\quad
\moment{J}_{\frac12,-\frac12}
=\frac{i}{24 \sqrt{3}} e^{2t} \left( a^2+3 {\bar a} {\bar b}\right)\ ,\nn\\
\moment{J}_{\frac12,-\frac32}&=&\frac{i\,e^{2t}}{864 \sqrt{2}}
\left( 2 a^3+9 a {\bar a}
   {\bar b}-27 {\bar b} (8-8 i \alpha+b {\bar b})\right)\ .\quad
\end{eqnarray}
Substituting \eqref{ehamg2} in \eqref{g2X}, a tedious computation shows
that the holomorphic moment map, seen as an element of $\fg_{\IC}$, is
nilpotent of degree 2. Thus, the Swann space $\cS$ is isomorphic (at least locally)
to the complexified minimal nilpotent orbit of $G_{2(2)}$. In particular,
we note the classical identities
\bse
\bea
\moment{E}_{p^1}^2 + \sqrt{3} \moment{E}_{p^0}
\,\moment{E}_{q_1} - 2\sqrt{2} \moment{E}\, \moment{Y}_+  &=&0
\label{haeg21clas}\\
\moment{E}_{q_1}^2 - \sqrt{3} \moment{E}_{q_0}
\,\moment{E}_{p^1} - 2\sqrt{2} \moment{E}\, \moment{Y}_-  &=&0
\label{haeg22clas}\\
3 \moment{E}_{p^0} \moment{E}_{q_0}
+ \moment{E}_{p^1} \moment{E}_{q_1} - 4 \moment{E}\, \moment{Y}_0 &=&0
\label{haeg23clas}\\
9 C_2(\moment{L}) - C_2(\moment{R})
= 9 C_2(\moment{J}) - C_2(\moment{S}) &=& 0\ . \label{su2su2minclas}
\eea
\ese

\subsubsection{Some finite $K$-types \label{kdecg2}}

We now discuss the finite $K$-types of the principal series representation.
The full $K$-type decomposition can be obtained using Frobenius 
reciprocity\footnote{The Frobenius reciprocity theorem implies
that the number of occurrences of a $K$-type $\sigma$ in the
principal series we consider is equal to the number of singlets
in the decomposition of $\sigma$ under $K\cap M$, where $M$ is the
centralizer in $K$ of $A$, appearing in the Langlands decomposition $P=MAN$;
see e.g. \cite{MR1880691}.};
here we are interested in the explicit construction of some specific states.

As in Section \ref{kdecsu21}, we can construct the spherical vector $f_K$
by noting that the Heisenberg parabolic
subgroup $P$ acting on the two-form $e_4 \wedge e_6$ gives a 1-dimensional
representation, where $e_{i}$ is the $i$-th row of the matrix in \eqref{g2X},
and $K$ preserves the norm of each row; so we take
\be
f_K = \norm{e_4 \wedge e_6}^{-k/2}
\ee
where
\be
\norm{e_4 \wedge e_6}^2 =
\tilde I_4 + (I_6 + \alpha \hat I_4) -2 [ 1 + I_2 + (\alpha^2-I_4) ]^2
\ee
Here, $I_4$ is the quartic polynomial \eqref{i4g2}, invariant under
$SL(2,\IR)$, while $I_2,\tilde I_4,\hat I_4, I_6$
are homogeneous polynomials of respective degree $2,4,4,6$
in $\xi^I,\txi_I$, invariant under the maximal compact
subgroup $SO(2)\subset SL(2,\IR)$.
In terms of the variables \eqref{xitoab}
\bse
\bea
I_2&=&\frac{1}{12} (a {\bar a}+3 b {\bar b})\ ,\quad
\tilde I_4 =-\frac{1}{27} \left(b a^3+{\bar a}^3   {\bar b}\right)\ ,\quad
\hat I_4 = \frac{i}{27} \left(a^3 b-{\bar a}^3 {\bar b}\right)\\
I_4 &=& \frac{1}{1728} \left[ 4 ( b a^3+ \bar b {\bar a}^3 )
+{\bar a}^2 a^2 - 27 {\bar b}^2 b^2 +18 a {\bar a} b {\bar b}
\right]\\
I_6 &=& -\frac{1}{5832} \left[
2 {\bar a}^3 a^3+
54 {\bar a}^2 a^2 b {\bar b}  +
9 (a \bar a + 3 b \bar b) ( b a^3+ {\bar b} {\bar a}^3  ) \right]
\eea
\ese
Using \eqref{jhamg2}, one easily recognizes that
the spherical vector is related to the quadratic Casimirs of
$SU(2)_J$ and $SU(2)_S$ simply by
\be
e^{-k t}\,f_{K} = \left[ C_2(\moment{J}) \right]^{-k/4}
= \left[ C_2(\moment{S})/9 \right]^{-k/4}
\ee
which makes the $K$ invariance manifest.

Other $K$-types may be obtained by acting on $f_{K}$
with the non-compact generators $J_{\pm \frac12,\pm \frac32}$.
For example, a set of $J_+$-highest weights are obtained by acting with
symmetrized products of the raising operators $J_{\frac12,\pm \frac32},
J_{\frac12,\pm \frac12}$, which transform in the spin-3/2 representation of $SU(2)_S$ 
(antisymmetric combinations of these operators lead instead to
$J_+$-descendants); these generate the spectrum
\begin{equation} \label{symm-spectrum}
\bigoplus_n \left[\frac{n}{2}\right]_J \otimes S^n\left( \left[\frac{3}{2}\right]_S \right).
\end{equation}
To see this spectrum appearing more explicitly,
note that a class of $(J_+,S_+)$-highest weight states
can be obtained by acting on $f_K$ with
\bse
\bea
M_1 &=& J_{\frac12,\frac32}\\
M_2 &=& J_{\frac12,\frac12}^2 + i \sqrt{3} J_{\frac12,-\frac12}
J_{\frac12,\frac32}\\
M_3&=& J_{\frac12,\frac32}^2 J_{\frac12,-\frac32}+ i\,
J_{\frac12,\frac32}J_{\frac12,\frac12}J_{\frac12,-\frac12}
+ \frac{2}{3\sqrt{3}}
 J_{\frac12,\frac12}^3\\
M_4&=&  J_{\frac12, \frac32}^2 \, J_{\frac12, -\frac32}^2
+ \frac13 J_{\frac12, \frac12}^2 \, J_{\frac12, -\frac12}^2
+2i\, J_{\frac12, \frac32} \, J_{\frac12, \frac12} \,
J_{\frac12, -\frac12} \, J_{\frac12, -\frac32} \nn \\&&+
\frac{4i}{3\sqrt{3}} J_{\frac12, \frac32} \, J_{\frac12, -\frac12}^3 +
\frac{4}{3\sqrt{3}} J_{\frac12, \frac12}^3\, J_{\frac12, -\frac32}
\eea
\ese
which satisfy
\be
\label{condm}
[M_i, M_j]\equiv 0\ ,\quad M_1^2\, M_4 + \frac{4}{27} M_2^3 - M_3^2 \equiv 0
\ee
when acting\footnote{For example, the second equation of \eqref{condm} follows from
$ M_1^2\, M_4 + \frac{4}{27} M_2^3 - M_3^2=
\frac{416}{9}J_{\frac12,\frac32}^2 J_+^2
- 8 J_{\frac12,\frac32}^3 J_{\frac12,-\frac32} J_+
- \frac{56i}{3} J_{\frac12,\frac32}^2 J_{\frac12,\frac12}
J_{\frac12,-\frac12} J_+
- \frac{40\sqrt{3}}{9} J_{\frac12,\frac32} J_{\frac12,\frac12}^3 J_+$.
} on states annihilated by $J_+$.
A generating function for the $(S,J)$ spectrum obtained by acting
with the $M_i$ on $f_K$ is
\be
\label{partsym}
\Tr\ z^{S} q^{2J} = \frac{1-q^6 z^3}
{(1-q\,z^{3/2})\,(1-q^2\,z)\,(1-q^3\,z^{3/2})(1-q^4)}
\ee
where the denominator corresponds to the action of $M_i$
while the numerator reflects the constraint \eqref{condm}.
This generating function indeed agrees with \eqref{symm-spectrum}.\footnote{A 
simple check is obtained by
manipulating \eqref{partsym} to get $\sum_{n,S} (2S+1) q^{2J} = 1/(1-q)^4$, the partition function
of four free bosons.}

As an example, the action of $M_i$ on $f_{K}$ may be easily computed,
\begin{subequations} \label{m-action}
\bea
M_1\,f_{K} &=&  k\, \left( \frac{\moment{S}_3
\moment{J}_{\frac12,\frac32}-\frac{i}{\sqrt3}
\moment{J}_{\frac12,\frac12} \moment{S}_+}
{ C_2(\moment{J})} \right)\, f_{K}\\
M_2\,f_{K} &=& k(3k-2)\, \left( \frac{\moment{J}_+
\moment{S}_+}{ C_2(\moment{J})} \right)\, f_{K}\\
M_3\,f_{K} &=& k (9k^2-4) \,  \left( \frac{\moment{J}_+\,
\moment{J}_{\frac12,\frac32}}{ C_2(\moment{J})} \right)\,f_K \\
M_4\,f_{K} &=& k^2(9k^2-4)\, \left( \frac{\moment{J}_+^2}
{ C_2(\moment{J})} \right)\, f_{K}
\eea
\end{subequations}
where equalities hold up to unimportant numerical factors.

Acting with the $M_i$ on $f_K$ is not sufficient to construct 
all of the ($J_+$, $S_+$) highest weight vectors.  To see this explicitly, it is 
enough to note that there are other operators built 
from $J_{\pm \frac12,\frac32},J_{\pm \frac12,\frac12}$ which commute with $S_+$ and $J_+$,
for example
\be
P_2 = J_{\frac12,\frac32}\,J_{-\frac12,\frac12} - i \, J_{-\frac12,\frac32}\,
J_{\frac12,\frac12}
\ee
which acts on $f_{K}$ by
\be
P_2\, f_K = k (k-2)\, \left(\frac{\moment{S}_+^2}
{C_2(\moment{J})}\right)\, f_K
\ee
Nevertheless, we focus on $(J_+,S_+)$-highest weight states generated by
the action of $M_i$ only, of the form
\be \label{acting-with-m}
f_{p_1,p_2,p_3,p_4} =
\left( \moment{S}_3
\moment{J}_{\frac12,\frac32}-\frac{i}{\sqrt3}
\moment{J}_{\frac12,\frac12} \moment{S}_+\right)^{p_1}\,
\moment{J}_{\frac12,\frac32}^{p_2}\,
\moment{J}_+^{p_3}\,
\moment{S}_+^{p_4}\,
\left[ C_2(\moment{J}) \right]^{-\frac{k}{4}-p_1-\frac12(p_2+p_3+p_4)}
\ee
where $p_i$ are integers related to the $SU(2)_J\times SU(2)_S$ spin by
\be
j=\frac12(p_1+p_2+2p_3)\ ,\quad s=\frac12(3p_1+3p_2+4p_4)\,.
\ee
The set of such states is pictured in Figure \ref{fig-kg2}.

Now we recall some results of \cite{MR1421947}.
First, the $K$-type decomposition of $H^1(\cZ, \cO(-k))$
for $k\geq 3$ is given by
\be
\label{ktypedecqg2}
\bigoplus_{n=0}^{\infty} \left[\frac{k-2+n}{2}\right]_J \otimes S^n\left(\left[\frac{3}{2}\right]_S\right).
\ee
For $k \ge 5$ this gives a quaternionic discrete representation of $G$, 
which we expect to find as a submodule of the principal series.  Indeed,
when $k-2$ is a multiple of $4$, these $K$-types do appear in Figure \ref{fig-kg2}:
they are the ones reached by acting with the $M_i$ on $f_K$ and in particular using
$M_4$ at least $\frac{1}{4}(k-2)$ times.

The paper \cite{MR1421947} also describes a pattern of other representations which would 
be expected to appear as submodules of the principal series for 
general quaternionic groups.  While that analysis does not directly apply to $G = G_{2(2)}$
we describe its naive extrapolation here, and explain how the expected $K$-type decompositions
can be naturally obtained by acting with the operators $M_i$ on an appropriate lowest $K$-type.

The representations in question are supposed to correspond to the orbits of $SL(2,\C)$ acting on 
the complexified spin-$\frac{3}{2}$ representation ($\C^4$).  Their $K$-type decompositions would be
of the form
\be
\bigoplus_{n=0}^{\infty}  \left[\frac{k-2+n}{2}\right]_J \otimes A_n(X)
\ee
where $A(X)=\sum_{n=0}^{\infty} A_n(X)$ is the algebra of functions
on an orbit $X$ of the action of $SL(2,\C)$ on $\C^4$, considered as a 
representation of $SU(2)_S$.  There are three examples:

\begin{enumerate}
\item A representation $\pi'_1$ at $k=1$, corresponding to the orbit $X$ defined by $I_4=0$.
The $K$-type decomposition is obtained by
removing the contribution of the operator $M_4$ from \eqref{partsym}, 
and acting on a highest weight vector with $(J,S) = (0,\half)$ instead 
of $f_K$: this gives
\be
\label{partsymX}
\Tr_{\pi'_1}\ z^{S} q^{2J} = z^\half \frac{1-q^6 z^3}
{(1-q\,z^{3/2})\,(1-q^2\,z)\,(1-q^3\,z^{3/2})}
\ee
This decomposition is multiplicity-free.
$\pi'_1$ has Gelfand-Kirillov dimension 4, and $K$-types
contained in a wedge --- see the black dots on Figure \ref{fig-kg2} (shifted by $\Delta S = \half$).
It appears on the list of unitary representations of $G_{2(2)}$ in \cite{MR1253210}.

\item A representation $\pi'_{2/3}$ at $k=2/3$, corresponding to the
locus where $I_4 = dI_4 = 0$.\footnote{The correspondence between
orbits and representations in this case is somewhat degenerate,
because actually this orbit is the same as the minimal orbit we
discuss next (unlike what happens for other $G$, where the orbit
defined by $dI_4 = 0$ is different from that defined by $d^2 I_4 =
0$).}  Its $K$-type decomposition can be obtained by acting on a
highest weight vector with $(J,S) = (0,2)$ with the operators $M_2$
and $M_1$, but imposing the requirement that $M_2^2 = 0$.  It is
represented in Figure \ref{fig-kg2} by the two leftmost ``Regge
trajectories'' of slope $3$ (shifted by $\Delta S = 2$).

\item A representation $\pi'_{4/3}$ at $k=4/3$, corresponding to 
the locus where $I_4 = dI_4 = d^2 I_4=0$.
This is the  minimal or ``ladder'' representation of
$G_{2(2)}$, with $K$-type decomposition
\be
\label{ladder}
\bigoplus_{m=0}^{\infty}  \left[\frac{m}{2}\right]_J \otimes
\left[\frac{3m+2}{2}\right]_S
\ee
The highest weight states can be obtained by acting on the 
highest weight of a $K$-type with $J = 0$, $S = 1$ by the 
operator $M_1$ only; it is represented in Figure \ref{fig-kg2} by
the leading ``Regge trajectory'' of slope $3$ (shifted by $\Delta S = 1$).

\end{enumerate}

A direct approach to the construction of the submodules $\pi'_k$ inside 
the principal series will be discussed in Section \ref{submodg2}.

\EPSFIGURE{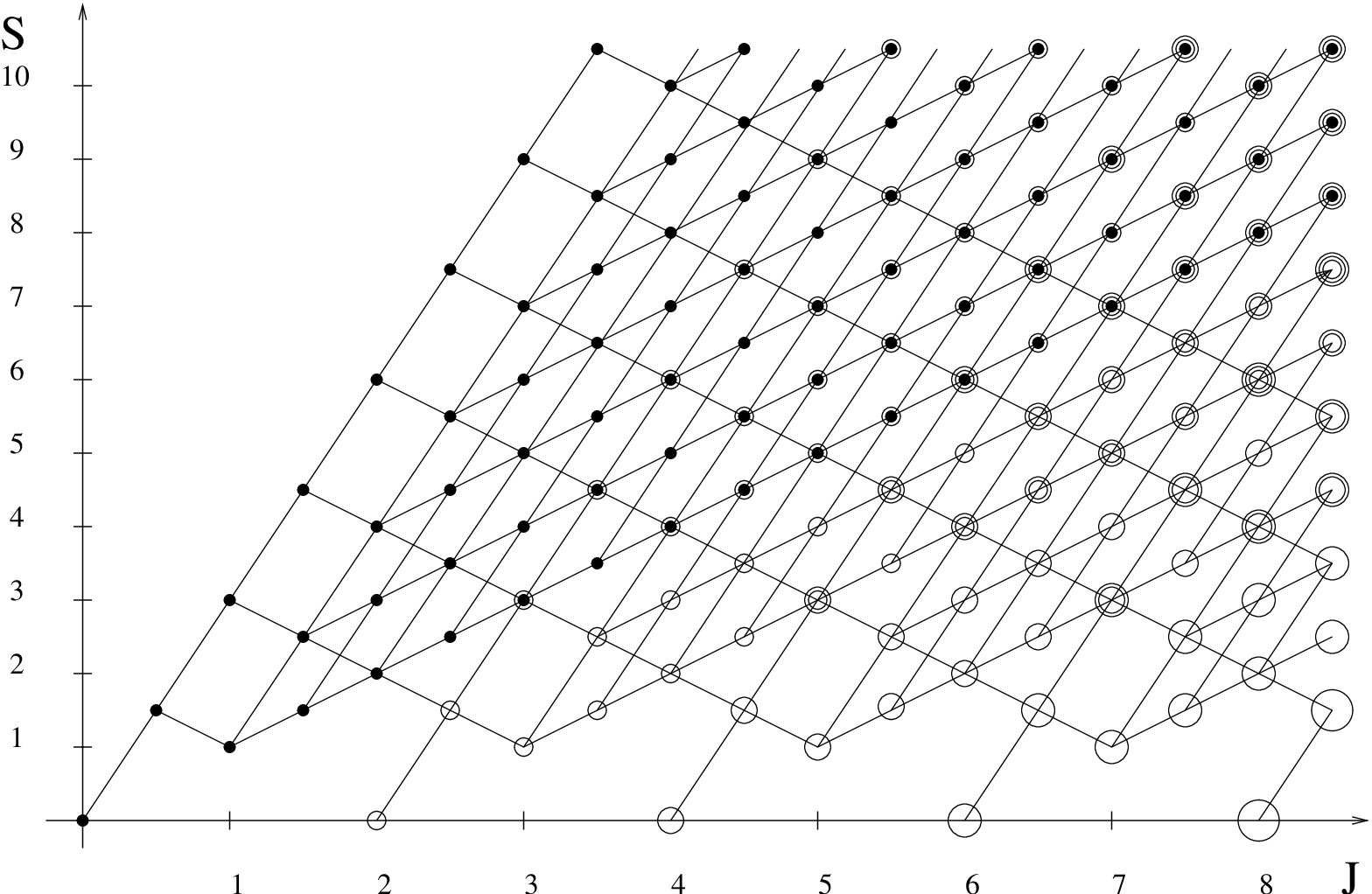,height=11cm,angle=90}{
The states \eqref{acting-with-m} obtained by acting with the $M_i$ on $f_K$. 
Multiplicities are indicated
by the number of concentric circles. The radius of the circle indicates
the number of powers of $M_4$ that need to be applied to $f_K$ 
in order to reach the state. \label{fig-kg2}
This figure also represents the set of highest weight vectors one could obtain by 
acting with the $M_i$ on some other ground state; in this case the labels 
$J$ and $S$ are shifted by the quantum numbers of the ground state,
and in some cases one gets only a subset of the states pictured.  In particular,
the modules $\pi'_1, \pi'_{2/3}, \pi'_{4/3}$ discussed in the text
correspond to the wedge spanned by the black dots,
to the first and second ``Regge trajectories'' of slope $3$, and
to the first ``Regge trajectory'' respectively.
}

% force displaying all floating objects
\clearpage

\subsection{Minimal representation}
The minimal representation of $G_{2(2)}$, of functional dimension 3,
was first constructed in \cite{MR0342049}, and further analyzed in
\cite{MR1421947}. As we just recalled, it can be obtained as a 
submodule of a degenerate principal series representation \cite{MR1421947}.
According to the orbit philosophy, it arises by quantizing the
minimal nilpotent orbit of $G_{2(2)}$, or equivalently by
holomorphic quantization of the Swann space $\cS$.
We start by recalling two different realizations of the minimal
representation by differential operators, the first one acting
on functions $f(y,x_0,x_1)$ of real variables,
the second on functions $f(x,a,b)$ of complex variables.

\subsubsection{Real polarization}
In the real polarization used in \cite{Kazhdan:2001nx},
\bea
\label{kpwg2}
E &=& -\frac{i y}{2}\ ,\quad
Y_0 = -\frac12 \left( x_1 \pa_1 + 3 x_0 \pa_0 +2 \right)\ ,\quad
H = 2 y \pa_y + x_0 \pa_0 + x_1 \pa_1 + 2 \nn\\
E_{p^0} &=& -3\sqrt{3} y \pa_0 \ ,\quad
F_{p^0} = -\frac{2}{3\sqrt{3}} \left( x_0 \pa_y - \frac{i x_1^3}{y^2} \right)
\nn\\
E_{q^0}&=& -\frac{i}{3\sqrt{3}} x_0\ ,\quad
F_{q^0} = -\frac{2}{3\sqrt{3}} \left[ - y \pa_1^3
+27 i \left(2 + x_0 \pa_0 + x_1 \pa_1 + y \pa_y \right)\pa_0 \right]\nn\\
E_{p^1} &=& y \pa_1\ ,\quad
 F_{p^1} = 2 x_1 \pa_y + \frac{4 x_1^2 \pa_1 + 4 x_1}{3y}
+\frac{2i}{9}  x_0 \pa_1^2\nn\\
E_{q^1}&=& i x_1  \ ,\quad
F_{q^1} = -\frac{6 x_1^2}{y}\,\pa_0+ 2i \left(
\frac43 + \frac13 x_1\pa_1 + x_0 \pa_0 + y \pa_y \right) \pa_1 \nn\\
Y_+ &=& \frac{1}{\sqrt{2}} ( i y \pa_1^2 + 9 x_1 \pa_0 )\ ,\quad
Y_- = -\frac{1}{\sqrt{2}}
\left( -\frac13 x_0 \pa_1 - i \frac{x_1^2}{y}\right) \\
F&=&
-\frac{2}{27} x_0 \pa_1^3
-\frac{2}{y^2}  x_1^3 \pa_0
-\frac{2i}{9y}   \left(3  x_1^2 \pa_1^2
+  6 x_1 \pa_1 + 2 \right)
-2 i \left(2 + y \pa_y +x_0 \pa_0 + x_1 \pa_1\right) \pa_y\nn
\eea
It is easy to check that the principal symbols associated to these generators
agree with the holomorphic moment maps \eqref{ehamg2}, upon identifying
\be
\label{vwtoxy}
v^\flat = -2 y \ ,\quad v^0=2 x_0 \ ,\quad v^1=2 x_1 \ ,\quad
w_\flat = -\frac{i}{2} p_y \ ,\quad w_0=\frac{i}{2} p_0\ ,\quad
w_1=\frac{i}{2} p_1
\ee
The minimal representation in the real polarization
is unitary under the inner product
\be
\langle f_1 | f_2 \rangle = \int dy\, dx_0\, dx_1\, f_1^*(y,x_0,x_1)
f_2(y,x_0,x_1)
\ee
Alternatively, one may redefine $q_0=-x_0/\sqrt{27 y}, q_1=x_1/\sqrt{y},
x=\sqrt{y}$ \cite{Pioline:2002qz}, leading to the polarization used
by \cite{Gunaydin:2001bt,Gunaydin:2006vz},
\bea
\label{kpwg2x}
E&=&-\frac{i}{2}x^2 \ ,\quad F=-\frac{i}{2}\pa_x^2 -\frac{3i}{2x}\pa_x
+\frac{1}{18 x^2}I_4 \nn\\
E_{p^0}&=&x \pa_{0} \ ,\quad
F_{p^0}=\frac{1}{9x}\left( 2i\sqrt3 q_1^3 - 9 q_0 ( q_0\pa_0+
q_1 \pa_1)\right)+ q_0 \pa_x \nn\\
E_{p^1}&=&x \pa_{1} \ ,\quad F_{p^1}=
\frac{1}{3x}\left(4 q_1 +  q_1^2 \pa_1
+2 i \sqrt3 q_0 \pa_1^2-3 q_0 q_1 \pa_0\right)+q_1\pa_x\nn\\
E_{q^0}&=&i x q_0     \ ,\quad F_{q_0}=
\frac{1}{9x}\left(2\sqrt3 \pa_1^3+9i (
q_0\pa_0^2+q_1\pa_0\pa_1+3 \pa_0)\right)+i \pa_0\pa_x \nn\\
E_{q^1}&=&i x q_1     \ ,\quad F_{q_1}=
\frac{1}{3x}\left(5i\pa_1-i q_1 \pa_1^2+2\sqrt3
q_1^2\pa_0+3iq_0\pa_0\pa_1\right)+i \pa_1\pa_x\nn\\
Y_+&=&\frac{i}{\sqrt2}\left( \pa_1^2 + i \sqrt3 q_1 \pa_{0}\right) \ ,\quad
Y_-=\frac{1}{\sqrt2}\left(-i q_1^2 + \sqrt3 q_0 \pa_1\right)\nn\\
Y_0&=&-\frac12( 3 q_0\pa_{0}+q_1\pa_{1}+2)\ ,\quad
H= x\pa_x + 2 
\eea
where
\be
I_4(q_\Lambda,\pa_\Lambda) = 4\sqrt3 q_0 \pa_1^3+4\sqrt3 q_1^3\pa_0
-3i q_1^2\pa_1^2 + 18 i q_0 q_1 \pa_0 \pa_1 + 9 i q_0^2 \pa_0^2
+3iq_1\pa_1+27 i q_0 \pa_0-8i
\ee
In accord with irreducibility, the quadratic Casimir evaluates to
a constant
\be
C_2 = -\frac{14}{9}
\ee
which coincides with that of the quaternionic discrete representation for
$k=4/3$ or $k=-14/3$. In addition, the minimal representation
is annihilated by the Joseph ideal, e.g.
\bse
\bea
4 Y_+^2 + 3\sqrt{3} ( E_{p^1} F_{q_0} - E_{p^0} F_{q_1} )&=&0 \\
E_{p^1}^2 + \sqrt{3} E_{p^0} \,E_{q_1} - 2\sqrt{2} E\, Y_+  &=&0
\label{haeg21}\\
E_{q_1}^2 - \sqrt{3} E_{q_0} \,E_{p^1} - 2\sqrt{2} E\, Y_-  &=&0
\label{haeg22}\\
3 E_{p^0} E_{q_0} + E_{p^1} E_{q_1} - 4 E\, Y_0 &=&0
\label{haeg23}
\eea
\ese
These last three identities can be shown to imply the
holomorphic anomaly
equations satisfied by the topological amplitude in the one-modulus model
with prepotential $F=-(X^1)^3/X^0$ \cite{Gunaydin:2006bz}. Other identities
will be discussed in Section \ref{decsu2su2} below.

For later reference, we note that the vector
\be\label{g2fp}
f_P(y,x^0,x^1)=  (x^0)^{-2/3} \exp\left[ -i \frac{(x^1)^3}{y x^0} \right]
\ee
transforms as a one-dimensional representation of the Heisenberg parabolic
$P$ (more specifically, it is annihilated by $Y_+,Y_0,Y_-,F_{p^I},F_{q_I},F$
and carries charge $4/3$ under $H$). In particular, it is invariant
under the Weyl reflection $S$ with respect to the root $E$
and therefore under Fourier transform over $x^0,x^1$.
The power $(P^0)^{-2/3}$ is
consistent with the semi-classical analysis in \cite{Pioline:2005vi}.

\subsubsection{Complex polarization}

It is also useful to consider a different realization of the
minimal representation \cite{Gunaydin:2004md}, on functions
$f(x,a^\dagger,b^\dagger)$ such that
\be
R_0 = i \left(  \frac12 a^\dagger a - \frac32 b^\dagger b - \frac12 \right)
\ ,\quad
R_+ = \frac{i}{\sqrt{2}} \left( a^\dagger a^\dagger + \sqrt{3}~ a b \right)
\ ,\quad
R_- = \frac{i}{\sqrt{2}} \left( a^2  + \sqrt{3}~ a^\dagger b^\dagger \right)
\nn\ee
\be
L_0 = \frac{i}{4} ( x^2 - \pa_x^2 )+ \frac{i}{9 x^2} I_4(a,b) \ ,\quad
L_\pm = -\frac{i}{4\sqrt{2}}
\left[ (\pa_x \mp x)^2 +\frac{4}{9 x^2} I_4 \right]
\ee
Here
\be
I_4(a,b) = -\sqrt{3} \left( a^3 b + a^{3\dagger} b^{\dagger} \right)
-\frac92 N_a N_b - \frac34 N_a^2 + \frac94 N_b^2 - 3 N_a - \frac{41}{16}
\ee
where $a,b,a^\dagger,b^\dagger$ are bosonic oscillators with
\be
[a,a^\dagger]=1\ ,\quad [b,b^\dagger]=1\ ,\quad
N_a=a^\dagger a, \quad N_b = b^\dagger b
\ee
so that $a\equiv \pa_{a^\dagger},b\equiv\pa_{b^\dagger}$;
we shall refer to this presentation as the ``upside-up'' complex
polarization. Alternatively, we could consider functions
$f(x,a^\dagger,b)$ and represent $b^\dagger=-\pa_{b}$: this
will be termed as the ``upside-down'' complex polarization,
and will turn out to be the most convenient one to compute
the lowest $K$-type. Irrespective of the choice of polarization
for the oscillator algebra, the generators in the non-compact basis read
\bea
\label{GPminrepnc}
E= - \frac{i}{2} x^2\ &,&\quad
F = -\frac{i}{2} \left( \pa_x^2 - \frac{4}{9x^2} I_4 \right)\\
E_{p^0} = \frac{i x}{2\sqrt{2}}
\left[\sqrt{3}(a^\dagger  + a) - (b^\dagger +b ) \right] \ &,&\quad
E_{p^1} = -\frac{x}{2\sqrt{2}}
\left[(a-a^\dagger ) +\sqrt{3} (b -b^\dagger) \right] \nn\\
E_{q_1} = -\frac{i x}{2\sqrt{2}}
\left[(a^\dagger  + a) +\sqrt{3} (b^\dagger +b ) \right]\ &,&\quad
E_{q_0} = \frac{x}{2\sqrt{2}}
\left[\sqrt{3}(a^\dagger  - a) - (b^\dagger -b ) \right] \nn
\eea
while the compact generators are given by
\bse
\label{GPminrepc}
\bea
J_+ &=& -\frac{i}{2} \left[
\left( \pa_x - x - \frac{N_a-N_b+\frac12}{x} \right) b
- \frac{2}{3\sqrt{3}} \frac{a^{\dagger 3}}{x} \right]\\
J_- &=& \frac{i}{2} \left[
\left( \pa_x + x +\frac{N_a-N_b-\frac12}{x} \right) b^\dagger
+\frac{2}{3\sqrt{3}} \frac{a^3}{x} \right]\\
S_+ &=& \frac{i\sqrt{3}}{2} \left[
\left( \pa_x - x - \frac{\frac13 N_a + N_b+\frac56}{x} \right) a
- \frac{2}{\sqrt{3}} \frac{a^{\dagger 2}b^\dagger}{x} \right] \\
S_- &=& -\frac{i\sqrt{3}}{2} \left[
\left( \pa_x + x + \frac{\frac13 N_a + N_b+\frac12}{x} \right) a^\dagger
+ \frac{2}{\sqrt{3}} \frac{a^2 b}{x} \right]
\eea
\ese
This presentation of the minimal representation is related to
\eqref{kpwg2} by a Bogoliubov transformation which is the
quantum version of the canonical transformation \eqref{xitoab}.
It can be decomposed into (i) a Fourier transform with respect to $x_0$
\be
f(y,x_0,x_1) = \int dp_0 \, e^{-i p_0 x_0/y}\, f(y,p_0,x_1)
\ee
(ii) a change of variables
\be
p_0 = -\frac{x\sqrt{3}}{18}( w - \sqrt{3} v) \ ,\quad
x_1=-\frac{x}{2} (\sqrt{3} w + v )\ ,\quad y=x^2\ ,
\ee
and finally (iii) a standard Bogolioubov transform
\be
f(x,v,w) = \int \exp\left[
\frac12 [ w^2 + (b^\dagger)^2-v^2 - a^2]
+\sqrt{2} ( a v - b^\dagger w) \right] f(x,a,b^\dagger)
\ee
implementing the change from real to oscillator polarization
$a+a^\dagger=v \sqrt{2}$, $b+b^\dagger=w \sqrt{2}$. One
may check that this sequence of unitary transformations in fact
implements the Cayley transform \eqref{cayleyg2}.

\subsubsection{$K$-type decomposition\label{decsu2su2}}
Let us now discuss the $K$-type decomposition \eqref{ladder} of the minimal
representation in more detail. We note that the correlation between the
two spins is a straightforward consequence of the identity in the
Joseph ideal\footnote{A similar identity holds in the non-compact
basis, $9 C_2(L) - C_2(R) + 2=0$.}
\be
9 C_2(J) - C_2(S) + 2 = 0\ . \label{su2su2min}
\ee
The fact that the lowest $K$-type is a $SU(2)_J$ singlet and
$SU(2)_S$ triplet\footnote{This is an exception among
quaternionic Lie groups; the lowest $K$-type of the minimal representation
is usually a singlet of the Levi factor of $P$, while carrying a non-zero $SU(2)_J$ spin \cite{MR1421947}.}
is less obvious, and will be further discussed below.

We also note that the $K$-type decomposition \eqref{ladder}
is consistent with the known decomposition of
the minimal representation under $SL(3,\IR)$
and  $SU(2,1)$ \cite{MR1253210}: indeed, the minimal
representation is an irreducible representation
in the non-spherical principal series of $SL(3,\IR)$.
The $K$-type decomposition of this representation, worked
out in \cite{MR0383989} (Eq. 7.8),
\be
[1]+[2]+[3]^2+[4]^2+[5]^3+[6]^3+[7]^4+[8]^4+\dots
\ee
is consistent with the diagonal
embedding of the maximal compact of $SL(3,\IR)$ inside
$SU(2)_J \times SU(2)_S$. Under $SU(2,1)$, the minimal
representation decomposes as a sum of three irreducible principal series
representations of $SU(2,1)$ with infinitesimal characters
$(0,\frac13,-\frac13), (\frac13,0,-\frac13),
(\frac13,-\frac13,0)$.
These three representations
correspond to the three supplementary series at $p=q=-2/3$ in
the terminology of \cite{Bars:1989bb}, and transform with different
characters of the center $\Z_3$ of $SU(2,1)$.
The $K$-type decomposition of these representations is given on
Figure \ref{su2fig83}, and is also in agreement with \eqref{ladder}.
We also compute the Casimirs,
\bse
\be
C_2[SL(3,\R)]=-\frac89\ ,\quad C_3[SL(3,\R)]=0
\ee
\be
C_2[SU(2,1)]=-\frac89 \ ,\quad C_3[SU(2,1)]=0\ ,\quad p=q=-2/3
\ee
\ese

\EPSFIGURE{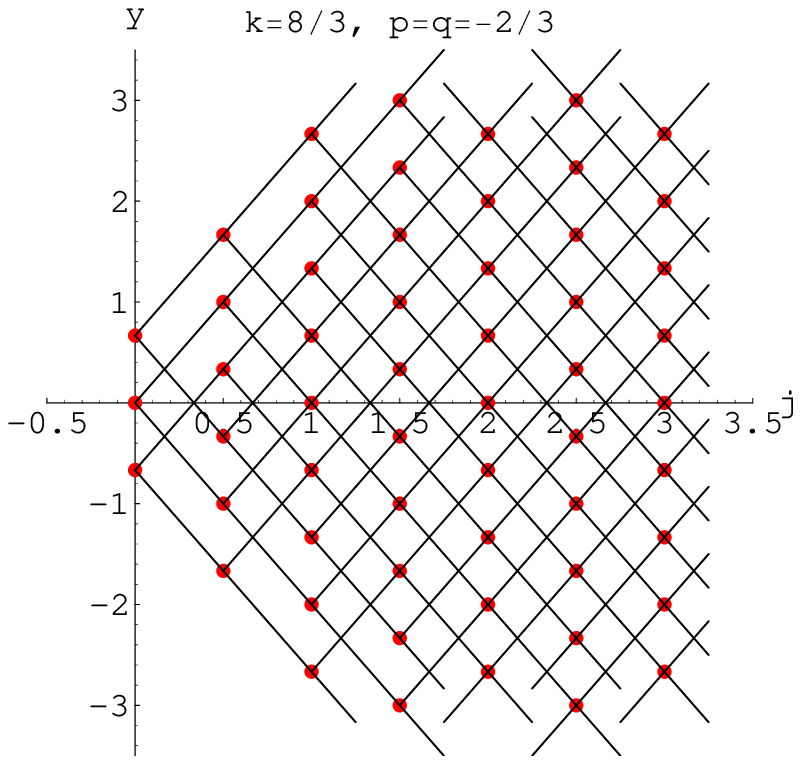,height=11cm}{
$K$-type decomposition of the three supplementary series representations
of $SU(2,1)$ at $p=q=-2/3$, in the $(j,y=2/3s)$ plane.\label{su2fig83}}

Let us now further analyze \eqref{su2su2min}, by rewriting them
in terms of the generators of $G$ acting in the minimal representation.  Using
\bse
\bea
C_2(J) &=& - J_3^2 -\frac12 \left( J_+ J_- + J_- J_+ \right)
=-J_3(J_3 \pm i) - J_\mp J_\pm \\
C_2(S) &=& - S_3^2 -\frac12 \left( S_+ S_- + S_- S_+ \right)
=-S_3(S_3 \pm i) - S_\mp S_\pm
\eea
\ese
we see that a normalizable eigenmode $f$ of $J_3,S_3$ satisfying
the highest weight condition $J_+ \ f = S_+\ f=0$ necessarily has
$(J_3,S_3) \in \frac{i}{2} \IN$.
{} From \eqref{su2su2min}, we have
\be
9 J_3(J_3+i) - S_3(S_3+i) +2 = 4 (2 R_0 + i) (\tilde R_0 + i) = 0
\ee
where
\be
R_0 = \frac12 (3 J_3 - S_3) \ ,\quad \tilde R_0 = \frac12 (3 J_3 + S_3)
\ee
Thus it is either an eigenmode of $R_0=-i/2$, or of $\tilde R_0 = -i$.
The second option is inconsistent with the highest weight condition,
so $R_0=-i/2$. With similar arguments we conclude that
\bse
\bea
\label{jpspr0}
J_+ \ f = S_+\ f=0 \quad &\Rightarrow& \quad
J_3 = \frac{i}{2} m\ ,\quad S_3 = \frac{i}{2} (3m+2)\ ,
\quad R_0 = -\frac{i}{2} \\
J_+ \ f = S_-\ f=0 \quad &\Rightarrow& \quad
J_3 = \frac{i}{2} m\ ,\quad S_3 = - \frac{i}{2} (3m+2)
\ ,\quad \tilde R_0 = -\frac{i}{2} \\
J_- \ f = S_+\ f=0 \quad &\Rightarrow& \quad
J_3 = -\frac{i}{2} m\ ,\quad S_3 =  \frac{i}{2} (3m+2)
\ ,\quad \tilde R_0 = \frac{i}{2} \\
J_- \ f = S_-\ f=0 \quad &\Rightarrow& \quad
J_3 = -\frac{i}{2} m\ ,\quad S_3 = - \frac{i}{2} (3m+2)
\ ,\quad R_0 = \frac{i}{2}
\eea
\ese
In the following, we analyze the consequences of these
equations in various polarizations.

\subsubsection{Lowest $K$-type in the complex polarization}

It turns out that the form of the lowest $K$-type is simplest
in the complex polarization, where the generator $R_0$ takes a simple form
\be
R_0 = i \left(  \frac12 a^\dagger a - \frac32 b^\dagger b - \frac12 \right)\\
\ee
The following linear combinations of \eqref{GPminrepc}
lead to  $x$-independent equations:
\bse
\bea
T_{++} &=& a \sqrt{3} J_+ + b S_+   = -\frac{i}{x}
(a^{\dagger 2} + \sqrt{3} a b) ( R_0 + \frac{i}{2}) \label{spp}\\
T_{--} &=& \sqrt{3} a^\dagger J_- + b^\dagger S_-  =
-\frac23{x} ( a^2 + \sqrt{3} a^\dagger b^\dagger )
(R_0 - \frac{i}{2}) \label{spm} \\
T_{+-} &=&  a^\dagger\ \sqrt{3} J_+ - b S_-\ ,\quad
T_{-+} =  a\ \sqrt{3} J_- - b^\dagger  S_+
\eea
\ese
where the first two lines are consistent with \eqref{jpspr0}.

Now we restrict to a highest weight $S_+ f=0$, which is
a singlet of $SU(2)_J$, i.e. $J_+\ f=J_-\ f$. It turns out
that it is convenient to
work in the ``downside-up polarization'' where $b=\pa_{b^\dagger},
a^\dagger=-\pa_a$. The constraint $R_0=-i/2$ requires
\be
\label{anabd}
f(x,a,b^\dagger) = (b^\dagger)^{-1/3}\ f_1(x,z)\ ,
\quad z=\frac{1}{3\sqrt{3}} \frac{a^3}{b^\dagger}
\ee
The constraint $T_{++}$ is automatically obeyed, however $T_{-+}$
leads to a second order ordinary differential equation in $z$ only,
\be
\label{eqf1}
\left[z \pa_z^2 -2 (z-\frac13) \pa_z + (z+x^2- \frac23)  \right] f_1(x,z) =0
\ee
The solution is
\be
f_1(x,z) = x^{1/3}\ z^{1/6}\ e^z\
\left[ K_{1/3}\left( 2 i x \sqrt{z} \right)\  f_2 (x) +
I_{1/3}\left( 2 i x \sqrt{z} \right)\  \tilde f_2 (x) \right]
\ee
In the following, we assume that normalizability forces $\tilde f_2(x)=0$.
It is one of the drawbacks of the complex polarization that normalizability
is difficult to check -- at any rate, it is straightforward to generalize the
computation below to include both solutions.
Requiring the action of $S_+$ on \eqref{anabd} to vanish, we find
\be
\left[6 x\pa_x - 12 z \pa_z^2 + 2(6z-4) \pa_z + (1-6 x^2) \right]\ f_1(x,z)=0
\ee
Combining this with \eqref{eqf1}, we can produce a first order
partial differential equation
\be
\left[x\pa_x - 2 z\pa_z + \left(2z + y^2 - \frac76\right)\right] f_1(x,z) =0
\ee
whose solution is
\be
f_1(x,z) = x^{7/6} \ e^{-x^2/2 + z }\ f_3(x^2 z)
\ee
This uniquely determines
\be
f_2(x)=x^{-7/6} \ e^{-x^2/2}\ ,\quad f_3(u)
= K_{1/3}\left( 2 i \sqrt{u} \right)
\ee
It is now easy to check that the resulting function
is annihilated
by $S_+, J_+, J_-, J_3$, and is an eigenmode of $S_3=i$. The remaining
two states in the triplet may be obtained by acting with $S_-$.
Altogether, we find that the lowest $K$-type is
\bse
\label{g2minreplkc}
\bea
f_{0,1} &=& ( a x^3 / b^\dagger)^{1/2}\ e^{z-x^2/2}\
K_{1/3}\left( 2 i x \sqrt{z} \right)\\
f_{0,0} &=& 2\ 3^{-1/4}\ x^{3/2} \ a\ (b^\dagger)^{-1}\ e^{z-x^2/2}\
K_{2/3}\left( 2 i x \sqrt{z} \right)\\
f_{0,-1} &=& -\frac{2}{\sqrt{3}} \ ( a x / b^\dagger)^{3/2} \ e^{z-x^2/2}\
K_{1/3}\left( -2 i x \sqrt{z} \right)
\eea
\ese
The highest weights in the higher $K$-types can be obtained
by acting with the raising operators $J_{1/2,3/2}$. For example,
\be
f_{\frac12,\frac52}
=\frac13 x^{3/2}\ a^{1/2}\ (b^\dagger)^{-1}\
e^{z-x^2/2} \left[
K_{\frac43} + i\ 3^{\frac14}\ a^{\frac32} x K_{\frac23}
+ 3 \sqrt{b^\dagger}\ (2x^2-3) K_{\frac13} \right]
\ee
where the argument of the Bessel function is as in $f_{0,1}$.
We note that semi-classically, all $K$-types behave as
\be
\label{semclask}
\exp\left[ - \frac{x^2}{2} + \frac{1}{3\sqrt{3}} \frac{a^3}{b^\dagger}
+ \frac{2}{3^{3/4}} i x \sqrt{ \frac{a^3}{b^\dagger}} \right]
\ee
As explained in \eqref{claslag}, and further at the end of the next subsection,
the argument of the exponential (or ``classical action'') provides
the generating function for a complex Lagrangian cone inside the
hyperk\"ahler cone $\cS$.

\subsubsection{Lowest $K$-type in the real polarization}

According to the above, the lowest $K$-type
should be a triplet of $SU(2)_S$, singlet under $SU(2)_J$. Thus we impose
the conditions
\be
J_+ = J_3 = J_- = S_+ = 0
\ee
Note in particular that the generator $F_{p^0}-E_{p^0}$ of rotations
in the $(y,x_0)$ plane, which was as the start of the KPW solution for
the spherical vector in the split case, no longer annihilates the
state, so we need a different strategy.
Our approach is to find a linear combination of the operators
$J_+, J_3, J_-,S_+$ which involves first order differential operators
only.  The one of interest is
\be
S_3 - 3 J_3 + 3i\sqrt{\frac32} \frac{y}{x_0} ( J_+ -2 J_- - S_+ )
\ee
which allows to rewrite $S_3$
as a first order differential operator in three variables,
\be
S_3 = \alpha_y \pa_y + \alpha_0 \pa_0 + \alpha_1 \pa_1 + \beta
\ee
where
\be
\alpha_y = y \left( i - 9 \frac{x_1}{x_0} \right)\ ,\quad
\alpha_0 = -9 x_1 - \frac{27 i}{2x_0} y^2\ ,\quad
\alpha_1 = \frac{x_0}{3} + \frac{-12 x_1^2 +9 y^2}{2x_0}\ ,\quad
\ee
\be
\beta = \frac{x_1(-6 y + x_1^2 + i x_0 x_1 )}{x_0 y}
\ee
The standard way to solve this equation is to integrate the flow
\be
\frac{dy}{ds}=\alpha_y, \quad
\frac{dx_0}{ds}=\alpha_0, \quad
\frac{dx_1}{ds}=\alpha_1, \quad
\frac{df}{ds} = (\beta - s_3) f
\ee
Using inspiration from the split case \cite{Kazhdan:2001nx},
it is easy find one constant of motion
along the flow,
\be
\label{defz}
z=\frac{\left( y^2 + \frac{2}{27} x_0^2 + \frac{2}{3} x_1^2 \right)^{3/2}}
{ y^2 + \frac{2}{27} x_0^2}
\ee
To find the second constant of motion, and integrate the flow completely,
we go to polar coordinates in the $(y,x_0)$ plane,
\be
\label{defrth}
y =\sqrt{2}\ r \cos\theta, \quad x_0=3\sqrt{3}\ r\sin\theta
\ee
We now change variables to
\be
\label{inspi}
f(y,x_0,x_1)= r^{-2/3} \ h(z,r,\theta)\
\exp\left(-2i \frac{x_0 x_1^3}{y(2 x_0^2+27 y^2)}\right)
\ee
The action of $S_3$ on $h(z,r,\theta)$
is still a first order differential operator, but now
involving only two variables $r,\theta$:
\be
S_3 = -i \cot\theta \ \pa_\theta - \frac{3r}{\sin\theta}
\left[\left( \frac{z}{r \sqrt{2}} \right)^{2/3} - 1
\right]^{1/2}\ \pa_r
\ee
Again, the way to solve this equation is to integrate the flow
\be
\frac{d\theta}{ds}=-\frac{i}{\tan\theta}\ ,\quad
\frac{dr}{ds}=  - \frac{3r}{\sin\theta}
\left[\left( \frac{z}{r\sqrt{2}} \right)^{2/3} - 1 \right]^{1/2}\ ,\quad
\frac{dh}{h ds} = s_3
\ee
The ratio of the first two equations gives $d\theta/dr$, and
produces a second constant of motion (in addition to $z$),
\be
t= \frac{(1-i\ e^{i\theta})^2
\left(1+i \sqrt{\left( \frac{z}{r\sqrt{2}} \right)^{2/3} - 1 }\right)}
{(1+i\ e^{i\theta})^2
\left(1-i \sqrt{\left( \frac{z}{r\sqrt{2}} \right)^{2/3} - 1 }\right)}
\ee
The third equation then constrain the $\theta$ dependence to
\be
\label{inspi2}
h(z,r,\theta) = [\cos\theta]^{-i s_3}\ h_1( z , t)
\ee
where $s_3$ is the eigenvalue under $S_3$, which we
independently know to be $s_3=i$ for the lowest $K$-type.

We now express the action of the other generators $J_\pm, S_+, R_0$
on our ansatz
\be
f(y,x_0,x_1)= \cos\theta \,r^{-2/3} \, h_1(z,r)\,
\exp\left(-2i \frac{x_0 x_1^3}{y(2 x_0^2+27 y^2)}\right)
\ee
For this purpose, we express $r,\theta,z,t$ in terms of $y,x_0,x_1$ using
\eqref{defz},\eqref{defrth} and
\be
t=\left[
1+\frac{4}{27} \frac{x_0^2}{y^2} + \frac{2\sqrt{2}}{27}
\frac{x_0}{y^2} \sqrt{2 x_0^2 + 27 y^2} \right]\,
\frac{1+ \frac{ 3 i \sqrt{2} x_1}{\sqrt{2x_0^2+27 y^2}}}
{1-  \frac{ 3 i \sqrt{2}x_1}{\sqrt{2x_0^2+27 y^2}}}\ ,
\ee
and act with the original differential operators. We then set $x_1=0$,
and revert to $z,t$ variables using
\be
x_0 = \sqrt{27 ( r^2 - \frac12 y^2 )}\ ,\quad
r= \frac{z}{\sqrt{2}}\ ,\quad
y= \frac{2\sqrt{t}}{1+t} z
\ee
The last two identities are only valid at $x_1=0$, but that is sufficient
to get the full action on $h_1(z,r)$. We find, in particular,
\bea
R_0 + \frac{i}{2} &=&
-\frac{i \sqrt{t}}{9 (t+1)^2 z^{5/3}}
\left(2t \left(3 t z-3 z+8 \sqrt{t}\right)
  \pa_t +4 \sqrt{t} \left(4 t^2\pa_t^2
-3 z \pa_z \right) -3 (t+1) z\right) \nn\\
J_+ + J_- &= &
\frac{t}{18 (t+1) z^{5/3}} \left(2 \left(3 t z+3 z+4 \sqrt{t}\right)
\pa_t -24 \sqrt{t} z \pa_z \pa_t -3 (t-1) z \right)
\eea
A lengthy but straightforward analysis allows to determine the expansion
of the solution at $z\to \infty$,
\be
\begin{split}
\label{asympz}
h_1(z,t) \sim  & \frac{(1+\sqrt{t})^2}{\sqrt{t}} \,z^{1/6}\,
e^{-z/2}  \left[
1+ \left( \frac{2\sqrt{t}}{3(1+\sqrt{t})^2} -\frac{5}{36} \right) z^{-1}
\right.\\
&\left.
+  \left( -\frac{35\sqrt{t}}{54(1+\sqrt{t})^2} +\frac{385}{2592}
\right) z^{-2}
+  \left( \frac{5005\sqrt{t}}{3878(1+\sqrt{t})^2} +\frac{-85085}{279936}
\right) z^{-3}
+\dots \right]
\end{split}
\ee
This suggests the Ansatz
\be
h_1(z,t) = z^{1/6} \left[ h_2(z) \,
+ \, \frac{(1+\sqrt{t})^2}{\sqrt{t}} \, h_3(z) \right]
\ee
Indeed, we find that all constraints reduce to two ordinary differential
equations for $h_2$ and $h_3$,
\bse
\bea
(1+3 z) h_2 + 6 z (h_2' + 4 h_3' + 2 h_3 ) &=& 0 \\
(4+15 z) h_2 + (54 z-2) h_3 + 12 z ( 2 h_2' + 9 h_3') &=& 0
\eea
\ese
which decouple into
\bse
\bea
(1+18 z+ 9 z^2) h_2 - 36 z (h_2' + z h_2^{''} ) &=& 0 \\
(9 z^2 - 5) h_3 - 36 z^2 h_3^{''} &=& 0
\eea
\ese
Each of them has two independent solutions, but we keep only the one
decaying at $z\to\infty$ in order to ensure normalizability, leading to
\be
h_1(z,t) = \frac{(1+\sqrt{t})^2}{\sqrt{t}} z^{2/3} \, K_{1/3}\left( z/2 \right)
\,+\,\frac{2\sqrt{\pi}}{3}\, z^{1/3}\, U_{\frac76,\frac43,7}(z)
\ee
which correctly reproduces the subleading terms in \eqref{asympz}.
In total, the final answer for the lowest $K$-type in the real
polarization is
\be
\label{g2minreplkr}
\begin{split}
f_{0,1}(y,x_0,x_1)=& (\cos\theta) \,r^{-2/3} \,
\left[ \frac{(1+\sqrt{t})^2}{\sqrt{t}} z^{2/3} \, K_{1/3}\left( z/2 \right)
\,+\,\frac{2\sqrt{\pi}}{3}\, z^{1/3}\, U_{\frac76,\frac43,7}(z) \right]\\
&\times \, \exp\left(-2i \frac{x_0 x_1^3}{y(2 x_0^2+27 y^2)}\right)
\end{split}
\ee
In the semi-classical limit, where $y,x_0,x_1$ are scaled uniformly to
$\infty$, it reduces to
\be
f_{0,1}(y,x_0,x_1) \sim r^{-2/3}\,
 \frac{(1+\sqrt{t})^2}{\sqrt{t}} \,z^{1/6}\, \cos\theta \, e^{-S}
\ee
where $S$ is the ``classical action''
\be
S =
\frac{\left(\frac{2 x_0^2}{27}+\frac{2 x_1^2}{3}+y^2\right)^{3/2}}
{2 \left(\frac{2  x_0^2}{27}+y^2\right)}
-\frac{2 i {x_0} x_1^3}{y \left(2 x_0^2+27 y^2\right)}
\ee
The same reasoning as in Section \ref{claslag} implies that $S$ is the
generating function of a holomorphic Lagrangian cone inside the
hyperk\"ahler cone $\cS$, invariant under
the maximal compact $K$. The precise identification of the variables
$y,x_0,x_1$ in the minimal representation and the complex coordinates
on $\cS$ was given in \eqref{vwtoxy}.
One may indeed check that the holomorphic moment maps of the compact
generators vanish identically on the locus $w_I = \pa_{v^I} S$.

After rescaling $x_0\to x_0\sqrt{27/2}, x_1\to x_1/\sqrt{2}$,
we find that $S$ can be cast in the same form as in Eq. 4.72
of \cite{Kazhdan:2001nx},
\be
S = \frac12 \| X \| -i \frac{x_0}{y\sqrt{x_0^2+y^2}}{\cal F}
\ee
where $X$ is an $Sp(2n+4)$ extension of the
usual $Sp(2n+2)$ symplectic section appearing in the special geometry
description of the $c$-map,
\be
X=\left( y ,x_0,x_1 ; \tilde y, \tilde x^0, \tilde x^1 \right)\ ,\quad
\| X \|^2 = y^2 +  x_0^2 + x_1^2 +\tilde y^2 + (\tilde x^0)^2
+ (\tilde x^1)^2
\ee
\be
\tilde y= \pa_y {\cal F}\ , \quad \tilde x^i =\pa_{x^i} {\cal F}\ ,
\quad {\cal F}=\frac{x_1^3}{3\sqrt{3}\sqrt{y^2+x_0^2}}
\ee
It would be interesting to see whether such a $Sp(2n+4)$ invariant
description also exists for non-symmetric $c$-map spaces, or even for
general hyperk\"ahler manifolds, and to investigate
whether the lowest $K$-type of the minimal polarization bears any
relation to the topological string amplitude of the corresponding
magical supergravity theory as discussed in \cite{Gunaydin:2006bz}.

\subsection{Small submodules of the principal series\label{submodg2}}

In this section, we study the construction of ``small'' submodules of the
principal series representation of $G_{2(2)}$ by imposing constraints directly, 
similarly to the discussion of the minimal representation for $SU(2,1)$ 
in Section \ref{emsu21}.

\subsubsection{The minimal submodule}

We start by considering the submodule of the principal series
annihilated by the ``holomorphic anomaly'' relations
\bse
\label{chaeg2}
\bea
C_+ \equiv E_{p^1}^2 + \sqrt{3} E_{p^0} \,E_{q_1} - 2\sqrt{2} E\, Y_+ &=& 0\\
C_- \equiv E_{q_1}^2 - \sqrt{3} E_{q_0} \,E_{p^1} - 2\sqrt{2} E\, Y_- &=& 0
\eea
\ese
in the Joseph ideal (see \eqref{haeg21},\eqref{haeg22}).
In terms of the differential operator realization \eqref{g2qcg},
these conditions reduce to
\bse
\label{consg2}
\bea
-3 (P^1)^2 +  P^0 Q_1 &=&0\\
(Q_1)^2 + 9 Q_0 P^1 &=&0
\eea
\ese
where
\be
\label{defPQ}
P^I = \pa_{\tzeta_I} - \zeta^I \pa_\sigma \ ,\quad
Q_I = \pa_{\zeta^I} +  \tzeta_I \pa_\sigma\ ,
\ee
are covariant derivatives commuting with $E_{p^I},E_{q_I}$,
analog of $\nabla,\bar\nabla$ in \eqref{defnabla}. For $k=4/3$,
this subspace is invariant under the action of $G$;
indeed, the commutators of the constraints with the lowest negative root $F$
can be rewritten as
\bse
\bea
\left[F,C_+\right] &=& -2 (\sigma + \frac23 D_+) C_0 + \frac23 D_0 C_+ \\
\left[F,C_-\right] &=& -2 (\sigma - \frac29 D_-) C_0 - \frac23 D_0 C_-
\eea
\ese
where $D_0, D_\pm$ is the $SL(2)$ triplet 
\be
D_+ = 3 (\zeta^1)^2+\zeta^0 \tzeta_1\ ,\quad
D_0 = 3 \zeta^0 \tzeta_0 + \zeta^1 \tzeta_1\ ,\quad
D_- = (\tzeta_1)^2-9 \zeta^1 \tzeta_0
\ee
At the level of differential symbols, the constraints \eqref{consg2}
are solved by
\be
Q_0 = - (P^1)^3 / (P^0)^2\ ,\quad Q_1 = 3 (P^1)^2 / (P^0)
\ee
It is therefore natural to go to a polarization where $P^I$ and
$E$ act diagonally,
\be
f(\zeta,\tzeta,\sigma) = \int dp\, dK
\, \exp\left( - i K \sigma -i p^I \tzeta_I \right) \
g(\zeta^I, p^I , K)
\ee
In this polarization, the generators become
\bea
E&=& i K \ ,\quad Y_0=\frac12(3p^0\pa_{p^0}+3\zeta^0\pa_{\zeta^0}+p^1\pa_{p^1}
+\zeta^1\pa_{\zeta^1}+4)\\
E_{p^0}&=&-i(p^0+K\zeta^0)\ ,\quad
E_{q_0}=-\pa_{\zeta^0}-K\pa_{p^0}\nn\\
E_{p^1}&=&-i \sqrt{3} (p^1+K\zeta^1)\ ,\quad
E_{q_1}=-(\pa_{\zeta^1}+K\pa_{p^1})/\sqrt{3}\nn\\
Y_+&=&\frac1{\sqrt2}( 6 i p^1 \zeta^1+p^0 \pa_{p^1}+\zeta^0 \pa_{\zeta^1})\ ,\quad
Y_-=-\frac{1}{3\sqrt2}(9 p^1\pa_{p^0}+2 i \pa_{p^1}\pa_{\zeta^1}+9
\zeta^1\pa_{\zeta^0})\ ,\dots\nn
\eea
while the constraints are
\bse
\bea
C_+&=& i (p^0-K \zeta^0)(\pa_{\zeta^1}-K \pa_{p^1})-3(p^1- K \zeta^1)^2\\
C_-&=& -3i(p^1-K \zeta^1)(\pa_{\zeta^0}-K \pa_{p^0})
+\frac13 (\pa_{\zeta^1}-K \pa_{p^1})^2
\eea
\ese
An invariant set of solutions can be found by restricting to functions
\be
g(\zeta^I,p^I,K)
=(P^0-2K\zeta^0)^{-2/3}\,
\exp\left[i \frac{(P^1-2K\zeta^1)^3}{2 K (P^0-2K\zeta^0)} \right]\,
h(P^I,K)
\ee
where $P^I=p^I+K\zeta^I$. The quasi-conformal action on $f(\zeta^I,\tzeta_I,\sigma)$
 at degree $k=4/3$ restricts to an action on $h(P^I,K)$ given by
\bea
E&=&-\frac{i}{2}y\ ,\quad
E_{p^0}=-\frac{i}{3\sqrt{3}} x^0\ ,\quad E_{q_0} = 3\sqrt{3} y \pa_0 \ ,\quad
E_{p^1}=i x^1\ ,\quad   E_{q_1} = - y \pa_1 \ ,\quad \nn\\
H&=&2y\pa_y+x^0\pa_0+x^1\pa_1+2 \ ,\quad
Y_0 = \frac12 ( 3 x^0\pa_0+x^1\pa_1+2)\nn \\
Y_+&=&-\frac{1}{3\sqrt{2}}\left[ x^0\pa_1+3i\frac{(x^1)^2}{y} \right]\ ,\quad
Y_- = \frac1{\sqrt{2}}\left[ i y \pa_1^2 + 9 x^1 \pa_0 \right]\ ,\dots
\eea
where we have redefined
\be
x^0 = 3\sqrt{3} P^0\ ,\quad x^1= - \sqrt{3} P^1\ ,\quad y=-2 K
\ee
We recognize this as the minimal representation in the
polarization \eqref{kpwg2}, after applying a Weyl reflection
$S$ with respect to
the highest root $E$, which has the effect
of exchanging $E_{p^I}$ with $E_{q_I}$ and $Y_+$ with $Y_-$.

We conclude that the minimal representation can be embedded into
the principal series at $k=4/3$, via
\bea
\label{embmrg2}
f(\zeta^I,\tzeta_I,\sigma) &=& \int dP^I\, dK\,
f_P^*(p^I-K\zeta^I,K)\,h(p^I+K\zeta^I,K)\, e^{-i K (\sigma + \zeta^I \tzeta_I)
-i p^I \tzeta_I}\nn\\
&=& \langle f_P | e^{-\sigma E - \zeta^I E_{q_I}
+\tzeta_I E_{p^I} } | h \rangle
\eea
where
\be
f_P(K,P^0,P^1)=  (P^0)^{-2/3} \exp\left[- i \frac{(P^1)^3}{2 K (P^0)} \right]
\ee
is the $P$-covariant vector introduced in \eqref{g2fp}.

\subsubsection{Intermediate representations}

Above we described two representations $\pi'_1$ and $\pi'_{2/3}$
which we suggested (following the pattern of \cite{MR1421947}) 
should be obtained as submodules of the
principal series for $k=1$ and $k=2/3$, respectively.  Now we verify that
one can indeed find invariant subspaces by imposing the appropriate constraints
at these values of $k$.

Keeping the same notation is in \eqref{defPQ}, we find that
\bea
I_4 &=& \frac{4}{27} P^0 (Q_1)^3 - 4 Q_0 (P^1)^3 + (Q_0)^2(P^0)^2
-\frac13 (Q_1)^2 (P^1)^2+2 Q_0 P^0 Q_1 P^1 \nn\\
&&- 6 Q_0 P^0
- \frac23 Q_1 P^1 -\frac59 E^2
\eea
commutes with all generators for $k=1$, in particular
\be
\left[F,I_4\right] = - 4 E\, I_4
\ee
Imposing the constraint $I_4=0$ on the representation space
of a principal series representation we expect to find the
representation $\pi'_1$ with Gelfand-Kirillov dimension~4.

Similarly, we find that the constraints\footnote{While 
$dI_4=0$ implies $I_4=0$ classically (at the level of the moment maps), 
this is no longer true quantum mechanically; 
in fact $Q_I \pa^I I_4 + P^I \pa_I I_4 
= 4 I_4 - \frac{28}{9} $.}
\bse
\bea
\pa^0 I_4 &\equiv& \pa I_4/\pa Q_0 =
-4 (P^1)^3 + 2 Q_0 (P^0)^2 + 2 P^0 Q_1 P^1 - \frac{32}{3} E P^0 \\
\pa^1 I_4 &\equiv& \pa I_4/\pa Q_1 =
\frac49 P^0 Q_1^2 -\frac23 Q_1 (P^1)^2 + 2 Q_0 P^0 P^1 -\frac{16}{3} E P^1 \\
\pa_1 I_4 &\equiv& \pa I_4/\pa P^1 =
-12 Q_0 (P^1)^2 -\frac23 Q_1^2 P^1 + 2 Q_0 P^0 Q_1 + 4 E Q_1 \\
\pa_0 I_4 &\equiv& \pa I_4/\pa P^0 =
\frac{4}{27} Q_1^3 + 2 Q_0^2 P^0 + 2 Q_0 Q_1 P^1 - \frac43 E Q_0
\eea
\ese
commute with the generators for $k=2/3$. In particular,
\bse
\bea
\left[F,\pa^0 I_4\right] &=&
(D_0-3\alpha) \pa^0 I_4 - 2 D_+ \pa^1 I_4 \\
\left[F,\pa^1 I_4\right] &=&
(\frac13 D_0 - 3 \alpha)\pa^1 I_4 - \frac49 D_+ \pa_1 I_4
- \frac29 D_- \pa^0 I_4
\eea
\ese

\subsection{$3$-step radical and $7$-grading \label{3stepsec}}
All the constructions in this paper so far hinged on the existence of
Heisenberg parabolic $P$ with unipotent radical $N$ of order 2, and the
resulting 5-grading of $\fg$. In section \ref{3stepind}, we construct a
representation of $G$ which relies on the 7-grading by the Cartan generator
$Y_0$ (the vertical axis in Figure \ref{g2root}, left),
corresponding to a parabolic subgroup $P_3$ with
unipotent radical $N_3$ of degree 3. In Section \ref{3stepind2},
we give a new polarization of the minimal representation appropriate
to this parabolic subgroup. Physically, we expect these representations
to become relevant in describing black holes in five dimensions.

\subsubsection{Induced representation from 3-step radical\label{3stepind}}
It is possible to construct a different representation using the
7-grading by the non-compact generator $Y_0$: we decompose
\be
g = p_3 \cdot n_3 = p_3 \cdot
e^{\sqrt{\frac32} ( m F_{q^1} + n E_{p^1} )}\cdot
e^{\sqrt{2} t Y_+}\cdot
e^{\frac{1}{\sqrt{2}} ( k F_{q^0} +l E_{p^0})}
\ee
where $p$ is an element of the parabolic subgroup $P_3$ opposite
to the 3-step nilpotent $N_3$. Acting on functions
$f(m,n,t,k,l)$ transforming with a character $\exp(j Y_0)$ of
$P$, we find
\begin{eqnarray}
F_{q_0} &=& \sqrt{2} \pa_{k}\ ,\quad
E_{p^0} =\sqrt{2} \pa_{l} \ ,\quad
Y_+ = \frac{1}{\sqrt{2}} \pa_t\ ,\quad
E= k \pa_l + m \pa_n\ ,\quad F=l \pa_k + n \pa_m \nn\\
F_{q^1} &=& \sqrt{\frac23} \left[
\pa_{m} - n \pa_t + (3t - m n) \pa_{k} - n^2 \pa_{l} \right]\nn\\
E_{p^1} &=& \sqrt{\frac23} \left[
\pa_{n} + m \pa_t + (3t + m n) \pa_{l} + m^2 \pa_{k} \right]\nn\\
H &=& m \pa_{m} - n \pa_{n} + k \pa_{k} - l \pa_{l}\ ,\quad
Y_0 = -\frac12 ( m \pa_{m} + n \pa_{n})
- t \pa_t - \frac32 ( k \pa_{k} + l \pa_{l}) + j\nn\\
E_{q_1} &=& -\sqrt{\frac23} \left[
m^2 \pa_{m}+(2t+m n)\pa_{n}
+(2m t-k)\pa_t+2m^2 t \pa_{k}
+t(3t+2m n) \pa_{l} - j m \right]\nn\\
F_{p^1}
&=&
\sqrt{\frac23} \left[
(2t-m n)\pa_{m}-n^2 \pa_{n}
+(l-2n t) \pa_t
+t(3t-2mn) \pa_{k}-2 n^2 t \pa_{l} +  j n\right]\nn\\
%\end{eqnarray}
%\begin{eqnarray*}
Y_- &=& \frac{1}{\sqrt{2}} \left[
(k + t m) \pa_{m} +
(l + t n) \pa_{n}
+ ( t^2 + l m- k n) \pa_t \right.\nn\\
&&\left.
+ [m(l m-k n) + 3 k t]\pa_{k}
+ [n(l m-k n) +3 l t]\pa_{l} - 2 j t\right\}\nn\\
F_{p^0} &=& -\sqrt{2} \left\{
\left[ ( k- t m) n  +t^2 \right] \pa_{m}
+n ( l- t n) \pa_{n}
+ t (l -t n) \pa_t \right. \nn\\
&& \left.
+  [ k l + t^2 (t-m n) ] \pa_{k}
+  (l^2-n^2 t^2) \pa_{l}
- j (l - n t) \right\}\nn\\
E_{q^1} &=& -\sqrt{2} \left\{
m ( k- t m) \pa_{m}
+ \left[ ( l- t n) m  +t^2 \right] \pa_{n}
+ t (k -t m) \pa_t\right. \nn\\
&& \left.
+  (k^2-m^2 t^2) \pa_{k}
+  [ k l - t^2 (t+m n) ] \pa_{l}
+ j (l - n t) \right\}
\end{eqnarray}
The quadratic Casimir is
\be
C_2 = \frac13 j(j+5)
\ee
The spherical vector is easy to determine: in the fundamental representation,
the first row of $g$ transforms in a 1-dimensional representation of the parabolic $P$.
Hence,
\be
f_K=\left[ 1+ m^2+n^2 + 2 t^2 + (k-m t)^2
+(l - n t)^2 + (t^2+ k n-l m) \right]^{j/2}
\ee

\subsubsection{A 3-step polarization for the minimal representation\label{3stepind2}}
We now present a third polarization for the minimal representation,
obtained by diagonalizing the generators $Y_-, E_{q^0}, F_{p_0}$,
which correspond to the height 2 and 3 generators of the 3-step nilpotent.
We shall denote by $i Q, i L, i J$ the corresponding eigenvalues.
In the real polarization \eqref{kpwg2x}, the eigenmodes of these
three commuting generators are given by
\be
f_{Q,J,L}(q_0,q_1,x)=
\exp\left[ i \left( \frac{q_1^3}{3\sqrt3 q_0}
+ \frac{\sqrt6}{3} Q \frac{q_1}{q_0} + J \frac{x}{2q_0}  \right)\right]
\delta(q_0 x - L)
\ee
Thus, we set
\be
f(q_0,q_1,x) = \int dQ\,dJ
\exp\left[ i \left( \frac{q_1^3}{3\sqrt3 q_0}
+ \frac{\sqrt6}{3} Q \frac{q_1}{q_0} + J \frac{L}{2q_0^2}  \right)\right]
g(Q,J,L=q_0 x)
\ee
The 3-step polarization is therefore related to the
polarization \eqref{kpwg2} by
multiplying by $e^{F(q_i)}$, and Fourier transforming over $q_1/q_0 \to Q$
and $1/q_0^2\to J/L$. The resulting generators are
\bea
E &=& L \pa_J\ ,\quad  Y_-= i Q\ ,\quad F_{p^0}= i J \ ,\quad E_{q_0}=i L\nn\\
E_{p^0}&=&2i(3\pa_J+ L \pa_{J}\pa_{L} + J \pa^2_{J} + Q \pa_Q \pa_J)
- \frac{L}{2\sqrt{2}} \pa_Q^3\nn\\
E_{p^1}&=&-2\sqrt{\frac{2}{3}} Q \pa_J - i\frac{\sqrt3}{2} L \pa_Q^2 \ \quad
E_{q_1}=-\sqrt{\frac32} L \pa_Q \nn\\
Y_+&=&-\frac{i}{2}\left( Q\pa_Q^2+3 J\pa_Q\pa_J + L\pa_Q \pa_L+7 \pa_Q \right)
\frac{2\sqrt2}{3L} Q^2 \pa_J\nn\\
Y_0&=&-\frac12\left( 2Q\pa_Q+3J\pa_J+3 L\pa_L+7\right)\ ,\quad
H=L\pa_L-J\pa_J+1\nn\\
F_{p_1}&=&-\frac{\sqrt6}{2} J \pa_Q -\frac{4\sqrt3}{9 L} Q^2  \ ,\quad
F_{q_1}=i\frac{\sqrt3}{2} J \pa_Q^2 -
\frac{4}{3}\sqrt{\frac23} \frac{Q^2}{L} \pa_Q -
2\sqrt{\frac23} Q \pa_L - \frac{10}{3}\sqrt{\frac23}
\frac{Q}{L}\nn\\
F_{q_0}&=&\frac{1}{2\sqrt2}\pa_Q^3 +\frac{8\sqrt2}{27 L^2} Q^3 \pa_J
+\frac{2i}{3L} Q^2\pa_Q^2+2i Q\pa_Q\pa_L
+\frac{10 i}{3L} Q\pa_Q+2 i J \pa_J\pa_L \nn\\
&&+2i\left(\frac{1}{L} J\pa_J + L \pa_L^2 + 4 \pa_L\right)+\frac{40i}{91}\nn\\
F&=&J\pa_L + \frac{J}{L} - \frac{4 i \sqrt2 Q^3}{27 L^2}
\eea
It would be interesting to determine the lowest $K$-type in this polarization.
This construction was generalized to all quaternionic-K\"ahler symmetric
spaces in \cite{Gao:2008hw}, where it was argued that the generalized 
topological string amplitude in this polarization computes the degeneracies
of five-dimensional black holes with electric charge $Q$, angular momentum
$J$ and NUT charge $L$.

%\bibliography{combined}
%\bibliographystyle{utphys}

\providecommand{\href}[2]{#2}\begingroup\raggedright\endgroup

\end{document}